\documentclass[npg, manuscript]{copernicus}
\nolinenumbers

\usepackage{hyperref}

\begin{document}

\title{Boosting Ensembles for Statistics of Tails at Conditionally Optimal Advance Split Times}

\Author[1,2][jfinkel@uchicago.edu]{Justin}{Finkel}
\Author[1]{Paul A.}{O'Gorman}

\affil[1]{Department of Earth, Atmospheric and Planetary Sciences, Massachusetts Institute of Technology, 77 Massachusetts Ave, Cambridge, MA 02139, United States}
\affil[2]{Current affiliation: Department of Geophysical Sciences and the Data Science Institute, University of Chicago, 5801 S. Ellis Ave, Chicago, IL 60637, United States}




\runningtitle{BEST COAST}

\runningauthor{Finkel and O'Gorman}

\received{}
\pubdiscuss{} 
\revised{}
\accepted{}
\published{}


\firstpage{1}

\maketitle

\begin{abstract}
    Climate science needs more efficient ways to study high-impact, low-probability extreme events. Ensemble boosting, a form of rare event sampling, offers a novel strategy to extract more information from those occasional simulated events, by perturbing them slightly to probe alternative scenarios immediately instead of waiting many simulation-years for the next event. But statistical accuracy and efficiency depend on the perturbation details. In particular for sudden and transient events like precipitation, performance of boosting depends sensitively on the \emph{advance split time} (AST), which must be long enough before the event to let the ensemble diversify, but not so much as to destroy the event. In pursuit of principled guidelines, we study the effect of AST for sampling tracer fluctuations in a quasigeostrophic flow, an idealized but informative model of midlatitude storm track dynamics. We formulate AST selection as an optimization problem for statistical fidelity with a ground truth. Since ground truth is not known in practice, we propose a proxy objective function of \emph{thresholded entropy}, which rewards ensembles with both a high mean and a large spread. We show that ensemble boosting, when given a well-chosen AST and equipped with methods to estimate probabilities, can accurately sample extremes at long return periods. We furthermore find evidence that thresholded entropy successfully identifies an optimal AST, which is roughly 1-3 eddy turnover timescales in the quasigeostrophic system. Moreover, this proxy captures the \emph{variation} of AST with the target location of the tracer within the flow field, suggesting generalizability to climate models. Large-scale deployment of our method will require further development in adaptive optimization strategies, but our work here is an essential first step for establishing what must be optimized.  

\end{abstract}


\section{Introduction}
\label{sec:introduction}

%

\subsection{Background and motivation}

The outsize impact of extreme weather events, and the need to understand the physical processes that cause them, have driven substantial research interest in the tails of climatological probability distributions. The fundamental challenge is scarcity of data: the historical record is too short to enable robust estimation of extremes rarer than a few times per century, even if the climate were stationary. Different modeling paradigms have developed to confront the issue. The most straightforward is direct numerical simulation (DNS), whereby a climate model is integrated extensively and the extreme events tallied, either as a single long run with stationary forcing \citep[e.g.,][]{Huang2016estimating,OGorman2009scaling} or as an ensemble with non-stationary forcing \citep[e.g.,][]{Thompson2017high,John2022quantifying}. This increases the sample size of extreme events, and reduces the relative error (mean/standard deviation) of an empirical estimate $\hat{p}=\frac{\#\text{ extremes}}{N=\#\text{ total samples}}$, but at a slow rate of $\frac{\sqrt{\mathbb{V}[\hat{p}]}}{\mathbb{E}[\hat{p}]}=\frac{\sqrt{p(1-p)/N}}{p}\sim(Np)^{-1/2}$ for $p\ll1$ \citep{Zuev2015subset}. For example, estimating the probability of a once-per-century storm ($p=0.01$ year$^{-1}$) to within 10\% relative error would take roughly $N=\frac1{0.01}(0.1)^{-2}=10^4$ model years. Most of that simulation time is wasted, just waiting for the next event. 

Rare event sampling (RES) takes a shortcut by repurposing that time to generate more extremes instead, perturbing simulations in a targeted way to favor extreme behavior---with the tradeoff of having to account for bias properly. The need to track probabilities makes rare event \emph{sampling} distinct from \emph{optimization}, i.e., finding the most extreme event possible (or plausible) given physical constraints. That problem has been attacked successfully with constrained optimization algorithms by \citet{Farazmand2017variational} and \citet{Blonigan2019are} for extreme dissipation events in turbulence, and in AI-based weather forecasting by \citet{Whittaker2025constructing} for extreme heat waves. RES can benefit from these techniques, but aims to represent the entire tail \emph{distribution} of extremes with statistical fidelity and not just the maximum. 

RES was first developed for nuclear safety assessment \citep{Kahn1951estimation}, and has since been generalized for diverse applications including structural reliability engineering \citep{Au2001estimation}, molecular dynamics \citep{Zuckerman2017weighted}, and more recently climate and weather \citep[e.g.,][]{Ragone2018computation,Webber2019practical,Baars2021application}.  RES stands in contrast to many other strategies which, in one way or another, replace the expensive physical model with a cheaper approximation. Extreme value theory gives principles for parametrically estimating distributions tails \citep{Coles2001introduction}, but its asymptotic assumptions are not always justified by the finite datasets available, and it is best suited to model univariate distributions (e.g., average temperature over a region) rather than full spatiotemporal processes like storms, although spatial extreme value modeling is steadily progressing \citep{Huser2020advances,Huser2025modeling}. Hybrid statistical/physical models aim to parameterize physical processes rather than the final output statistics, and include linear inverse models \citep{Penland1993prediction}; stochastic weather generators based on analogues or Markov state models \citep{Dool1989new,Ghil2011extreme,Yiou2020simulation,Finkel2023revealing,Pons2024simulating}; empirical downscaling \citep{Vandal2017deepsd,Saha2024statistical,Rampal2025reliable}; statistical (including machine-learned) emulation \citep{Tebaldi2020emulating,Boulaguiem202modeling,Mahesh2024hens1,Mahesh2024hens2}; and generative modeling \citep{Watt2024generative,Sundar2024taudiff,Giorgini2024response}. Machine learning models in particular are proliferating at a dizzying pace, and they can indeed generate new samples at low cost, but their ability to represent physics outside their training data---perhaps the most essential requirement for extreme event modeling---is rightly regarded with suspicion. 

In light of these options, modelers have several tools to help deal with the tradeoff between bias (incorrect physics or limited resolution) and variance (erratic statistical estimates due to limited sample size). The methods are not mutually exclusive, with many interesting synergies possible \citep[e.g., as conceptualized in][]{Lucente2022coupling}, but RES in particular is our focus here as an under-utilized and under-developed strategy to reduce variance without incurring extra bias.  

\subsection{Rare event sampling: promise and pitfalls} 

The generic RES procedure can be summarized as follows. We denote the full state vector by $\mathbf{x}(t)\in\mathbb{R}^d$, and the measure of \emph{severity} by $R^*$: some functional of a trajectory $\mathbf{x}$ that is user-defined, e.g., rainfall averaged over any time interval and spatial region of interest.

\begin{enumerate}
    \item Generate an ensemble of initial conditions to serve as candidate extreme events. Call these ``ancestors''.
    
    \item Select a subset of ancestors with high propensity to produce extreme events (large $R^*$), discarding the others. Apply small perturbations to this subset to generate ``descendants'': new simulations likely to generate large $R^*$ like their parents, but to do so in diverse ways.
    
    \item Adjust the probability weights downward on these selected ancestors, spreading their weight across their descendants to correct for the over-sampling.
    
    \item Repeat steps 2-3 multiple times on the new, extreme-skewed population, until hitting a termination criterion.
    
    \item Estimate any climatological statistics of interest by taking weighted averages of all the simulations.
\end{enumerate}

This template must be specialized for the kind of target event. Diffusion Monte Carlo (DMC), as applied to season-long hot extremes \citep[with a variant called ``GKTL'' after its inventors;][]{Ragone2018computation} and tropical cyclones \citep[with a variant called ``QDMC'' that applies quantile mapping to intensity values;][]{Webber2019practical}, performs the split/kill operation at a chronological sequence of time points, extending the timespan of surviving members while aborting discarded members before they can run to completion---thus, before their $R^*$ values can even be measured. This is appropriate when the propensity for a \emph{future} extreme $R^*$ is well-approximated by some property $R(\mathbf{x}(t))$ measurable at the \emph{present}: for example, if $R^*$ is the mean temperature from June to August, $R(\mathbf{x}(t))=$ (running average temperature from June 1 to $t$) is a good splitting criterion \citep{Ragone2018computation}. If $R^*$ is peak wind speed over a tropical cyclone's lifetime, $R(\mathbf{x}(t))=$ (minimum sea-level pressure in the eye) is a good splitting criterion \citep{Webber2019practical}. 

\begin{figure}
\includegraphics[width=0.98\linewidth,trim={0cm 0cm 10cm 0cm},clip]{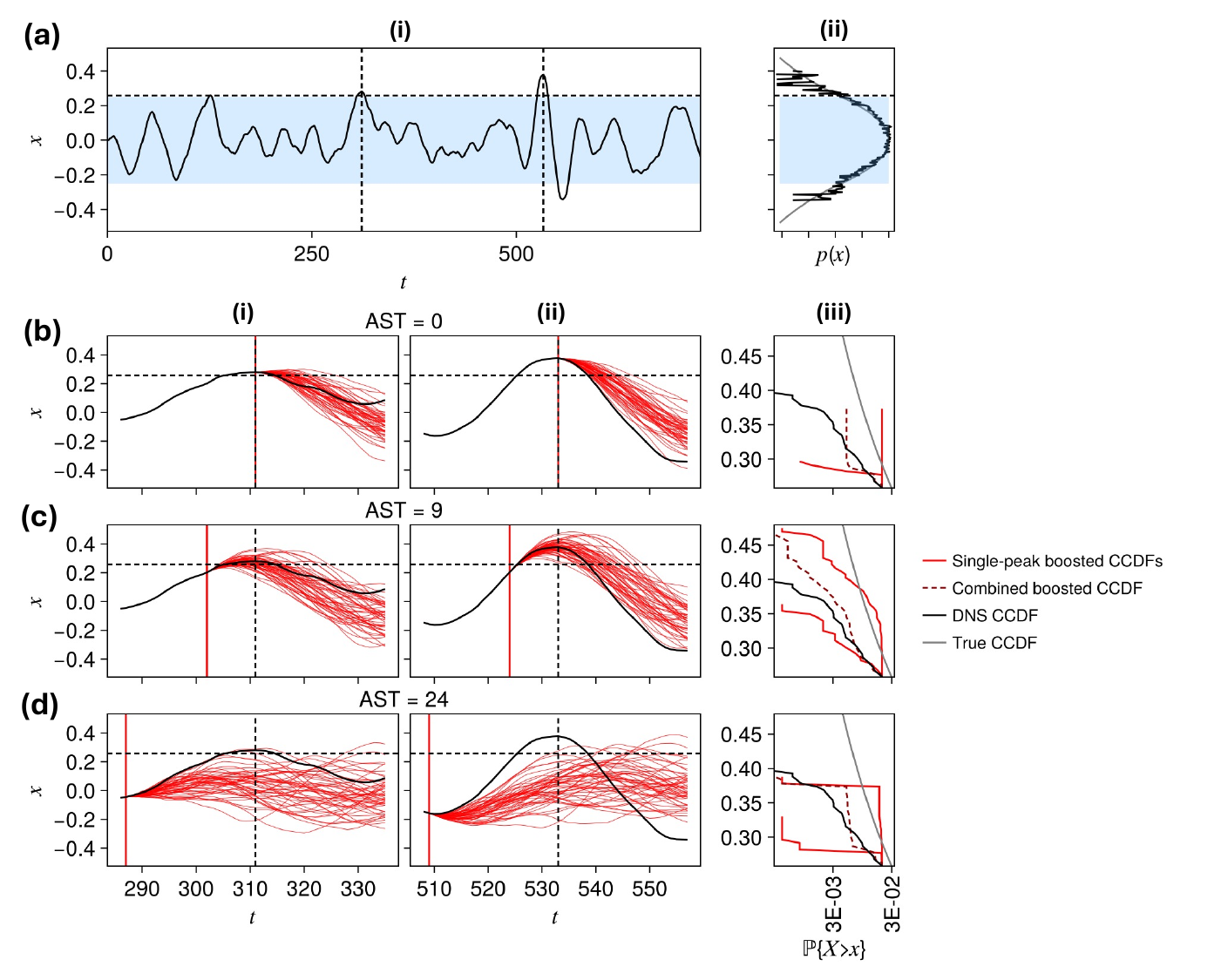}
    \caption{Schematic summarizing the ensemble boosting and tail estimation procedure, using a simple Langevin dynamics with a potential that is quadratic for $x\in(-0.25,0.25)$---the blue-shaded region in (a)---and logarithmic outside this range. Appendix A specifies the system completely. The position variable $X(t)$ exhibits intermittent, transient extremes (a.i) and power law tails $\mathbb{P}\{|X|>|x|\}\sim|x|^{-3.1}$ (a.ii). We set a threshold for severity (horizontal black dashed line) at roughly the minimum probability estimable from a relatively short (duration 1600) timeseries (see the black empirical PDF in a.ii and the black empirical complementary CDFs (CCDFs) in (b,c,d).iii, as compared with the true PDF and CCDF in gray). We then identify the peaks over the threshold (marked by vertical black dashed lines in a.i), and perturb the simulation in advance of these peaks. Three choices of advance split time (AST) are shown in rows b,c,d, marked by vertical red lines, each resulting in ``boosted'' peak ensembles, shown as red curves in (b,c,d).(i,ii) and summarized by CCDFs shown in light red in (b,c,d).(iii). Combining these conditional CCDFs together using the ``MoCTail'' estimator introduced later in Eq.~(\ref{eq:moctail}) gives the dark red dashed line, which is meant to approximate the ground truth (gray line) better than the short DNS alone can do, including by going to higher values of $x$. The intermediate AST (c) is best among the three for this task, and our goal is to formulate and characterize this optimal AST more generally.} 
    \label{fig:schematic}
\end{figure}

But suppose that no good predictor exists. In particular, assume that the severity function $R^*$ of a simulation is the maximum over the event's timespan of a user-defined observable $R(\mathbf{x}(t))$, such as the accumulated rainfall over a small region between $t-1$ day and $t$, which we generically call the \emph{intensity} function. Assume further that no better predictor for $R^*$ is known besides $R$ itself at the present time. In this case, a better choice of RES algorithm might be adaptive multilevel splitting  \citep[AMS;][]{Cerou2007adaptive}, or more general versions such as ``anticipated AMS'' \citep{Rolland2022collapse} and ``trying-early'' AMS (TEAMS), which we previously introduced in \citet{Finkel2024bringing}---itself a special case of subset simulation \citep{Au2001estimation} from engineering---in which every ensemble member runs to completion and produces an actual value of $R^*$, not some proxy for it. Descendants are then spawned from the ancestor at some \emph{advance split time} (AST) $A$ before $R^*$ is achieved, to give them enough time to diversify and perhaps exceed their ancestor's severity, but not so much time to forget their ancestor's special initial conditions. Fig. \ref{fig:schematic} illustrates this tradeoff when selecting AST in the context of a simple stochastic system, namely Langevin dynamics \citep{Pavliotis2014stochastic} with a logarithmic potential which is specified in Appendix A, but the picture alone conveys the essential phenomenon of an \emph{optimal} AST. The existence of a nontrivial (i.e., strictly positive) optimum is obvious when looking at isolated events, but its precise value is subtle to quantify when our purpose relates to \emph{climatological} statistics, i.e., averages over many events.  

There is no general procedure for selecting AST and other hyperparameters, which impedes the application of RES methods to arbitrary target events and models. We have shown empirically in \citet{Finkel2024bringing} the existence of an optimal AST---in the sense of accuracy of long return period estimates---that is roughly approximated by the time until $\frac38$ of error saturation. But this result might be specific to the Lorenz-96 system and a number of choices made in \citet{Finkel2024bringing}, in particular relating to
\begin{enumerate} 
    \item The target variable defining intensity (energy density, $x_k^2$, with site index $k=0$, though for Lorenz-96 all sites are statistically equivalent).
    \item The spatial and temporal scale for averaging the target variable (we simply studied the instantaneous maximum at a single site, $k=0$)
    \item The stochastic parameterization (smooth in space, white in time) 
    \item The metric in which to measure distances between ensemble members (Euclidean distance, $D(\mathbf{x},\mathbf{x}')=\sqrt{\frac1K\sum_{k=1}^K(x_k-x_k')^2}$)
\end{enumerate}

Practitioners working with models more complicated than Lorenz-96 face a vast menu of choices in all four domains, the first two falling under the purview of domain science and the last two falling under algorithm design. If the physical model or the choice of target variable changes, it stands to reason that the choice of metric should also change, and any single prescription of AST (like the $\frac38$-saturation time) is unlikely to work for all cases. Indeed, in our recent application of TEAMS to extremes of temperature and daily precipitation in a general circulation model, we found that the $\frac38$ rule provided some guidance but underestimated the optimal AST for both temperature and precipitation \citep{Finkel2026rare}. Error norms incorporating global information will be less relevant than local norms around the target region, which tend to saturate more slowly \citep{Finkel2026rare}. 

Our primary goal in this study is to establish a general principle for optimizing AST for intermittent extreme events in meteorologically relevant dynamical systems. To balance computational economy with physical realism, we select a system of intermediate complexity between Lorenz-96 and a moist GCM: a 2-layer quasigeostrophic (QG) flow with a passive tracer. Since its original formulation in \cite{Phillips1956general}, the 2-layer QG model has served as a useful paradigmatic minimal model for baroclinic instability and associated jets, waves, and vortices in the atmosphere and ocean. It has been augmented in many ways to study specific processes, for example by \cite{Lapeyre2004role}, who coupled in a moisture component and found the resulting precipitation and latent heating to strongly affect the balance between waves and vortices in the underlying flow. However, even the simpler addition of a \emph{passive} tracer---one without feedback through latent heating---is enough to advance the algorithmic questions we pursue here. Passive tracer dynamics is physically interesting in its own right, as seen by many studies of \emph{intermittency} and heavy-tailed tracer statistics in turbulent flows \citep{Castaing1989scaling,Gollub1991fluctuations,Pumir1991exponential}. In climate science too, extremes of pollution concentration and temperature can be captured partially through passive tracer advection \citep{Bourlioux2002elementary,Neelin2010long,Linz2020framework}.

Our choice of the 2-layer QG model as a test system is thus a major upgrade in physical relevance as well as algorithmic difficulty from Lorenz-96, which resembles QG dynamics only loosely via its Hopf bifurcation structure \citep{Kekem2018wave}. This path up the model hierarchy has been trodden before by \citet{Qi2016predicting,Qi2018predicting}, who added passive tracers to Lorenz-96 and a QG model respectively and studied extreme fluctuations in the tracer's Fourier modes. Also,  \citet{Galfi2017convergence} quantified extreme value statistics---including local and global statistics---of QG wind fields themselves. All these works have inspired and guided this one, but we focus distinctly on the link between \emph{short-time perturbation dynamics} and \emph{long-term climate statistics}.

The QG model has enough ``space'' to explore the effects of all four decision axes listed above on optimal AST. In principle, one can do this with an exhaustive suite of experiments: for every target region (location, size) and every version of stochastic input (e.g., perturbation magnitude and spatial scale) of interest, run TEAMS with a wide range of AST parameters, measure the skill of each AST in matching a reference ground truth distribution, and select the optimal AST. In practice, this exhaustive procedure is not feasible, in part because of the huge number of potential targets, but more  fundamentally because TEAMS' performance is \emph{highly subject to randomness}. Measuring the effect of any parameter change on the algorithm's performance requires many repetitions---several dozen at least---to average out the variability inherent in Monte Carlo. Moreover, other hyperparameters related to ``population management'' exist within TEAMS and other rare event algorithms: the number of initial ensemble members, how many of them to kill and clone at every iteration, and the termination criterion, to name a few. Randomness appears not only as physical forcing, but also in selecting which members to clone, thus interacting tightly with the population hyperparameters. One can think of this as confounding due to sampling bias, which further blurs the imprint of AST itself on performance. 

So instead of using TEAMS for our investigation, we turn to a related method of ensemble boosting \citep{Gessner2021very,Fischer2023storylines}. The idea of ensemble boosting is simple: identify some extremes from an initial climatic timeseries, and re-simulate them with perturbed antecedent conditions to generate unrealized but physically plausible (and possibly more extreme) scenarios. 
By focusing on a limited set of ancestor events to boost, we avoid the additional randomness that occurs in TEAMS as the level is raised and additional ensemble members are stochastically added, which simplifies our investigation. In addition, \citet{BloinWibe2025estimating} has developed an approach to estimate probabilities based on the boosted ensembles, and we have also been developing such an estimator that is introduced below. With the addition of an ability to estimate probabilities, ensemble boosting may now be viewed as an RES algorithm.

We suspect that the optimal AST is closely related to a physically intrinsic quantity that is not particular to a given algorithm. Analogously to Lyapunov exponents, which encode the timescale for small perturbations to double, the optimal AST should encode the timescale for \emph{extreme values of some target variable} to \emph{maximize in variability}. This statement is heuristic, and a primary goal here is to propose some quantities that are very close to the optimal AST and that, like Lyapunov exponents, are intrinsic to the system and don't depend on arbitrary algorithmic choices. We propose and evaluate several candidates, including \emph{entropy} and \emph{expected improvement}: two functionals of ensemble distributions which are drawn from reinforcement learning.

We have three major contributions. First, we develop a new estimator for low probabilities of extreme fluctuations from boosted ensembles, similar to the estimator of \cite{BloinWibe2025estimating} but distinct in the aggregation step. Our approach includes an optional parametric fit of the response function to perturbations (applicable to both estimators), a simple quadratic regression model that imposes regularity on the resulting severity distribution. 
Second, we use the two estimators to measure the quality of a range of ASTs across a range of target events (tracer concentration at different target locations), finding evidence for an entropy-based optimality principle. Third, and most importantly from a practical perspective, we demonstrate that both estimators successfully approximate low probabilities when the ensembles are launched from a good AST, which the optimality principle can help to select efficiently. Our goal here is not to demonstrate a performant rare event algorithm---only to elucidate a necessary ingredient (AST) to be optimized in future algorithms---but even when comparing statistical errors at equal cost, we find (and report at the end of the analysis, in Fig. \ref{fig:mixture_ccdfs}) that our boosted ensembles are already competitive with an equal-cost DNS.

The rest of the paper is organized as follows. Sect. \ref{sec:sampling_estimation_methodology} details the procedure of generating samples and estimating tail statistics, at a model-agnostic level, and proposes several candidate indicators of measuring ensemble dispersion that may help select an optimal AST. Sect. \ref{sec:model} specifies the QG system, its numerical simulation, and its extreme value statistics. Sect. \ref{sec:ensemble_design} specifies the perturbed-ensemble design at a model-specific level.  Sect. \ref{sec:conditional_severity} visualizes some examples of perturbed events, and how the AST selection criteria behave on these examples. Sect. \ref{sec:climatological_severity_distributions} reports the performance of different AST choices, and visualizes the overall ``optimization landscape''. Sect. \ref{sec:conclusion} concludes with an outlook and proposed roadmap for subsequent research---theoretical, algorithmic, and applied. 

\section{Sampling and estimation methodology}
\label{sec:sampling_estimation_methodology}

Our methodology can be separated into three parts, summarized here and expounded in three subsections. For a given target variable and location defining the extreme event, we
\begin{enumerate}
    \item run a relatively short direct numerical simulation (``short DNS''), identify the extreme events within it, and generate a dataset of boosted ensembles for each event at a range of ASTs;
    \item estimate tail distributions, conditional on the event and the AST; 
    \item combine the conditional tails into an unconditional (``climatological'') tail, using the estimators specified below, for a range of ASTs, and select the optimal AST based on the skill of the corresponding tail estimate in reproducing the tail of a ``long DNS''. 

\end{enumerate}

We then display the results of applying this procedure to a range of target locations in the model flow domain.

\subsection{Generating the dataset of boosted ensembles}

There are many design choices in ensemble boosting \citep{Gessner2021very}: how to select extreme events to boost, how many boosts to generate, when to launch them, etc. This subsection details the choices used here. 

\label{sec:generating_dataset_boosted_ensembles}

We run a direct numerical simulation (``short DNS'') $\{\mathbf{x}(t):0\leq t\leq T_{\text{short}}\}$, long enough to generate some extremes but not enough to estimate probabilities smaller than $1/(\epsilon^2T_{\text{short}})=100/T_{\text{short}}$ for a relative error tolerance of $\epsilon=0.1$. The premise of RES, and ensemble boosting, is that the extremes it does generate might have been even worse, perhaps just a butterfly flap away from the more intense extremes one would see with a ``long DNS'' of duration $T_{\text{long}}\gg T_{\text{short}}$. We generate such a long DNS as well to serve as a ground-truth for validation. Following the ensemble boosting methodology laid out in \citet{Gessner2021very,Gessner2022physical,Fischer2023storylines} and \citet{Noyelle2024statistical}, we first identify a threshold $\mu$ with exceedance probability $q(\mu)$ that is moderate enough to estimate precisely with the short DNS. In other words, $\mu$ is the $[1-q(\mu)]$th quantile, or ``$q(\mu)$th complementary quantile''. Equivalently, $q(\mu)$ is the \emph{complementary cumulative density function} (CCDF) of the random variable $R$, evaluated at $\mu$. In line with the \emph{peaks-over-threshold} procedure \citep{Coles2001introduction}, we take cluster maxima of exceedances above $\mu$ as the ``ancestral'' extreme events. Concretely, a cluster maximum is a state from the DNS,  $\mathbf{x}^*=\mathbf{x}(t^*)$, such that 
\begin{equation}
    R^*=R(\mathbf{x}(t^*))=\max\Big\{R(\mathbf{x}(t)):t^*-A_{\text{max}}\leq t\leq t^*+B\Big\}>\mu.
\end{equation}
where $A_{\text{max}}$ and $B$ are buffer times longer than the mixing timescale of the dynamics (i.e., how long two perturbed simulations need to become independent), ensuring that two consecutive events $\big(\mathbf{x}(t_n^*),\mathbf{x}(t_{n+1}^*)\big)$ are genuinely independent from each other. $A_{\text{max}}$ is an upper bound on the ASTs used for boosting. 

We collect all such peaks occurring in the short DNS, 
\begin{align}
    \{\mathbf{x}^*_n=\mathbf{x}(t_n^*):n=1,\hdots,N_{\text{short}}\},
\end{align}
and for a sequence of increasing ASTs $\{A_j:j=1,\hdots,J\}$ bounded between 0 and $A_{\text{max}}$, launch an ensemble of descendants $\{\mathbf{x}^*_{n,j,m}:m=1,\hdots,M_{n,j}\}$ by applying $M_{n,j}$ different perturbations to the DNS at time $t_n^*-A_j$, and running each simulation to time $t_n^*+B$. Note that $M_{n,j}$ could in principle vary between ancestors $n$ and lead times $j$, which is not needed for our exhaustive sweeps in this paper, but certainly would be needed in an ``online'' rare event sampling procedure that iteratively homes in on a subset of the most extreme-ogenic ancestors $\{n\}$ and ASTs $\{j\}$ to draw more samples from. 

A bit more notation helps clarify how the perturbing is done, abstractly at first and concretely in Sect. \ref{sec:model} when we specialize to the QG system. For each $(n,j,m)$, we draw a random sample $\omega_{n,j,m}$ from some sample space $\Omega$. Denoting by $\Phi^{\Delta t}:\mathbb{R}^d\times\Omega\to\mathbb{R}^d$ the flow map that integrates the perturbed dynamics forward by a time interval $\Delta t$, the $(n,j,m)$th descendant's trajectory through state space $\mathbb{R}^d$ can be written
\begin{align}
    \mathbf{x}_{n,j,m}(t)
    =
    \begin{cases}
        \mathbf{x}(t) & \text{ for }t_n^*-A_{\text{max}}\leq t\leq t_n^*-A_j \\
        \Phi^{t-(t_n^*-A_j)}\Big(\mathbf{x}(t_n^*-A_j), \omega_{n,j,m}\Big) & \text{ for }t_n^*-A_j<t\leq t_n^*+B.
    \end{cases}
\end{align} 
In words, the descendant shares its ancestor's past up until the time of perturbation $t_n^*-A_j$, after which it diverges.

There are two main forms of commonly used perturbation. An \emph{impulsive} perturbation is a kick applied at a single time (which is used in ensemble boosting), in which case $\Omega=\mathbb{R}^k$ or $\mathbb{C}^k$, typically with $k\ll d$, and a sample $\omega$ is transformed to spate space via a function $G:\mathbb{R}^k\to\mathbb{R}^d$ (e.g., a low-rank matrix multiplication). Then, the perturbed dynamics can be written $\Phi^{\Delta t}(\mathbf{x},\omega)=\Phi^{\Delta t}(\mathbf{x}+G(\omega))$, where $\Phi^{\Delta t}$ with only one argument is the unperturbed dynamics. We also use the convention that $G(0)=0$, i.e., $\omega=0$ corresponds to no perturbation. 

The other common case is where $\mathbf{x}(t)$ is a stochastic process, e.g., an It\^o diffusion forced by white noise, as we used in \citet{Finkel2024bringing} as well as the schematic in Fig. \ref{fig:schematic}. In that case, $\omega$ is a white noise process sampled at discrete times, whose dimensionality scales with the number of timesteps. In the QG experiments, we adhere to impulsive perturbations for three reasons: it introduces fewer arbitrary parameters, it is less disruptive to the system's intrinsic dynamics, and it keeps the dimensionality of the random space low. If, as we conjecture, even low-dimensional butterfly flaps are sufficient to excite the more extreme fluctuations, it would make deterministic search methods---which should always be preferred over Monte Carlo---more viable.

Following the perturbation, the descendant drifts away from the parent and achieves its own severity $R^*$ (peak of its intensity function $R$) at some time $t_{n,j,m}^*$ possibly different from its ancestor's peak time $t_n^*$:
\begin{align}
    R_{n,j,m}^*
    =
    R(\mathbf{x}_{n,j,m}(t_{n,j,m}^*))
    =
    R_{n,j}^*(\omega_{n,j,m})
\end{align}
where the latter notation emphasizes dependence on $\omega$, while recognizing that each $(n,j)$ induces a different severity function $R^*$ because perturbations may be felt differently depending on the initial condition. 

If the perturbation is small, the descendant's peak time $t_{n,j,m}^*$ will be close to the ancestor's peak time $t_n^*$. However, if the intensity function $R(\mathbf{x}(t))$ tends to oscillate, e.g., with each passing Rossby wave crest, a large-enough perturbation might cause the next wave crest after $t_n^*$ to outgrow the original peak, misappropriating the imposed perturbation to fuel a different event than the original target. Tersely, $t^*=\text{argmax}_tR(\mathbf{x}(t))$ might be a discontinuous function of $\omega$, and $R^*(\omega)$ a non-differentiable function of $\omega$, which is a nuisance for our goal to optimize over $\omega$ and, more importantly,  complicates the causal chain between perturbation and response. We explicitly prohibit this behavior by restricting the range of $t_{n,j,m}^*$ as follows. 
\begin{itemize}
    \item Set an ``argmax drift'' parameter $\delta t^*$ based on physical timescales, e.g., half an oscillation period. Initially set $t_{n,m,j}^*=\text{argmax}\{R(\mathbf{x}_{n,j,m}(t)):t_n^*-\delta t^*\leq t\leq t_n^*+\delta t^*\}$. 
    \item If $t_{n,j,m}^*$ is a local maximum in $R$, then don't change it.
    \item Otherwise, shift $t_{n,j,m}^*$ backward (if at the beginning of the interval) or forward (if at the end of the interval) until it is at a local maximum.
\end{itemize}
Although it is ad-hoc, this adjustment aims to uphold the core idea of ensemble boosting to \emph{augment existing events}, while preserving their basic identity, rather than \emph{discover totally new events}---which may as well be done by extending the DNS. In general this is a nontrivial condition to impose, as multiple spikes in a sequence may be dynamically correlated to each other, but we use only this simple strategy as demonstration.

\subsection{Estimating conditional and climatological probabilities from boosted ensembles}

Assume now there is a probability measure $\mathbb{P}^\Omega$ on $\Omega$ with associated density function $p^\Omega(\omega)$, which might for example place higher weight on smaller kicks. The $\Omega$ superscript will generally relate to statistics over this conditional probability measure, to distinguish it from long-term climatological statistics. A major aim of this paper is to show how they relate to each other. Each ensemble of descendants at each lead time gives rise to its own conditional severity distribution:
\begin{align}
    \label{eq:conditional_severity}
    Q_{n,j}^\Omega(r)
    =
    \mathbb{P}^\Omega\{R_{n,j}^*>r\}
    =
    \int_\Omega \mathbb{I}\{R_{n,j}^*(\omega)>r\}p^\Omega(\omega)\,d\omega,
\end{align}
which can be estimated from the samples $\{R_{n,j,m}^*:m=1,\hdots,M_{n,j}\}$. 
Here \emph{conditional} means starting with a perturbation of the $n$th ancestor's particular initial condition at time $t_n^*-A_j$ and running forward until time $t_n^*+B$. By contrast, we refer to the \emph{climatological} severity distribution as that resulting from a long DNS.

Integrals of the form (\ref{eq:conditional_severity}) arise in many diverse risk analysis tasks, such as reliability engineering, where $\Omega$ often represents wind, waves, or tremors buffeting a built structure \citep{Au2001estimation,Mohamad2018sequential}, and is therefore \emph{high-dimensional}. The default strategy for high-dimensional sampling is vanilla Monte Carlo, whose infamously slow convergence has motivated more efficient workarounds. A particular class of ``variational'' \citep{Dematteis2019extreme,Tong2021extreme} and ``first- and second-order reliability'' methods \citep{Breitung2021sorm} approximate the sampling by constrained optimization, relying on the large-deviation principle that increasingly rare events have a shrinking space of possible pathways, concentrating around a single point of $\Omega$. We could certainly make use of those methods here, but there is a crucial distinction: in our setting, the perturbation space is an arbitrary design choice aiming at an indirect goal (climate estimation), rather than some externally imposed distribution (e.g., a Gaussian process model for ocean bathymetry in \cite{Dematteis2019extreme} and \cite{Tong2021extreme}). Therefore, nothing stops us here from deliberately choosing low-dimensional perturbations instead of high-dimensional ones as in \citet{Ragone2018computation,BloinWibe2025estimating}. This enables numerical quadrature instead of Monte Carlo or elaborate large-deviation approaches, and saves on cost by allowing sample re-use across different input distributions. 

It is possible that higher-dimensional spaces are more effective for exciting extreme fluctuations, which would make the above-cited methodologies very useful for our purpose in future research. They can also be useful when conditional risk estimation (for near-term weather forecasting) is the end goal, as well as the previously-mentioned optimization methods demonstrated in \citet{Farazmand2017variational,Blonigan2019are}, and \citet{Whittaker2025constructing}. But our first goal is to determine whether our chosen low-dimensional kicks can suffice for climatological estimation.

Based on the samples drawn from $\Omega$, we fit a regression model $\widehat{R}_{n,j}^*(\omega;\theta)$ with parameters $\theta$, in our case coefficients for linear and quadratic polynomials. In general $\widehat{R}_{n,j}$ could be a more elaborate parametric model, e.g., a Gaussian process or neural network with learned weights $\theta$, as often used in modern uncertainty quantification \citep{Kabir2018neural,Sapsis2020output,Pickering2022discovering}. Then the integral over $\Omega$ can be estimated, either analytically (if $p$ and $\widehat{R}^*$ take simple enough forms) or numerically by densely filling $\Omega$ with a grid of points, evaluating $\widehat{R}^*$ and $p$ at each point, and taking the inner product of $\mathbb{I}\{\widehat{R}^*>r\}$ with $p$ for any $r$. The result is an estimate $\widehat{Q}^\Omega_{n,j}(r)$ for the conditional CCDF, $\widehat{Q}^\Omega_{n,j}(r)$, obtained by replacing the $R_{n,j}^*$ with $\widehat{R}_{n,j}^*$ in Eq.~(\ref{eq:conditional_severity}). The final step is to estimate the \emph{tail} of the conditional CCDF,
\begin{align}
    \label{eq:conditional_tail_ccdf_ratio}
    Q^\Omega_{n,j}(r;\mu)
    =
    \mathbb{P}^\Omega\{R_{n,j}^*>r|R_{n,j}^*>\mu\}
    =
    \frac{Q_{n,j}^\Omega(r)}{Q_{n,j}^\Omega(\mu)},
\end{align}
which we could do just by putting hats $\widehat{(\cdot)}$ on the $Q$s on the right-hand side. However, this risks dividing by zero, because the fitted function $\widehat{Q}_{n,j}$ may imply zero probability of exceeding the threshold, particularly at long ASTs when descendants have enough time to decorrelate totally with their ancestor. This loss of ancestral ``wisdom'' is a more fundamental problem than the numerical issue of zero denominator, and we address it by implementing a continuous version of the ``accept-reject'' step of the TEAMS procedure in \citet{Finkel2024bringing}. Wherever the descendant severity $\widehat{R}_{n,j}^*(\omega)$ falls below $\mu$, we replace it with the ancestor severity, denoted $R_n^*$ (with no second subscript):
\begin{align}
    \widehat{Q}_{n,j}^\Omega(r;\mu)
    &:=
    \int_\Omega
    \left\{
    \begin{array}{ll}
        \mathbb{I}\{\widehat{R}_{n,j}^*(\omega)>r\}
        & \text{if }\widehat{R}_{n,j}^*(\omega)>\mu \\
        \mathbb{I}\{R_n^*>r\}
        & \text{ otherwise}
    \end{array}
    \right\}
    p^\Omega(\omega)\,d\omega
    \\
    &=
    \int_{\{\widehat{R}_{n,j}^*(\omega)>\mu\}}\mathbb{I}\{\widehat{R}_{n,j}^*(\omega)>r\}p^\Omega(\omega)\,d\omega
    +
    \int_{\{\widehat{R}_{n,j}^*(\omega)\leq\mu\}}\mathbb{I}\{R_n^*>r\}p^\Omega(\omega)\,d\omega
    \\
    &=
    \int_\Omega\mathbb{I}\{\widehat{R}_{n,j}^*(\omega)>r\}p^\Omega(\omega)\,d\omega
    +
    \mathbb{I}\{R_n^*>r\}\int_{\{\widehat{R}_{n,j}^*(\omega)\leq\mu\}}p^\Omega(\omega)\,d\omega
    \\
    &=
    \widehat{Q}_{n,j}^\Omega(r)+\mathbb{I}\{R_n^*>r\}\big[1-\widehat{Q}_{n,j}^\Omega(\mu)\big] \label{eq:conditional_severity_estimate}
\end{align}
($\widehat{Q}_{n,j}^\Omega(r)=0$ when $\widehat{Q}_{n,j}^\Omega(\mu)=0$ since $Q_{n,j}^\Omega$ is decreasing, hence the two terms in the last expression correspond to the two cases). 

This estimator can be extended to other expectations of interest conditional on the target variable being extreme. Denote by $\Phi[X_{n,j}]=:\Phi_{n,j}$ a generic function of the trajectory $X_{n,j}$ launched at AST $A_j$ ahead of ancestor $n$, such as time-averaged wind speed or air temperature. It is actually a random variable (a function of $\omega$, $\Phi_{n,j}(\omega)$) and its mean can be estimated by multiplying each $\mathbb{I}\{R^*(\omega)>\mu\}$ with $\Phi(\omega)$ to obtain
\begin{align}
    \mathbb{E}[\Phi_{n,j}|R^*_{n,j}>\mu]\approx\widehat{\Phi}_{n,j}=\int_\Omega\Phi_{n,j}(\omega)\mathbb{I}\{\widehat{R}_{n,j}^*(\omega)>\mu\}p^\Omega(\omega)\,d\omega+\Phi_{n,j}(0)\big[1-\widehat{Q}_{n,j}^\Omega(\mu)\big].
\end{align}
The first term collects statistics of the part of the $W$-disc that contributes to the tail $\{R^*>r\}$, and the second term just moves the remaining (``rejected'') probability mass back onto the ancestor at $\omega=0$. We don't explore the properties of this estimator for different $\Phi$s, but note it could be important to an applied study with RES.

Another heuristic way to justify the accept-reject expression for $\widehat{Q}_{n,j}(r;\mu)$ in Eq.~(\ref{eq:conditional_severity_estimate}) is to stipulate that we care about approximating \emph{only the extreme part of the boosting distribution}, i.e., those $\omega$s near enough to 0 that $R^*(\omega)>\mu$, excluding the descendants bound to fall below $\mu$.
We re-allocate the probability mass in the ``non-extreme'' region of the disc (where $R^*(\omega)\leq\mu$) to the very center of the disc (the ancestor, where $R^*>\mu$ by construction). This rearrangement 
ensures that $\widehat{Q}^\Omega(\mu)$ is close to 1, justifying a Taylor series expansion in $1-\widehat{Q}^\Omega(\mu)$
\begin{align}
    Q_{n,j}^\Omega(r;\mu)
    &=
    \frac{Q_{n,j}^\Omega(r)}{Q_{n,j}^\Omega(\mu)}
    \\
    &=
    \frac{Q_{n,j}^\Omega(r)}{1-[1-Q_{n,j}^\Omega(\mu)]}
    \\
    &\approx
    Q_{n,j}^\Omega(r)+[1-Q_{n,j}^\Omega(\mu)]Q_{n,j}^\Omega(r)
    \\
    &\approx
    \widehat{Q}_{n,j}^\Omega(r)+\big[1-\widehat{Q}_{n,j}^\Omega(\mu)\big]\mathbb{I}\{R_n^*>r\}\\
    &=:
    \widehat{Q}_{n,j}^\Omega(r;\mu)
\end{align}
The crux of our hypothesis is that these conditional distributions from boosting can be aggregated across ancestors to approximate the climatological distribution $Q^\Theta(r,\mu) = P(R^\ast >r | R^\ast > \mu)$, where $\Theta$ is used to 
denote the ground truth that
would be obtained from a long DNS. We specifically propose to aggregate the conditional CCDFs as a uniform mixture over ancestors, selecting one representative AST $A_{j_n}$ from each ancestor $n$ to best represent its alternate realities according to some selection rule (different rules will be evaluated thoroughly for the QG system in Sect. \ref{sec:climatological_severity_distributions}). We write the mixture as
\begin{align}
    \label{eq:moctail}
    \widehat{Q}^{M}(r;\mu)=\frac1{N_\text{short}}\sum_{n=1}^{N_\text{short}}\widehat{Q}_{n,j_n}^\Omega(r;\mu),
\end{align}
and call it the ``MoCTail'' estimator of $Q^\Theta(r,\mu)$, for ``Mixture of Conditional Tails.'' 

The recent works \citet{Noyelle2024statistical} and \citet{BloinWibe2025estimating} formulate a different estimator, which makes for an interesting comparison. Rather than summing $N_{\text{short}}$ tail CCDFs, each approximating a ratio of the form (\ref{eq:conditional_tail_ccdf_ratio}), they construct a single ratio by summing $N_\text{short}$ numerators and $N_\text{short}$ denominators.Translated into our own notation, this becomes
\begin{align}
    \label{eq:popt}
    \widehat{Q}^P(r;\mu)
    =
    \frac
    {\sum_{n=1}^{N_\text{short}}\widehat{Q}_{n,j_n}^\Omega(r)}
    {\sum_{n=1}^{N_\text{short}}\widehat{Q}_{n,j_n}^\Omega(\mu)}.
\end{align}
We call this the ``PoPTail'' estimator of $Q^\Theta(r,\mu)$, for ``Pool of Perturbed Tails.'' \citet{BloinWibe2025estimating} do not model $R^*(\omega)$ parametrically, but instead use a standard Monte Carlo estimate $\widehat{Q}_{n,j}^\Omega(r)=$ (fraction of descendants exceeding $r$), which is probably necessary for their high-dimensional perturbations. However, we can convert the PoPTail estimator to our parametric version just by thinking in terms of CCDFs, hence the formulation in Eq.~(\ref{eq:popt}). The more important difference is that PoPTail avoids the potential degeneracy $\widehat{Q}^\Omega(\mu)=0$ by ``pooling'' non-extreme descendants together with extreme ones in the denominator.

One could argue for either estimator based on the validity of its underlying assumptions which are challenging to rigorously verify. Here we adopt a more openly empirical perspective in testing the skill of both.

An important advantage of both estimators is \emph{extensibility} with respect to the dataset: if the variance is too high, one can always either generate new ancestors by extending the short DNS, or extend the range of ASTs sampled, or enlarge the ensemble at any ASTs deemed promising, without discarding the laborious samples already generated. This is unfortunately not the case with an algorithm like AMS, TEAMS, GKTL, or QDMC: because of the random rules by which ancestors are selected and new members generated, a completed run cannot be enlarged while retaining its estimation properties unless we are willing to do an entirely new additional run and combine estimates from multiple runs as was done in \citet{Ragone2018computation,Webber2019practical,Finkel2024bringing}.  This results in waste during the fine-tuning process of calibrating TEAMS. 
For example, one might decide in retrospect that a TEAMS run was too aggressive in killing non-extreme simulations and raising the threshold and we can't easily extend the run with a new set of hyperparameters. With boosting, we can simply go back, perturb those less-extreme simulations, and incorporate them into the dataset, without needing to re-generate everything. To make boosting competitive at sampling the highest levels of severity, we suspect it will be necessary to augment our current scheme with an iterative level-raising schedule, like TEAMS, but with less restriction on the sampling procedure.

\subsection{Evaluating performance: statistical accuracy and computational cost}
\label{sec:evaluating_performance}

We evaluate the MoCTail and PoPTail estimators $\widehat{Q}^M$ and $\widehat{Q}^P$ by comparing to the ground truth $Q^\Theta$ as estimated from a long DNS. DNS is in fact a trivial special case of ensemble boosting with $M=0$ (no descendants), reducing each summand of Eq.~(\ref{eq:moctail}) and the numerator of Eq.~(\ref{eq:popt}) to $\mathbb{I}\{R_n^*>r\}$ and the denominator of Eq.~(\ref{eq:popt}) to $N_{\text{short}}$. Both estimators reduce to the same vanilla empirical CCDF in this case, and this is what we use to estimate $Q^\Theta$. 

We use $\chi^2$-divergence to measure the disparity of $\widehat{Q}^M$ and $\widehat{Q}^P$ from $Q^\Theta$. This is estimated from a discrete histogram with a sequence of thresholds $\mu=r_0<r_1<r_2<\hdots<r_{K-1}<r_K=\infty$, and define the probability mass function $\Delta Q^\Theta_k=Q^\Theta_k-Q^\Theta_{k+1}$ as the probability contained in the $k$th bin (note that $Q^\Theta_K=0$ and so $\Delta Q^\Theta_{K-1}=Q_{K-1}$). As described further in Sec. \ref{sec:target_variable}, we select the $r_k$s as quantiles with consecutively halving exceedance probabilities, i.e., $Q^\Theta_k=(\frac12)^{5+k}$ for $0\leq k<K=11$. These quantiles change with latitude, as the tail is different for each. Note the same set of $r_k$'s based on the climatological distribution is used also for evaluating estimated distributions. The $\chi^2$-divergence of either estimator $\widehat{Q}\in\{\widehat{Q}^M,\widehat{Q}^p\}$ is then defined as
\begin{align}
    \label{eq:chi2div_defn}
    \chi^2(\Delta\widehat{Q}\lVert\Delta Q^\Theta)
    =
    \sum_{k=0}^{K-1} \frac{(\Delta Q^\Theta_k-\Delta\widehat{Q}_k)^2}{\Delta Q^\Theta_k}
\end{align}

We will compute both the MoCTail and PoPTail estimates on the same dataset, and find them numerically quite similar, both in terms of skill and in terms of individual bin estimates. It would be interesting to develop test cases where they differ more systematically, to clarify which (if either) is generally superior.  

Computational efficiency is another important consideration besides accuracy, as the entire goal of rare event algorithms is to improve efficiency or accuracy (or both) relative to DNS. For a boosting-like rare event algorithm to be useful, its error should decrease faster by perturbing existing ancestors (increasing $M$) than by extending DNS by generating new ancestors (increasing $N$ and not $M$), at least in some range of $N$ that samples the attractor broadly but not exhaustively. However, this paper does not present a complete rare event algorithm \emph{per se}, in the sense we don't yet stake our claim on a speedup. Rather, we ask a pre-requisite question: does increasing $M$ decrease the error \emph{at all}? Clearly boosting can increase the maximum severity, but that could happen in ways that don't respect the tail CCDF's shape, e.g., if perturbations tend to maximize the event's severity while bypassing moderate severities that carry significant statistical weight. 

We will thus make two comparisons between boosting and DNS: accuracy at fixed $N$, and accuracy at fixed cost (where DNS runs an additional length equal to the cost of simulating descendants, allocating its full budget to ``exploration'' rather than ``exploitation''). Specifically, we approximate the cost of the boosting approach for a given AST $A$ as
\begin{align}
    \label{eq:cost_per_ancestor}
    \text{Average boosting cost per ancestor}=M(A+\delta t^*)+\text{(mean return period)},
\end{align}
where $\delta t^*$, the ``argmax drift'' parameter, accounts for the extra time needed to run after the ancestor's peak to account for changes in peak timing. ``Mean return period'' is the average time between consecutive independent peaks over the threshold $\mu$, which will be longer than $1/(1-q(\mu))$ because of de-clustering. The dependence on $A$ is a complication, as each AST tried would merit a different-length DNS for cost comparison, and we don't want to penalize boosting too severely by summing over all ASTs because in practice we would not bother simulating the obviously sub-optimal ASTs. Rather, we optimistically estimate the cost if $A$ is already known. On the other hand, our chosen $M(=21)$ is likely more samples than necessary to fit a satisfactory parametric model, as we have deliberately sampled the perturbation space more generously than we would if chasing a speedup. We simplify the comparison by fixing $A$ to $\frac12A_\text{max}$ in Eq.~(\ref{eq:cost_per_ancestor}), which is close to or slightly greater than the optimal values that we found empirically. 

We will show (Fig. \ref{fig:mixture_ccdfs}b) that boosting is unambiguously more accurate than DNS when fixing the number of ancestors $N$, and similarly accurate with marginal improvements when fixing cost, though with variation across latitudes and AST criteria. Thus, we do achieve some speedup, even though it is not (yet) our main objective. Any fixed-cost performance gains we achieve here (not our main objective) should be viewed as a lower bound for future algorithms, which will benefit from the conceptual insights into AST that we glean presently. 

To emphasize the \emph{conditional} nature of the AST---its possible dependence on the ancestor $n$ due to initial condition-dependent predictability---we refer to $A_{j_n}$ as the ``conditional advance split time'' (CAST), and its optimal value (by $\chi^2$ or other criteria) as the ``conditionally optimal advance split time'' (COAST). Our goal is to define the COAST, calculate it given extensive sampling from boosted ensembles, and finally to suggest useful criteria to estimate it when sample size is limited, as in a real rare event algorithm deployment. 

\subsection{AST selection criteria}
\label{sec:ast_selection_criteria}

With a data-generating plan and an estimator in place, we return to our central question of interest: how to select the CASTs $\{A_{j_n}\}$? There are three natural kinds of criteria. 

\begin{enumerate}
    \item Choose a single uniform AST $A_{j_n}=A^{\text{U}}$ for all ancestors (${\text{U}}$ for ``uniform''). In this case, the CAST is not really ``conditional'' at all. In \citet{Finkel2024bringing}, we found the COAST for TEAMS by systematic grid search through candidate ASTs, and found \emph{post-hoc} an empirical relationship for the COAST: $A^{\text{U}}\approx\overline{t_{3/8}}$, where $t_\epsilon(x_0)$ is the time until an ensemble dispersing from initial condition $x_0$ (each member forced by a different noise realization) reaches a fraction $\epsilon$ of its asymptotic root-mean-squared-error (RMSE), and $\overline{t_\epsilon}$ is the average of $t_\epsilon(x_0)$ over different initial conditions $x_0$. In \citet{Finkel2024bringing}, we sampled $x_0$ from the stationary distribution; here, for computational expediency, we will repurpose the boosting ensembles for estimating $\overline{t_\epsilon}$, i.e., sampling $x_0$ from pre-peak antecedent conditions. 
    
    \item Choose the CAST $A_n$ separately for each ancestor $n$ such that that an ensemble launched at $t_n^*-A_n$ disperses to a pre-defined threshold at time $t_n^*$. One could measure dispersal in different ways like RMSE, but here we opt instead for a \emph{pattern correlation}, defined with respect to spatiotemporal fields $F_0$ (from the ancestor) and $F_m$ (from the $m$th ensemble member) as
    \begin{align}
        \rho[F_0,F_m]&:=\frac{\overline{f_0f_m}}{\sqrt{(\overline{f_0^2})(\overline{f_m^2})}}
        \text{ where }
        f:=F-\langle{F}\rangle, 
        \ \ 
        \langle{\cdot}\rangle=\text{ time-average (climatology), and }
        \overline{(\cdot)}=\text{ space-average.}
      \label{eq:eq:pattern_correlation}
    \end{align}
    Unless  noted otherwise, $\rho$ will refer to the average of $\rho[F_0,F_m]$ over all members $m=1,\hdots,M$. 
    The reason for subtracting time-averages is to fairly weight spatial regions with smaller background $\langle F\rangle$, e.g., poles if $F$ is temperature. Dividing by spatial standard deviations is simply a useful normalization that restricts $\rho$ to the range $[-1,1]$ by the Cauchy-Schwarz inequality. $\rho$ tends to decrease over time from 1 to 0 except for occasional negative values when $F_0$ and $F_1$ are similar up to translation (but this effect usually disappears when averaging large-enough ensembles). We then choose some threshold $\rho^{\text{U}}\in(0,1)$, and select the corresponding CAST $A_{j_n}=A_n^{{\text{PC}}}[\rho^{\text{U}}]$---a function of the threshold---as the smallest sampled AST $A_n$ for which $\rho$ decreases from 1 to $\rho^{\text{PC}}$ between the split time $t_n^*-A_n$ and the peak time $t_n^*$. (${\text{PC}}$ stands for for ``pattern correlation''). Note that the CAST varies with $n$, but the correlation threshold, denoted $\rho^{\text{U}}$, is uniform. Finding the COASTs $A_n^{\text{PC}}$ then boils down to finding the optimal value of $\rho^{\text{U}}$.
    
    The $\frac38$ rule from \citet{Finkel2024bringing}, which used Euclidean distance $D^2[F_0,F_m]=\overline{(F_0-F_m)^2}=\overline{(f_0-f_m)^2}$ as the dispersion indicator, can be approximately restated in terms of pattern correlation: 
    \begin{align}
        D^2&=\epsilon^2\langle D^2\rangle
        && \langle{D^2}\rangle=\text{saturation value of }D^2
        \\    \implies\overline{f_0^2}+\overline{f_m^2}-2\overline{f_0f_m}
        &=
        \epsilon^2(\langle{\overline{f_0^2}}\rangle+\langle{\overline{f_m^2}}\rangle) 
        &&
        \text{Using }\langle{\overline{f_0f_m}}\rangle=\langle{\overline{f_0}}\rangle\langle{\overline{f_m}}\rangle=0 
        \\
        \frac{(\overline{f_0^2}-\epsilon^2\langle{\overline{f_0^2}}\rangle)+(\overline{f_m^2}-\epsilon^2\langle{\overline{f_m^2}}\rangle)}{\sqrt{(\overline{f_0^2})(\overline{f_m^2})}}
        &=
        \frac{2\overline{f_0f_m}}{\sqrt{(\overline{f_0^2})(\overline{f_m^2})}}=2\rho(F_0,F_m)
        \\
        \frac{(1-\epsilon^2)\langle{\overline{f_0^2}}\rangle+(1-\epsilon^2)\langle{\overline{f_m^2}}\rangle}{\sqrt{\langle{\overline{f_0^2}\rangle\langle\overline{f_m^2}}\rangle}}
        &\approx
        2\rho(F_0,F_m) 
        && 
        \text{Approximating }\overline{f^2}\approx\langle{\overline{f^2}}\rangle
        \\
        1-\epsilon^2&\approx\rho(F_0,F_m) && 
        \text{Using }\langle{\overline{f_0^2}}\rangle=\langle{\overline{f_m^2}}\rangle.
    \end{align}
    (The approximation invoked in the second-to-last step, $\overline{f^2}\approx\langle\overline{f^2}\rangle$, will hold when the spatial region is large enough that global fluctuations in the same direction are unlikely.)
    This calculation shows that the time until RMSE reaches $\frac38$ of its saturation value is roughly equivalent to the time at which pattern correlation drops to $1-(\frac38)^2=0.86$. We do not assume this threshold is optimal, but include it as a reference for comparison. And we stress that the $\frac38$ rule implemented in \citet{Finkel2024bringing} determines a uniform $A^{\text{U}}$, not a conditional $A^{\text{PC}}$, because their averaging was performed over the attractor, whereas here we will use $\rho$ as an initial condition-specific diagnostic. 
    
    \item Define the CAST as the solution to an optimization problem, where we seek to maximize a functional on the boosted severity distribution that favors both a high mean and high variability of the severity. This would implicitly favor intermediate ASTs, as short-AST ensembles have high mean but low variability while long-AST ensembles will have high variability but low mean (approaching the climatological distribution). We propose and evaluate two such functionals in this paper:
    \begin{enumerate}
        \item Expected improvement (EI): 
        \begin{align}
            \label{eq:expected_improvement}
            \mathbb{E}[(\Delta R^*)_+]=\int_\Omega p^\Omega(\omega)[R^*(\omega)-R^*(0)]_+\,d\omega,
        \end{align}
        where $(\cdot)_+:=\max(\cdot,0)$ and we recall that $\omega=0$ means no perturbation (i.e., the ancestor)
        
        \item Thresholded entropy (TE):
        \begin{align}
            \label{eq:thresholded_entropy}
            S[(R^*-\mu)_+]=-\sum_{k=0}^{K-1}\Delta Q_k\log\Delta Q_k,
        \end{align}
        where the bin boundaries $r_k$ start at $\mu$, and so only the tail part of the conditional CCDF contributes. The thresholded entropy is thus defined based on probability over discrete bins (with the bin boundaries $r_k$ set based on quantiles of the ground-truth distribution) and would change if the bins were changed.
    \end{enumerate}
    Where it doesn't cause confusion, we will also call the CASTs $A^{\text{EI}}$ and $A^{\text{TE}}$ themselves COASTs because they optimize something, although it is something different than $\chi^2$. We conjecture that that these two notions of optimality coincide: if each ancestor separately optimizes EI or TE, the resulting aggregate of distributions (via MoCTail or PoPTail estimators) will minimize $\chi^2$-divergence from the true climatological tail. Our results will approximately confirm the conjecture in the case of TE. 
\end{enumerate}
These criteria are each in turn more complex, but also more theoretically appealing. The correlation-based CASTs $\{A_n^{{\text{PC}}}\}_{n=1}^{N_{\text{short}}}$, unlike the synchronized AST $A^{\text{U}}$, can vary with $n$ to respect differences in predictability between different initial conditions, a well-recognized phenomenon in chaotic systems \citep{Maiocchi2024heterogeneity}, including the atmosphere \citep{Lucarini2020new}. Still, both $A^{\text{U}}$ and $A_n^{{\text{PC}}}$ require the user to set some arbitrary global threshold. 
The open question is whether optimizing $A_n^{\text{EI}}$ or $A_n^{\text{TE}}$ individually for each $n$ will also optimize the accuracy of the unconditional (MoCTail) climatological CCDF estimator against the ground truth climatological CCDF from a long DNS. 

\vspace{1cm}
\noindent\textbf{Main result}: Climatological tails are estimated more accurately with perturbed ensembles than with un-perturbed ancestors alone (fixed-$N$ comparison between DNS and boosting). This holds with few exceptions for both MoCTail and PoPTail estimators, for all COAST selection rules, and for all target spatial locations. At fixed cost, boosting and DNS are tied overall, but with some variation across latitudes and the value that cost is fixed to, suggesting that substantial speedups are possible with more highly optimized boosting-like algorithms. No single selection rule is superior across the board. The EI and TE criteria, however, have a distinct advantage of needing no arbitrary threshold choices. TE-based estimates strike a reasonable compromise between statistical error and arbitrariness, which is strong enough support that \textbf{we recommend TE as a generic AST selection rule}. 
\vspace{1cm}

The remainder of the paper demonstrates the theoretical framework above on the QG system. Sect. \ref{sec:model} specifies the dynamical model and its numerical simulation, displays some representative output, defines the target intensity functions of interest, and reports on their basic tail statistics. Sect. \ref{sec:ensemble_design} specifies the perturbation protocol (i.e., the space $\Omega$ and probability densities $p^\Omega(\omega)$) and visualizes representative examples of the system's response, providing motivation for our choices of AST selection criteria. Sect. \ref{sec:climatological_severity_distributions} compares the performances of all proposed AST selection criteria criteria in matching the climatological tail CCDF. Sect. \ref{sec:conclusion} concludes with a summary and outlook on important future lines of work. 

Throughout, we present more in-depth results for one select target latitudes just south of the domain center, and only summarize for the wider range of target latitudes, which reveals large-scale variations in extreme event predictability and representability across space. 

\section{The quasigeostrophic model}
\label{sec:model}

The model setup aims to distill some challenges we have encountered with rare event algorithms. We first recognized the need for advance splitting (or ``trying early'') to sample extreme precipitation in an aquaplanet GCM \citep{Finkel2026rare}. A minimal surrogate model replicating this challenge was found in Lorenz-96 \citep{Lorenz1998optimal}, which provided a testbed for the first working version of TEAMS and a recognition of an ``optimal advance split time'' \citep{Finkel2024bringing}. There is a huge gap in model complexity between Lorenz-96 and the GCM (see Table \ref{tab:hierarchy}), and we wish to test our idea in this middle ground where the target spatial location can have an effect. Lorenz-96, with a one-dimensional domain and homogeneous forcing, is too simple. For this reason, and to make closer contact with physics, we selected the two-layer QG model as a suitable intermediate between Lorenz-96 and the GCM. 

\begin{table*}[t]
	\caption{Three rungs on the model hierarchy. 
    Left: the Lorenz-96 system used in \citet{Finkel2024bringing} has a one-dimensional spatial domain (``longitude'') divided into discrete sites $k=0,\hdots,39$, on which generic meteorological variables $\{x_k\}$ evolve in time. Its state space dimension is 40. 
    Right: the aquaplanet model used in \citet{Finkel2026rare} has a three-dimensional spatial domain: latitude $\lambda$, longitude $\phi$, and pressure normalized by its surface value, $\sigma=p/p_s$. It has six prognostic fields: zonal wind $u$, meridioal wind $v$, temperature $T$, and humidity $q$ vary in all three dimensions, whereas surface pressure $p_s$ and precipitation rate $R$ vary only in the horizontal. 
    Center: the 2-layer quasigeostrophic model used in this study has two layers ($z=1,2$) of two dimensions each (longitude $x$, latitude $y$), and two dynamical fields: streamfunction $\psi$ which is discretized spectrally, and tracer concentration $c$ which is discretized on a grid. 
    }
	\begin{tabular}{c|c|c|c}
	Model & One-tier Lorenz-96 & 2-layer quasigeostrophic channel & Global aquaplanet \\
	\hline
	Domain & 
	$k\in\{0,\hdots,39\}$ & 
	$(x,y,z)\in[0,L)^2\times\{1,2\}$ & 
	$(\lambda,\phi,\sigma)\in[0,360)\times[-90,90)\times[0, 1)$ \\
	Fields & 
	$\{x_k\}$ & 
	$\{\psi_z,c_z\}(x,y)$ &
	$\{u,v,T,q\}(\lambda,\phi,\sigma)\cup\{p_s,R\}(\lambda,\phi)$ \\	
	\end{tabular}
	\label{tab:hierarchy}
\end{table*}

\subsection{Equations of motion and numerical simulation}

We implement a version of the QG model combining elements of several classic studies. Our numerical method and friction form follow \citet{Haidvogel1980homogeneous}, but on a smaller domain with weaker friction magnitude as in \citet{Panetta1993zonal} to contain only 1-2 more energetic zonal jets. We furthermore add bottom topography in the lower layer as in \citet{Thompson2010jet} to fix preferred latitudes for jets while still allowing them to temporarily split, merge, and meander. Thus climate statistics, and hence the COAST itself, can vary with latitude. Further, we augment the system with a passive tracer to represent a key component of precipitation dynamics, following the spirit of \citet{Bourlioux2002elementary} and \citet{Qi2016predicting,Qi2018predicting}  who used turbulent advection-diffusion as a paradigm for intermittency.

The model equations are as follows, in non-dimensionalized form using the deformation radius $\lambda$ as the length scale and a velocity scale $\mathcal{U}$. To make plain the role of the background shear, we define a non-dimensional wind $U$ as the ratio between the imposed upper-level zonal wind and $\mathcal{U}$. All non-dimensional parameter values are listed in Table \ref{tab:ctrl_params}. The horizontal coordinates $(x,y)$ each run from $0$ to $L$. The integer-valued vertical coordinate $z$ is an index for the layer (1 for the top and 2 for the bottom, appearing as a subscript). $\psi$ represents the streamfunction minus a background of $-Uy\delta_{z,1}$. 
$h$ is the bottom topography which is specified to vary sinusoidally with wavenumber 2 in latitude.
$q$ represents potential vorticity minus a background of $\beta y+h\delta_{z,2}$, due to planetary vorticity gradient and topography. $c$ represents the passive tracer field. 

\begin{align}
    \Big[\partial_t+(\partial_x\psi_z)\partial_y+(U\delta_{z,1}-\partial_y\psi_z)\partial_x\Big](q_z+h\delta_{z,2}+\beta y)
    &=
    -\kappa\delta_{z,2}\nabla^2\psi_z-\nu\nabla^6\psi_z
    \label{eq:qgmodel_dtq}
    \\
    \Big[\partial_t+(\partial_x\psi_z)\partial_y+(U\delta_{z,1}-\partial_y\psi_z)\partial_x\Big]c_z
    &=
    0
    \label{eq:qgmodel_dtc}
    \\
    \text{for }
    (x,y,z)
    &\in
    [0,L)^2\times\{1,2\} 
    \\
    \text{where }&
    \\
    q_z
    &=
    \nabla^2\psi_z+(-1)^z\bigg(\frac{\psi_1-\psi_2}{2}\bigg)
    \label{eq:qgmodel_pvinv}
    \\
    h(y)
    &=
    h_0\sin\bigg(2\cdot\frac{2\pi y}{L}\bigg)
    \label{eq:qgmodel_h}
\end{align}

For $\psi$, we impose doubly periodic boundary conditions and timestep with a pseudo-spectral method with 64 Fourier modes in each dimension and standard $\frac23$-dealiasing (hence, an effective maximum wavenumber of 20). We time-step linear terms with the trapezoid rule (Crank-Nicolson) and nonlinear and topographic terms with a predictor-corrector (Heun's) method. Meanwhile, boundary conditions on $c$ are periodic in $x$ and Dirichlet in $y$, with values $(0,1)$ at $y=(0,L)$. Together with a first-order upwind monotone finite-volume scheme, this setup guarantees that $0\leq c\leq 1$ everywhere, making clear that its probability distribution has compact support. Note there is no explicit dissipation for $c$, but the low-order discretization creates some effective diffusivity.

The number of degrees of freedom, or state space dimension, is
\begin{align}
    \label{eq:state_space_dimension}
    d=(2\text{ layers})\times(41^2\text{ Fourier modes for }\psi+64^2\text{ grid cells for $c$})=11554,
\end{align}
and we will sometimes refer to the full state vector as $\{\psi,c\}(x,y,z,t)=\mathbf{x}(t)\in\mathbb{R}^d$---not to be confused with the spatial coordinate $x$. For simplicity, we refer to one time unit as a day, which is $\sim\frac1{10}$ of an eddy turnover timescale (see Fig. \ref{fig:hovmoller}). The common timestep for $\psi$ and $c$ is 0.025 days, and the output frequency is once per day. The spatiotemporal resolution is coarse by modern standards, but we aren't seeking to calculate any real-world physical quantity: we are seeking a general rule that will help make the COAST clear for a wide class of models. 

\begin{table}[]
    \centering
    \begin{tabular}{l|c|l}
         Description & Symbol & Value\\
         \hline
         Coriolis gradient & $\beta$ & 0.25 \\
         Ekman friction coefficient & $\kappa$ & 0.05 \\
         Wind shear & $U$ & 1 \\
         Hyperviscosity & $\nu$ & $(0.292)^3$ \\
         Topography amplitude & $h_0$ & 0.25 \\
         Domain size & $L$ & $6\cdot2\pi$
    \end{tabular}
    \caption{Non-dimensional physical parameters used for the numerical simulation, similar to those chosen in \citet{Panetta1993zonal}.}
    \label{tab:ctrl_params}
\end{table}

\subsection{Baseline simulation and statistics}
\label{sec:eva}

We run a ``short DNS''  of length $T_{\text{short}}=4\times10^3$ days $\approx11$ years (after a 500-day spinup) to supply the pool of initially un-perturbed (``ancestral'') events. Then, to provide ``ground truth'' statistics, we run a control simulation, or ``long DNS'', of duration $T_{\text{long}}=16\times10^3\text{ days}\approx44$ years, which is $O(1600)$ eddy turnover times and $O(160)$ jet meandering times (see Fig. \ref{fig:hovmoller} caption for timescale definitions). However, in estimating climatological statistics from the long DNS, we take advantage of statistical zonal symmetry by concatenating the timeseries of all 64 longitudes, increasing the effective sample size by a factor of $\sim L/$(some typical correlation length). Conceptually, the short and long DNS are analogous to ``training'' and ``validation'' datasets in standard machine learning procedures, in the sense that we want to infer properties of the validation set using only information extracted from the training set (for example, by perturbing and re-simulating events seen in training). As we show below, simply counting events from the short DNS gives probability estimates that deterioriate at high levels of severity, which we aim to rectify with boosting. 

\begin{figure}
\includegraphics[width=0.98\linewidth,trim={0cm 3cm 16cm 0cm},clip]{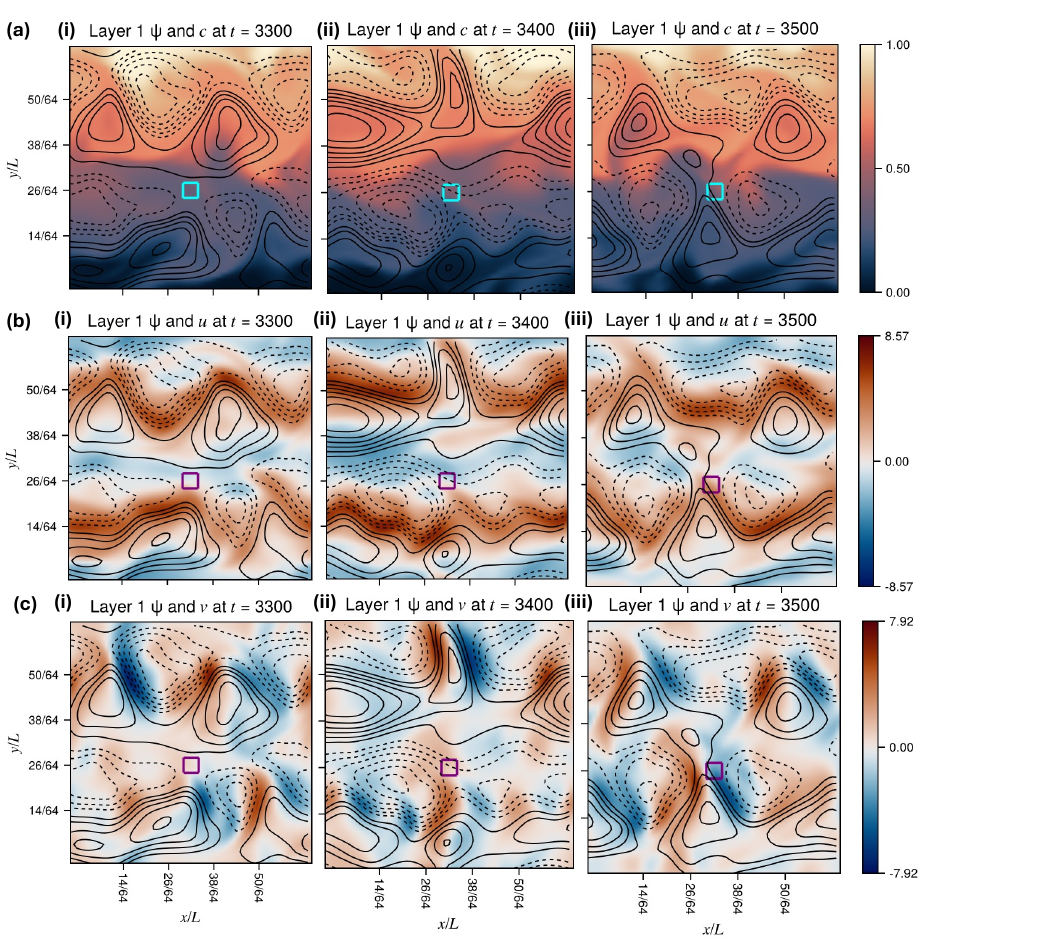}
    \caption{Snapshots of the QG system configuration in the upper layer. Contours indicate the anomaly streamfunction $\psi$, which varies over a non-dimensional range of approximately $\pm18$, dashed contours indicating negative anomalies. Colors indicate (a) tracer concentration $c$, (b) zonal wind velocity $u=U-\partial_y\psi$, where $U=1$ is the basic background shear, and (c) meridional velocity $v=\partial_x\psi$. The timestamps increase from left to right, and come from the long DNS. The small square represents an example target region in which to sample extremes of the local tracer concentration, in this case centered at $x_0=\frac12L,y_0=\frac{26}{64}L$ and extending $\pm\ell=\frac2{64}L$ in both meridional and zonal directions. This same region is the target used in the following results, and we consistently refer to the domain coordinates in fractions of 64 across all figures.}
    \label{fig:snapshots}
\end{figure}

\begin{figure}
\includegraphics[width=0.98\linewidth,trim={0cm 0cm 0cm 0cm},clip]{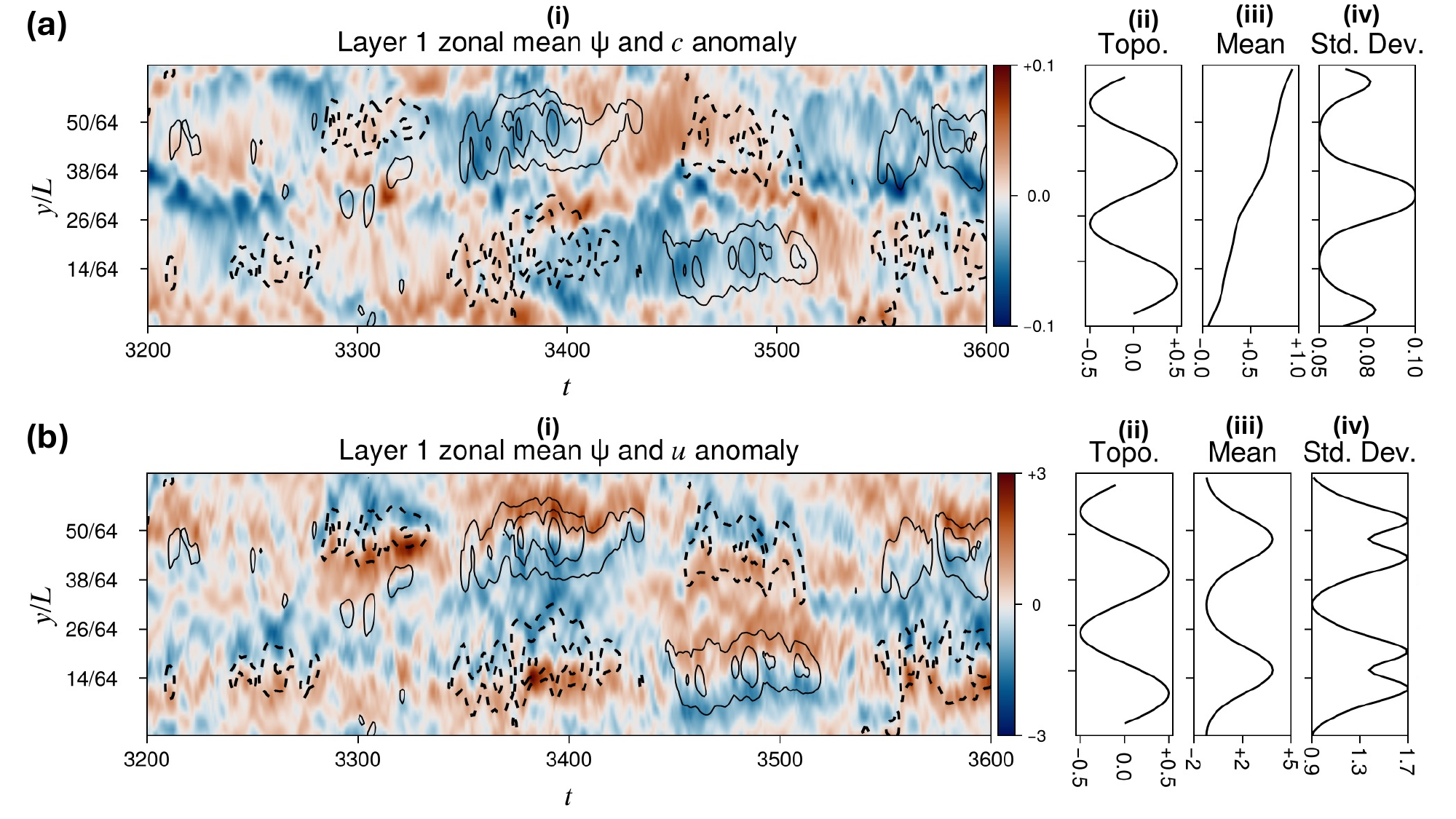}
    \caption{Hovm\"oller diagrams of anomalies (departures from time-means) of zonal-mean concentration (a.i) and zonal-mean zonal wind (b.i). Contours indicate zonal-mean streamfunction anomaly (range $\pm10$, negatives values dashed). Column (ii) shows bottom topography, which \emph{directly} affects the lower layer only, but indirectly sets the preferred jet positions in the upper layer as well. For the same quantities, column (iii) shows the zonal and time mean and column (iv) shows the zonal mean of the temporal standard deviation. 
    The Hovm\"oller diagrams give context to the snapshots of $u$ from Fig. \ref{fig:snapshots}b, which come from times (i) 3300, when the upper and lower jets are both shifted  south; (ii) 3400, when the jets are unusually far apart; and (iii) 3500, when the jets are unusually close together. These intermittent, discrete shifts in jet location happen every $\sim100$ days, which we call the ``jet meandering timescale''. During a typical 100-day timespan of stationary jet, the fields oscillate roughly 10 times (not shown here; see Fig. \ref{fig:spaghetti}); hence we assign the eddy turnover timescale a nominal value of 10 days. }
    \label{fig:hovmoller}
\end{figure}

Fig.~\ref{fig:snapshots} shows representative snapshots of three dynamical fields in the upper layer from the long DNS: tracer concentration $c$, zonal velocity $u=U-\partial_y\psi$, and meridional velocity $v=\partial_x\psi$. Fig.~\ref{fig:hovmoller} shows Hovm\"oller diagrams of zonal-mean anomalies of $c$ and $u$ (not $v$, since zonal-mean meridional velocity is zero), as well as their climatological means and standard deviations plotted alongside the topography. These are statistics of the grid-cell values, not zonal means, but depend only on latitude because so does topography. Two eastward jets are prominent in the snapshots Fig.~\ref{fig:snapshots}(b) and in the zonal mean profile Fig.~\ref{fig:hovmoller}b.iii, with preferred latitudes of $\sim\frac14L$ and $\sim\frac34L$. The Hovm\"oller diagram gives a sense of characteristic timescales: jets tend to remain roughly stationary for stretches of $\sim100$ days at a time before shifting, as seen by the group of closed contours of $\psi$ and associated dipole of $u$ centered at time $t=3400$. and persisting $\pm50$ days to either side. Within these stretches of quasi-stationarity, there are shorter undulations of duration $\sim10$, which we identify as the eddy turnover timescale.

The tracer statistics (Fig. \ref{fig:hovmoller}a.(iii,iv)) have some easily explainable large-scale patterns and some subtler small-scale patterns. The tracer time-mean $\langle c\rangle(y)$ increases linearly overall as $\frac{y}{L}$, in keeping with its Dirichlet boundary conditions. However, in the central region of the domain (inside the weak westward jet) the tracer mean varies more rapidly with latitude and has a larger standard deviation (see also dashed curves in Fig. \ref{fig:gpd}b,c). 

In the eastward jets, the tracer mean varies more slowly with latitude and has a smaller standard deviation. Comparison with the Hovm\"oller diagram (Fig. \ref{fig:hovmoller}a.i) suggests that the central region owes its high variance to short-lived anomalous pulses, both positive and negative, which are more intense than in surrounding regions. We won't try to explain these patterns from first principles, but simply state that the setup accomplishes our intention to provide a variety of statistical behaviors as a suite of test cases for our approach.

\subsection{Target variable}
\label{sec:target_variable}

We define the intensity function of interest $R(\mathbf{x})$ as the upper-level concentration, $c_1$ (henceforth, simply $c$), averaged over a small square box $[x_0-\ell,x_0+\ell]\times[y_0-\ell,y_0+\ell]$ of half-width $\ell=\frac{2}{64}$. This function is designed to capture the real-world considerations and algorithmic difficulties that originally motivated the AST: it describes \emph{localized} conditions, similar to concentrated pollution, high heat, or heavy rainfall over a region on Earth, and it is mediated by traveling baroclinic waves, and as a result it displays intermittency, with extreme spikes that come and go quickly. The choice of upper- instead of lower-level concentration is simply to weaken the impact of arbitrary aspects of the model setup like the surface topography. Real-world applications would of course refine this choice in many ways, but our choice is suitable for the QG level of model idealization. 

To investigate the effects of location-dependent flow regimes, we vary $y_0$ across 23 evenly spaced latitudes $y_0\in\big\{\frac{10}{64},\frac{12}{64},\hdots,\frac{54}{64}\big\}L$, restricted to the central region to avoid boundary effects. The central longitude $x_0$ is fixed to $L/2$, but by zonal homogeneity any longitude would be statistically equivalent. We also repeated the analysis with double the box length, and found results to be qualitatively similar, so we only show results for the smaller box size. The effect of spatial scale is worth considering in its own right with a wider range, which we postpone to future work. 

\begin{figure}
    \includegraphics[width=0.98\linewidth,trim={0cm 0cm 0cm 0cm},clip]{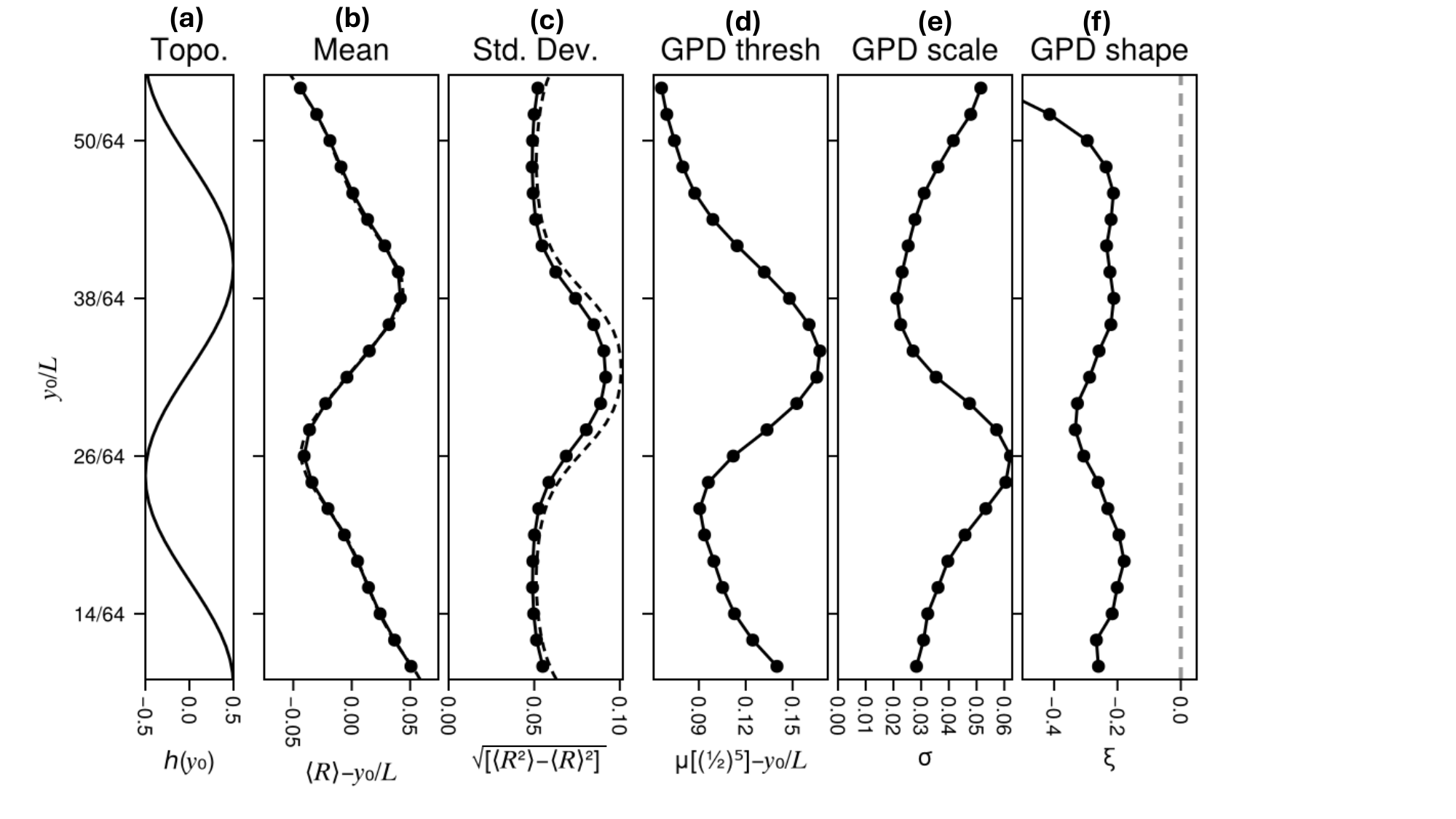}
    \caption{Summary statistics of latitude-dependent climatological tail distributions of local tracer concentrations, also called ``intensities'', which are denoted $R$ and defined as the average concentration $c$ over a box $(x,y)\in(x_0,y_0)+[-\ell,\ell]^2$. $x_0=\frac12L$ and $\ell=\frac1{32}L$ are fixed, while $y_0$ varies across the midlatitudes from $\frac{10}{64}L$ to $\frac{54}{64}L$. Panel (a) shows the lower-layer topography in this same range of middle latitudes, (b) shows the mean intensity $\langle R\rangle(y_0)$, after subtracting a nominal trend of $\frac{y_0}{L}$ to reveal a finer-scale structure that resembles the underlying topography, and (c) shows the standard deviation of intensity $\sqrt{\langle{R^2}\rangle-\langle{R}\rangle^2}$. Dashed curves in (b) and (c) indicate the mean and standard deviation, respectively, of the concentration field $c$ without box-averaging. Panels (d,e,f) summarize the distribution of intensities $R^*$ via the parameters of the generalized Pareto distributions (GPD), inferred by the peaks-over-threshold fitting procedure  (see section \ref{sec:target_variable} for details).
    The threshold is set to the $(\frac12)^5$-complementary quantile, denoted $\mu[(\frac12)^5]$ and shown in (d) with linear trend removed. Panels (e, f) display the estimated (scale, shape) parameters ($\sigma,\xi$). }
    \label{fig:gpd}
\end{figure}

Fig.~\ref{fig:gpd} displays some  summary statistics of $R(\mathbf{x}(t))$ as functions of the target latitude $y_0$: alongside (a) the topography for reference, we show (b) the meridionally de-trended time-mean $\langle{R}\rangle(y_0)-\frac{y_0}{L}$ and (c) the standard deviation $\sqrt{\langle{R^2}\rangle(y_0)-\langle{R}\rangle^2(y_0)}$. Note the restricted latitude range. In (a) and (b), dashed lines show the corresponding mean and standard deviation of $c$ itself, as in Fig. \ref{fig:hovmoller}(c,d), of which $R$ is a regional average: note that $R$ has the same mean as $c$ but a smaller standard deviation, and larger box sizes would reduce it even further. 

While the low-order moments capture ordinary behavior of intensities $R$, the intensity peaks---i.e., severities $R^*$, defined in Sect. \ref{sec:sampling_estimation_methodology}---are better viewed from an extreme value theory perspective, and summarized with the peaks-over-threshold procedure \citep{Coles2001introduction}. We set a threshold $\mu$ as the $(\frac12)^5$th complementary quantile of $R$, also denoted $\mu[(\frac12)^5]$, i.e., the level whose exceedance probability is $q(\mu)=(\frac12)^5$. Severities $R^*$ are extracted as cluster maxima above $\mu$, with buffer times $A_{\text{max}}=40$ days and $B=20$ days.  All cluster maxima from the long DNS are used as input data points to infer the parameters (scale $\sigma$, shape $\xi$) of a generalized Pareto distribution (GPD), using the maximum-likelihood routine of the \texttt{Extremes.jl} package \citep{Jalbert2024extremes}:
 
\begin{align}
    \mathbb{P}\{R^*>r|R^*>\mu\}
    \approx
    G_\mu(r;\sigma,\xi)
    =
    \begin{cases} 
    \big[1+\xi\big(\frac{r-\mu}{\sigma}\big)\big]_+^{-1/\xi} & \xi\neq0 \\
    \exp\big[-\big(\frac{r-\mu}{\sigma}\big)_+\big] & \xi=0
    \end{cases}
\end{align}
where $(\cdot)_+=\max(\cdot,0)$. Fig. \ref{fig:gpd}(d,e,f) display the threshold (detrended by $\frac{y_0}{L}$), scale parameter $\sigma$, and shape parameter $\xi$. Several characteristics are noteworthy. 

\begin{figure}
    \includegraphics[width=0.98\linewidth,trim={0cm 0cm 5cm 0cm},clip]{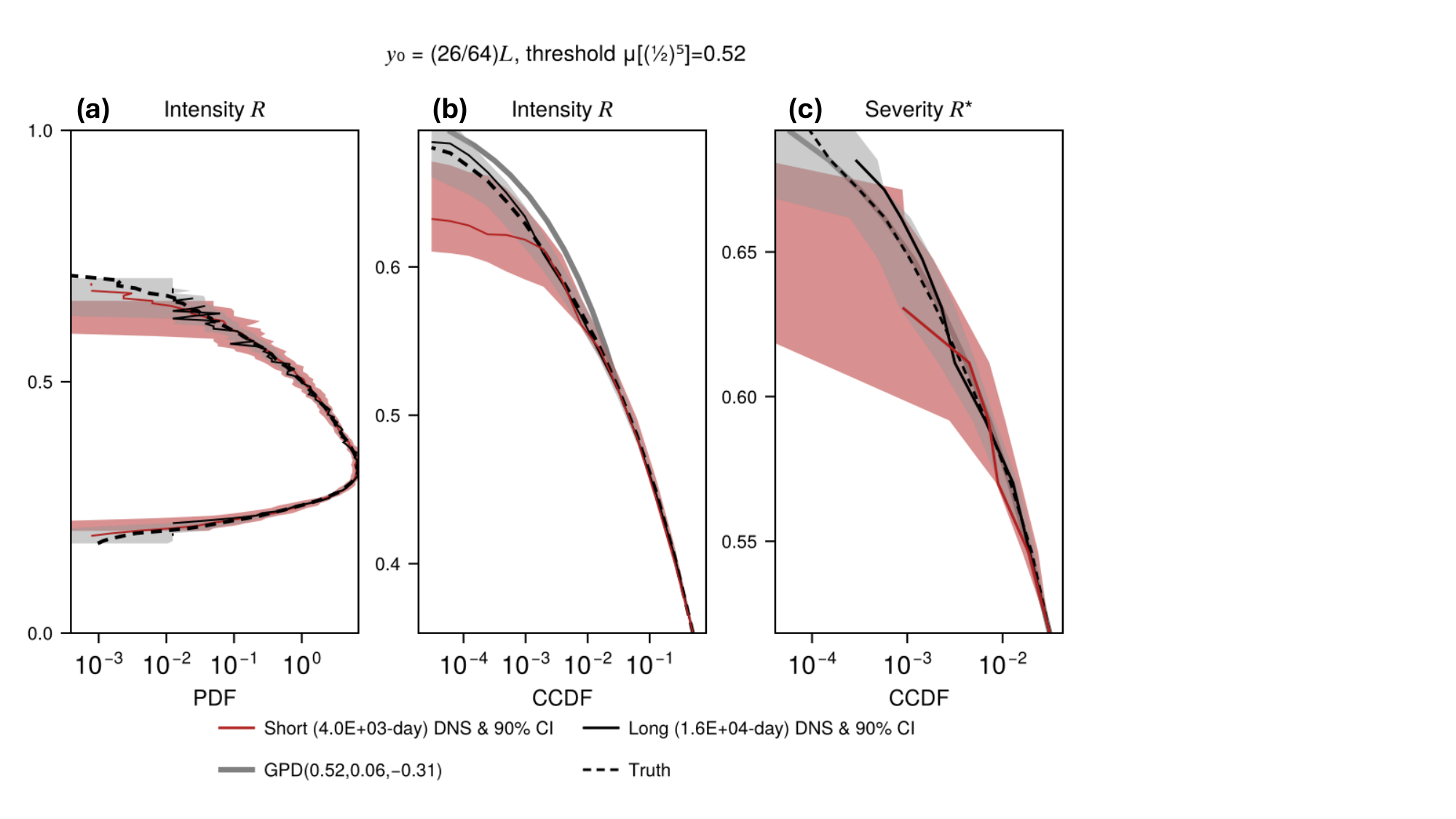}
    \caption{Probability distributions of local tracer concentrations at latitude $y_0=\frac{26}{64}L$ and averaged over a box of half-width $\ell=\frac{2}{64}L$. (a) The full PDF of intensity $R$. (b) The CCDF (tail integral) of intensity $R$, restricted to $R>\mu[\frac12]$. (c) Further zoomed-in CCDF of the severity $R^*$ (peaks of $R$ over $\mu[(\frac12)^5]$). 
    In all three panels, solid black and red lines represent estimates from long and short DNS, respectively, with shaded 90\% confidence intervals obtained by repeating the inference $64$ times, once for each possible longitudinal rotation of the dataset. Error bars become degenerate at levels experienced by $<5\%$ of longitudes. Black dashed lines show the mean over all longitudinal rotations, our best estimate of ground truth. The gray line in (b,c) represents the GPD fit to $R^*$ with $\mu=0.52$, $\sigma=0.06$, and $\xi=-0.31$, and this is a much better fit to the severities in (c) which makes sense given they are defined in terms of peaks. }
    \label{fig:tail_zoom}
\end{figure}

\begin{itemize}
    \item The detrended threshold $\mu-\frac{y_0}{L}$ has a maximum-over-minimum profile similar to the the detrended mean intensity $\langle{R}\rangle-\frac{y_0}{L}$, but shifted southward. The maximum of $\mu-\frac{y_0}{L}$ is close to the mid-channel maximum in the standard deviation of $R$, perhaps because extremes depend more on variability than on average behavior. 
    \item As expected for an upper bounded tail, we find $\xi<0$ at all latitudes (Fig. \ref{fig:gpd}f).
    \item The GPD scale parameter, $\sigma$, is anti-correlated with the (detrended) mean. The constraint $R^*\leq 1$ can explain this, as a higher distribution center leaves less room for an expansive tail. In addition, the threshold $\mu$ tracks approximately with the mean, and we can understand the anticorrelation mathematically through the non-uniqueness of GPD parameters: the same tail can be adequately described by two different choices of threshold $(\mu_1,\mu_2)$, and the two corresponding scale parameters will be related by $\sigma_2-\sigma_1=\xi(\mu_2-\mu_1)$. Only the shape parameter, $\xi$, is invariant with respect to $\mu$.  Seeing that $\xi$ is negative and varies only slightly with latitude, $\sigma$ and $\mu$ would vary inversely even if the tail itself were not changing.
    \item The mean appears odd-symmetric and the standard deviation appears even-symmetric about the midline (Fig. \ref{fig:gpd}b,c), which is not surprising given the tracer boundary conditions which transform as $c\mapsto1-c$ when $y\mapsto L-y$, negating the sign of fluctuations but leaving their absolute value constant (or perhaps disrupted slightly by topography). However, the GPD parameters are not symmetric (Fig. \ref{fig:gpd}d,e,f), because they describe the \emph{upper} tail of the local $R^*$ distribution, and the transformation $c\mapsto1-c$ swaps the lower and upper tails. The subsequent figures (\ref{fig:tail_zoom} and \ref{fig:spaghetti}) demonstrate pronounced skewness, so the upper and lower tails are markedly different. These partial symmetries will imprint upon the COAST's latitudinal variation seen later in Figs. \ref{fig:chi2_latdep} and \ref{fig:heatmaps_latdep}. 
\end{itemize}

We implemented the boosting and estimation procedures for every latitude separately, but for illustration focus the in-depth analysis on $y_0=\frac{26}{64}L$ (the small boxes in Fig. \ref{fig:snapshots}), an interesting location where the (detrended) mean and threshold $\mu[(\frac12)^5]$ are both low, the GPD scale $\sigma$ is large, and the GPD shape slightly more negative than in surrounding regions. Fig. \ref{fig:tail_zoom} displays the underlying probability distributions at $y_0=\frac{26}{64}L$ to show the nature of the tails of the distributions and also to help clarify the relationship between intensities, severities, and GPD parameters. The full PDF of intensity, in (a), has a positive skew and sub-Gaussian tail. Black and red solid curves are estimates obtained from the long and short DNS, respectively, and 90\% confidence intervals are obtained by longitudinal translation. Specifically, the shaded intervals are the 5th-95th percentile ranges of intensities at the same $y_0$, but with $x_0$ shifted from its base location of $\frac12L$ by $\frac0{64}L,\frac1{64}L,\frac2{64}L,\hdots,\frac{63}{64}L$. The dashed black curve is the mean of all 64 curves, our best available estimate of ground truth.  
The discrepancy between short and long DNS is most pronounced in the upper tail, which in panel (b) is magnified and integrated from the top, giving the CCDF. Gray lines mark the threshold, $\mu=0.52$, and its CCDF value $\frac1{32}\approx0.03$. Above this level, the short DNS becomes rapidly more uncertain (error bar widens), and severely underestimates probabilities smaller than $\sim0.005$. 

Both short and long DNS estimates diverge markedly from the GPD fit shown in gray in panel (b). This is where the distinction between intensity and severity comes into play: the GPD is fitted to \emph{peaks over the threshold} $\mu$---i.e., severities---whose distribution differs (most notably in the upward direction) from that of \emph{all} exceedances over $\mu$, which would include the clusters surrounding the peaks. Panel (c) confirms that the GPD fit is much more appropriate for severities $R^*$ than for intensities $R$, and thereby clarifies the distinction. If the threshold were raised, the clusters would shrink, the sequence of peaks would form a Poisson process, and the CCDFs of $R$ and $R^*$ would converge. For computational economy and because non-asymptotic extremes are of interest for climate risk, we keep the threshold at $\mu[(\frac12)^5]$ and formally define our goal with boosting as correcting the distribution of severities---not intensities. Hence, our measure of success will be whether the short-DNS severity CCDF in Fig. \ref{fig:tail_zoom}c, when augmented by boosting, will become closer to the long-DNS severity CCDF. 

\section{Ensemble design}
\label{sec:ensemble_design}

\subsection{Stochastic inputs}

We perturb the QG model with impulsive forcing, which we now specify as a concrete version of the generic form in Sect. \ref{sec:sampling_estimation_methodology}. The stochastic input $\omega$ lives in the complex plane $\mathbb{C}\ (=\Omega,\text{ the ``input space''})$, and the state-space perturbation $G(\omega)$ consists of a single Fourier mode to be added to $\psi$. We stress that our focus here is on optimizing AST, not the perturbation space $\Omega$, so the choice of mode is arbitrary so long as $\Omega$ remains low-dimensional. The optimal AST would probably change if $\Omega$ changes, e.g., if we perturbed a different mode or multiple modes at once; but the \emph{rule for choosing it} based on entropy may well generalize, which will have to be tested in follow-up research.

\begin{figure}
    \includegraphics[width=0.98\linewidth,trim={0cm 0cm 12cm 0cm},clip]{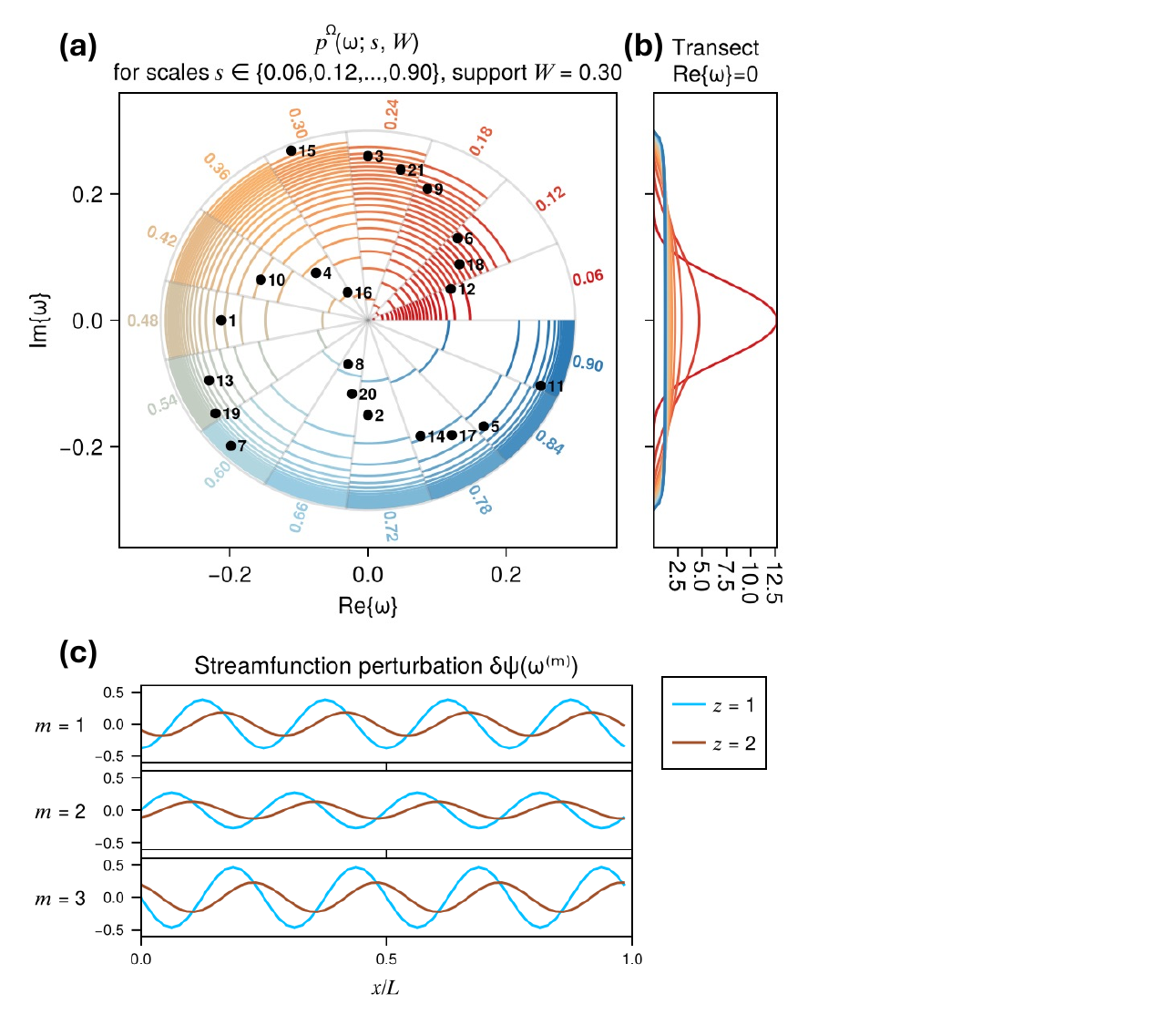}
    \caption{Structure of perturbations and their probability distribution. (a) Level sets of each considered input distribution from scales $s=0.06$ (red) to $s=0.9$ (blue), each scale restricted to $\frac1{15}$ of the circle each so that all scales may be seen. Labels on the outer edge of the circle indicate the corresponding scale. Dots show the 21 impulses used at each AST before each ancestor, sampled by quasi-Monte Carlo. (b) One-dimensional transects of $p(\omega; s, W)$ at each scale. (c) The shape of perturbations to the streamfunction corresponding to $\omega_1,\omega_2,\omega_3$. Note that the absolute amplitudes and phases vary, sampling the two degrees of freedom in the disc, but the relative amplitudes and phases of the upper and lower layers are fixed. }
    \label{fig:pert_struct}
\end{figure}

Bearing these caveats in mind, we select a mode  that is likely to grow fast, according to linear stability analysis, which is more easily explained as a procedure than as a closed formula: 
\begin{enumerate} 
    \item Decompose $\psi$ into a Fourier basis $\psi_z(x,y)=\sum_{k,\ell}\widehat\psi_z(k,\ell)e^{i(kx+\ell y)}$, and write the linearized dynamics (about the baroclinically unstable background state with vertical zonal wind shear and $\psi=0$,  and neglecting topography) into the abstract form
    \begin{align}
        C(k,\ell)\frac{d}{dt}
        \begin{bmatrix}
            \widehat\psi_1(k,\ell)\\\widehat\psi_2(k,\ell)
        \end{bmatrix}
        =D(k,\ell)
        \begin{bmatrix}
            \widehat\psi_1(k,\ell)\\\widehat\psi_2(k,\ell)
        \end{bmatrix}
    \end{align}
    where $C\in\mathbb{C}^{2\times2}$ represents the conversion from streamfunction to potential vorticity, and $D\in\mathbb{C}^{2\times2}$ represents the advection and linear dissipation terms (excluding topography). 
\item Calculate the eigenvalues and eigenvectors $\{(\lambda^{(m)}(k,\ell),\widehat\varphi^{(m)}(k,\ell)): m=1,2\}$ of the Jacobian matrix $C^{-1}(k,\ell)D(k,\ell)$, ordered by stability: $\text{Re}\{{\lambda^{(1)}}\}\geq\text{Re}\{{\lambda^{(2)}\}}$, and select $(k^*,\ell^*)=\text{argmax}_{k,\ell}\{\text{Re}\{{\lambda^{(1)}(k,\ell)}\}$, i.e., the linearly most unstable mode from the background state. Restrict the optimization to $(k,\ell)$ both nonnegative, and not both zero.
    \item For $z=1,2$, increment $\widehat\psi_z(k^*,\ell^*)$ by $\omega\widehat\varphi_z^{(1)}(k^*,\ell^*)$, and to maintain the solution's reality add the complex conjugate (c.c.) to $\widehat{\psi}_z(-k^*,-\ell^*)$. The perturbation can be written as a function of space, 
    \begin{align}
        \label{eq:g_of_omega} G(\omega)=\delta\psi_z(x,y)=\omega\widehat{\varphi}_z^{(1)}(k^*,\ell^*)e^{i(k^*x+\ell^*y)}+\text{c.c.},
    \end{align}
    which can have pointwise magnitudes up to $2|\omega|$. In the QG model, the mode we identify is $(k^*,\ell^*)=(4,0)$, and $G(\omega)$ is plotted in Fig.~\ref{fig:pert_struct}c for three different example $\omega$s, which correspond to the points labeled 1,2,3 in panel (a). All share the same inter-layer \emph{relative} phase and magnitude, as these are properties of $k^*,\ell^*$, and $\widehat{\varphi}_z^{(1)}(k^*,\ell^*)$, but differ in \emph{absolute} phase and magnitude. Note that points 2 and 3 are approximately diametrically opposed, and hence spatially $\sim180^\circ$ out of phase, whereas point 1 is approximately one-quarter revolution away and spatially $\sim90^\circ$ out of phase with both 2 and 3. Points (2, 3) are (closest to, farthest from) the center of the circle, and hence have the (smallest, largest)-magnitude spatial perturbations. of the three examples shown. 
\end{enumerate}

The steps above completely specify $G(\omega)$, a linear map from $\mathbb{C}$ to functions of ($x,y,z$), which can be easily computed offline before running any ensembles. One could argue for two obvious refinements of this choice: (1) specializing the linearization to the actual initial state, not just the background state, by linearizing the quadratic form $J(q,\psi)$ and including that in the calculation of $D(k,\ell)$; and (2) accounting for finite time horizons by using the leading singular vector of the \emph{linear propagator}, i.e., the initial \emph{infinitesimal} error whose magnitude amplifies the most over a given time horizon \citep{Farrell1996generalized1,Farrell1996generalized2} and which could be estimated by a bred vector approach \citep{Norwood2013lyapunov}. 

For this study, we stick to the simpler choice of the most unstable modes of the background shear, choosing to focus attention on the less-studied optimization of the advance split time given a fixed perturbation shape. There are several reasons that singular vectors may not be suitable for our goals. First, it is easier to compare different initial conditions, different advance split times, and even different topographies (which we don't do here) when they are all subject to precisely the same perturbation. Second, as our results will demonstrate, the COAST tends to lie beyond the time range where linearized error dynamics are appropriate, which is natural because we aim for finite-amplitude boosts in extreme events. Third, singular vectors are typically designed to optimize global errors, which might not be as relevant for local extremes. Fourth, such highly specialized perturbation shapes might not be accessible in a generic GCM. Nonetheless, sensitivity analysis with respect to perturbation shape leads the agenda for follow-up work. 

Having fixed a subspace $\Omega=\mathbb{C}$ for perturbations $\omega$, we need to specify an input distribution $p^\Omega(\omega)$ over that space. We design the PDF for $\omega$ as a radially symmetric, smooth, ``bump function'' which has compact support in order to prevent perturbations so large as to induce oscillatory transients. The PDF is parameterized by two scales: $W$ which is the maximum permissible magnitude of $\omega$, and $s$ which sets the typical perturbation strength:
\begin{align}
    p(\omega; s, W)\propto\exp\bigg[-\frac{|\omega|^2}{2s^2}\bigg(1-\frac{|\omega|^2}{W^2}\bigg)^{-1}\bigg]
    \text{ for $|\omega|<W$, and 0 for $|\omega|\geq W$.}
    \label{eq:bump_unnormalized}
\end{align}
When $s\ll W$, $p$ is approximately a bivariate Gaussian density with diagonal covariance $s^2I$. When $s\gtrsim W$, $p$ is approximately uniform over the $W$-disc $\{\omega:|\omega|\leq W\}$, with rapid (but mathematically smooth) transition to 0 on the boundary. We fix $W=0.3$, limiting the maximum possible perturbation amplitude to $|\delta\psi|\leq2W=0.6$ (see text after Eq.~(\ref{eq:g_of_omega})), which is small compared to the characteristic streamfunction amplitude of $|\psi|\sim10$. We include $s$ as a parameter to vary because there is no established principle to set the magnitude of impulses for the purpose of rare event sampling. In contrast, numerical weather prediction has an established (if heuristic) practice of tuning noise amplitude to match ensemble spread with model error \citep[e.g.,][]{Berner2015increasing}. Optimizing for climatological accuracy is a different, murkier goal calling for less prejudice with regard to perturbation magnitude. We therefore vary $s$ widely from $0.06$ to $0.9$ in increments of $0.06$ for 15 total values. $s$ is the impulsive-forcing analogue to the continuous-forcing amplitude that we called $F_4$ in \citet{Finkel2024bringing}, which strongly influenced the perturbation growth rate and therefore the optimal advance split time.

Fig.~\ref{fig:pert_struct}(a,b) depicts $p(\omega; s, W)$ in two ways: (a) two-dimensional level sets of the unnormalized density~(\ref{eq:bump_unnormalized}) logarithmically spaced from $e^{-4}$ to $e^{-0.01}$, each value of $s$ occupying one of 15 sectors of the circle; and (b) one-dimensional transects across $p(\omega; s, W)$ fixing $\text{Re}\{\omega\}=0$. To save the labor of drawing Monte Carlo samples from $p(\omega;s,W)$ separately and simulating the perturbed children for each value of $s$, we compute the MoCTail and PoPTail estimators using numerical quadrature over the $W$-disc using a single set of samples drawn by \emph{quasi}-Monte Carlo (QMC), and displayed as black dots in \ref{fig:pert_struct}a. QMC is a general strategy which places samples deterministically across the input space in a way that mimics properties of randomness, but with lower \emph{discrepancy} (fewer clumps and patches), thereby aiming to reduce variance in estimated statistics \citep{Leobacher2014quasi}. We specifically use the \texttt{LatticeRuleSampler} from the \texttt{QuasiMonteCarlo.jl} Julia library \citep{Rackauckas2023qmc} to distribute points $\{(U_m,V_m)\}_{m=1}^M$ quasi-uniformly on the unit square $[0,1]^2$, and transform them to the $W$-disc with the formula  
\begin{align}
    \omega_m=W\sqrt{U_m}\exp(2\pi iV_m).
\end{align}
Since $U_m$ is a ``quasi-random sample'' of the uniformly distributed random variable $U\sim\mathcal{U}([0,1])$, we have 
\begin{align}
    \mathbb{P}\{r_1\leq|\omega|\leq r_2\}
    =
    \mathbb{P}\{r_1^2\leq W^2U\leq r_2^2\}
    =
    \mathbb{P}\bigg\{\frac{r_1^2}{W^2}\leq U\leq\frac{r_2^2}{W^2}\bigg\}
    =
    \frac{r_2^2-r_1^2}{W^2}
\end{align}
which is the fraction of the $W$-disc between the radii $r_1$ and $r_2$. The phase $2\pi V$ is clearly $\mathcal{U}([0,2\pi])$. If $U$ and $V$ were independent random variables, we would immediately conclude $\omega$ is uniformly distributed over the $W$-disc; in QMC they are not independent, but the conclusion still holds true \citep{Leobacher2014quasi}. In all experiments to follow, $M=21$, corresponding to the 21 points plotted in Fig.~\ref{fig:pert_struct}a. While other sampling rules are possible, the \texttt{LatticeRuleSampler} enjoys a distinct advantage of being extensible: sampling 12 points at first and later deciding to add 9 more gives the same result as sampling 21 in one batch. 

\subsection{Sweeping over ancestors and advance split times} 

Following the procedure laid out in Sect. \ref{sec:sampling_estimation_methodology}, we apply each perturbation $\{\omega_m\}_{m=1}^M$ to a collection of ancestor events $\{\mathbf{x}(t_n^*)\}_{n=1}^N$ at a range of ASTs $\{t_n^*-A_j\}_{j=1}^J$. We set the number of ancestors, $N$ to whichever is smaller: the total number of cluster maxima (see Sect. \ref{sec:model}) in the short DNS, or 32. Considering all latitudes, the minimum $N$ was 14, the median was 22, and the maximum 32 was found at four latitudes including $y_0=\frac{26}{64}L$ which we consider in more depth. In the equal-cost comparisons to be shown later, we restrict $N$ to smaller values. The ASTs sampled are $\{A_j\}_{j=1}^{J=20}=\{2,4,\hdots,40\}$, with a two-day spacing chosen as roughly half the period of small fluctuations in $R(\mathbf{x}(t))$ (see Fig. \ref{fig:spaghetti}). 

\section{Results: conditional severity distributions}
\label{sec:conditional_severity}

In this section we present some case studies of conditional perturbed ensembles (from individual ancestors) and corresponding dispersion measures to be subsequently used in the MoCTail and PoPTail estimation. The results will add context and motivation to the protocols laid out above, and set the stage for the aggregation of results across ancestors.  

\subsection{Perturbed ensembles: case studies}

\begin{figure}
    \includegraphics[width=0.65\linewidth,trim={0cm 0cm 20cm 0cm},clip]{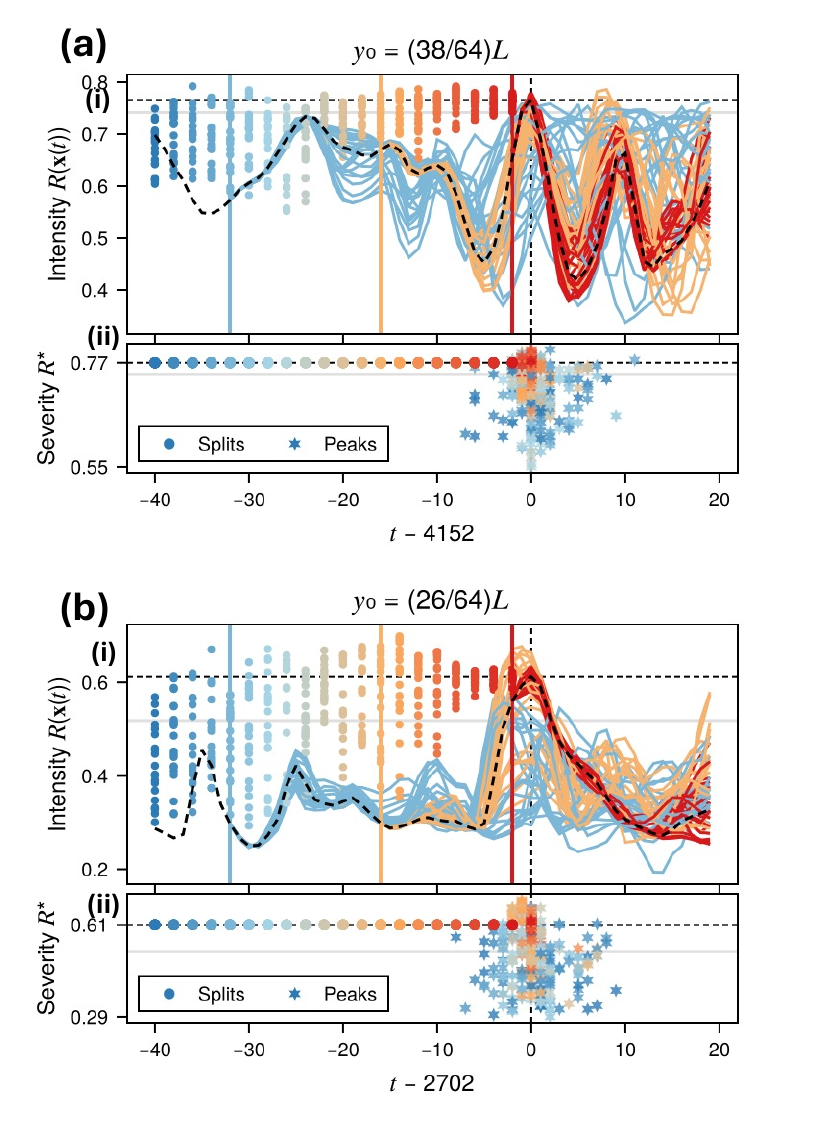}
    \caption{Boosted ensembles of two selected events: (a) time $t^*=4152$ at latitude $y_0=\frac{38}{64}L$, and (b) time $t^*=2702$ at latitude $y_0=\frac{26}{64}L$. These are times when the intensity function $R(\mathbf{x}(t))$ from the short DNS (dashed black curves) achieved a peak value (horizontal dashed black lines) above the threshold $\mu[(\frac12)^5]$ (horizontal gray lines). For each AST $A\in\{2,4,\hdots,40\}$, an ensemble of perturbed events (descendants) is launched at $t^*-A$, indexed by $m=1,\hdots,21$. For three selected ASTs $A=2,16,32$, the full timeseries $\{R_m(t)\}_{m=1}^{21}$ are shown in (a,b).i. The red-to-blue color scale indicates short-to-long ASTs. Each descendant achieves a different severity $R_m^*$ (peak intensity), indicated by circles in (a,b).i at $(-A,R_m^*)$ for all values of $A$. The peaks also occur at different times  $t_m^*$, indicated in (a,b).ii by stars at $(t_m^*-t^*,R_m^*)$, again for all $A$ and colored accordingly. }
    \label{fig:spaghetti}
\end{figure}

Fig.~\ref{fig:spaghetti} displays a small but representative sample of boosted ensembles at two target latitudes at the inner edges of the two eastward jets: (a) $y_0=\frac{38}{64}L$ and (b) $y_0=\frac{26}{64}$. The ancestors' intensity (black dashed curves) reach their respective peaks at times $t^*=4152$ for (a) and  $2702$ for (b). Note the differences in peak value and peak shape: the upper latitude has long-lasting, flat maxima and the lower latitude has brief, spiky maxima. The statistical properties at these two locations, both in Fig.~\ref{fig:spaghetti} and in Fig.~\ref{fig:hovmoller}, are approximately equivalent after reflection about $\frac12$ ($c\to1-c$), meaning the upper tail of one resembles the lower tail of the other. This can be understood by the approximate north-south symmetry of the tracer's dynamics imposed by Dirichlet boundary conditions.

We show the perturbed intensities launched from three ASTs $A\in\{2,16,32\}$, colored (red, orange, blue) respectively. Following the split time, the ensemble members spread apart from the parent and from each other, achieving their own peak values (severities) that differ in both amplitude and timing from the ancestor, the discrepancies increasing with $A$. The red curves ($A=2$) replicate the ancestral peak very closely; the orange curves ($A=16$) peak at substantially higher or lower levels, and up to $\sim2$ days earlier or later. Still, the orange peaks are clearly dynamically related to the ancestral peaks. This is no longer true for the blue curves ($A=32$), whose intensity peaks are widely scattered in time and systematically lower than the ancestors' peaks. 

Besides these three selected ASTs, each descendant is charted in  Fig.~\ref{fig:spaghetti}(a,b).i as a circle color-coded by AST, positioned vertically at its severity value and horizontally at its launch time. A corresponding star is plotted in Fig.~\ref{fig:spaghetti}(a,b).ii, positioned vertically at its severity value (on a zoomed-in scale) and horizontally at its peak timing (constrained by the ``argmax drift'' parameter $\delta t^*=5$ days $\approx$ half of an eddy turnover timescale, as explained in Sect. \ref{sec:generating_dataset_boosted_ensembles}). We can see the transition of the $R^*$ ensemble from tightly clustered (for short AST) to roughly independent and climatologically distributed (for long AST), and in between there is a golden window of opportunity where severities can be both large and diverse. The optimal AST must balance these two objectives, a task akin to the exploitation-exploration tradeoff in Bayesian optimization and reinforcement learning \citep[e.g.,][]{Yang2022output}. In this light, the two functionals defined in Eqs. (\ref{eq:expected_improvement}) and (\ref{eq:thresholded_entropy}) are candidate \emph{acquisition functions}.

\subsection{Relating severities to impulses: case studies}

We now construct ``severity response functions'' $\widehat{R}_{n,j}^*(\omega;\theta)$ mapping impulses $\omega\in\mathbb{C}$ to severities $R^*$, approximating the action of the flow map using some empirical parameters $\theta$. This will be needed to estimate conditional and unconditional probabilities through the MoCTail and PoPTail estimators (see Eq.~(\ref{eq:conditional_severity})), and will also help to understand the joint dependence between impulses $\omega\in\mathbb{C}$ and the times $\{t_n^*-A_j\}$ at which they are applied.

How should the response functions be parameterized? The simplest choice would be a linear model, often used in numerical weather prediction to optimize ensemble spread by perturbing in the most-effective directions, so-called singular vectors \citep{Diaconescu2012singular}. However, linear models are strictly valid only for infinitesimal perturbations, hence short lead times.
Similar logic should apply when optimizing for severity instead of ensemble spread, and indeed we demonstrate below that the COAST tends to lie beyond the range where a linear model $\widehat{R}^*$ is valid. We therefore construct a quadratic model as well, and it turns out that this minor upgrade is sufficient. Future work with more complex dynamics and objectives may call for more elaborate response functions (orthogonal polynomials, Gaussian processes, and neural networks for example), but we adhere to quadratic models in this study as a proof of concept that is easy to construct and interpret, which we do in the following two figures. 

\begin{figure}
    \includegraphics[width=0.98\linewidth,trim={0cm 0cm 0cm 0cm},clip]{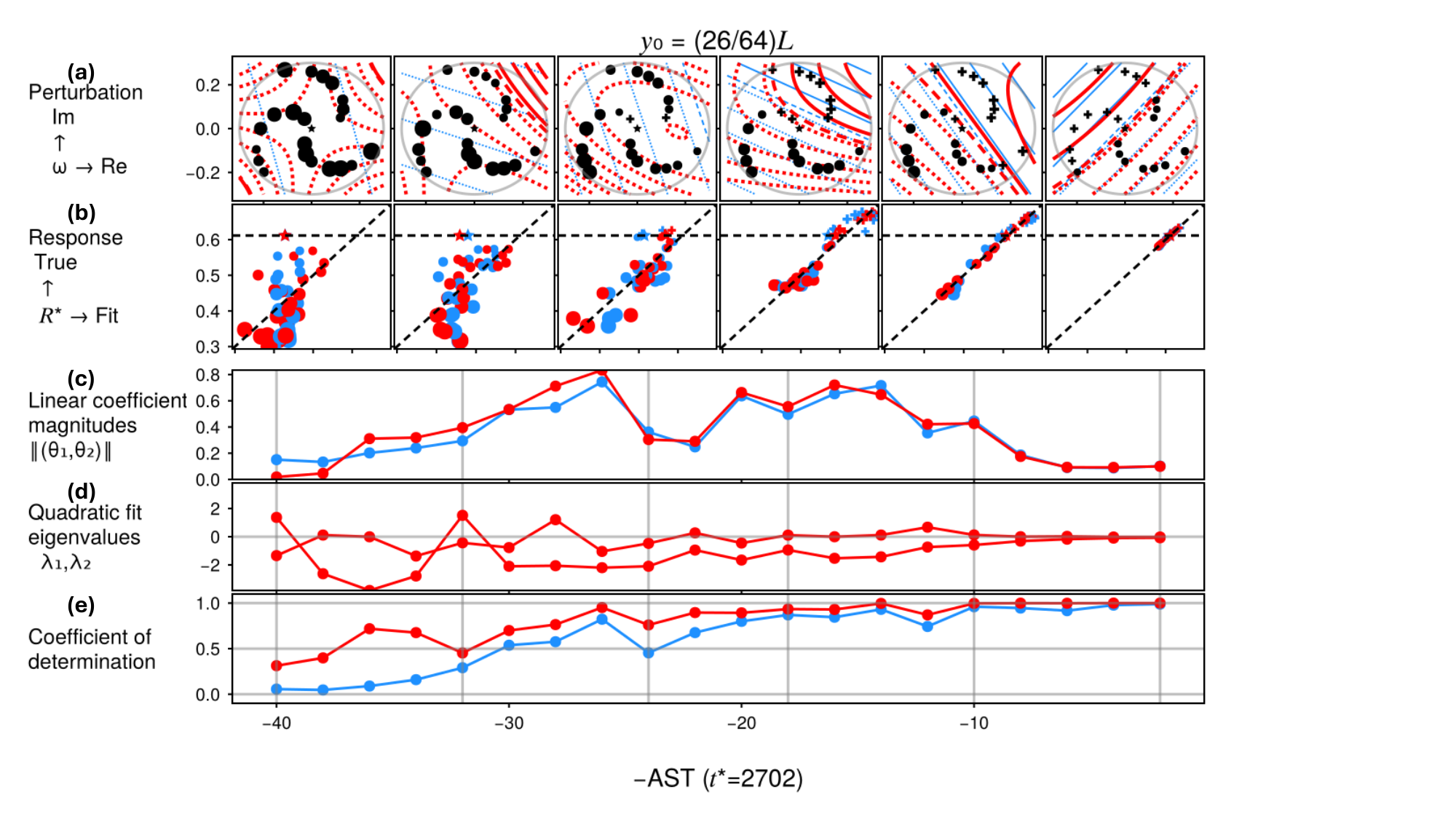}
    \caption{The response of an extreme event to perturbations: magnitude, phase, and timing. The event is the same as in Fig.~\ref{fig:spaghetti}b. Row (a) represents impulses as in Fig. \ref{fig:pert_struct}, but additionally shows the responses to them separately at six sampled ASTs (2, 10, 18, 24, 32, and 40 days, marked with vertical gray lines in c-e), which increase from right to left (launch time $t^*-A$ increases left to right). Horizontal and vertical scales are equal. At the shortest AST shown, $A=2$, the response function is clearly linear: the impulses above and left of center are marked by $+$, representing an increased severity, and those below and right of center are marked by $\bullet$, representing decreased severity, with marker sizes representing the magnitude of the change. Colored curves represent level sets of the fitted linear (blue) and quadratic (red) models, with (solid, dashed, dotted) contours to differentiate (positive, zero, negative) changes to $R^*$. Row (b) displays the quality of these models by plotting true vs. fit responses (again, horizontal and vertical scales are equal). As AST increases, the impulses causing higher and lower severities become more intertwined and less linearly separable, as the red contours progressively bend and separate from the blue contours. Accordingly, the modeled linear response ceases to correlate with the true response. The modeled quadratic response has a slightly longer range of good quality, but also fails for AST $\gtrsim26$ days. Row (c) shows that the linear components $\theta_1,\theta_2$ are estimated similarly (at least in magnitude) regardless of whether quadratic terms are also included. Row (d) shows that the quadratic model implies a local maximum (both eigenvalues nonpositive) for most of the range $A<26$, beyond which the landscape starts looking less like a hilltop and more like a saddle. Row (e) displays the coefficients of determination, conventionally denoted $R^2$ but not here so as to avoid confusion with intensity $R$.}
    \label{fig:response}
\end{figure}

The linear and quadratic response functions take the form 

\begin{align}
    \widehat{R}^*(\omega;\theta)
    &=
    \theta_0
    +
    \theta_1\text{Re}\{{\omega}\}
    +
    \theta_2\text{Im}\{{\omega}\} 
    &&
    \theta_0,\theta_1,\theta_2\text{ fitted for both linear and quadratic models}  
    \label{eq:response_linear}
    \\
    &+
    \theta_3\text{Re}\{{\omega}\}^2
    +
    \theta_4\text{Re}\{{\omega}\}\text{Im}\{{\omega}\}
    +
    \theta_5\text{Im}\{{\omega}\}^2 
    &&
    \theta_3,\theta_4,\theta_5\text{ fitted for quadratic model only.}
    \label{eq:response_quadratic}
\end{align}
We use ordinary least squares regression on the $M=21$ sampled impulses $\{\omega_m\}_{m=1}^M$ and associated severities $\{R_{n,j,m}^*\}$, in addition to the non-perturbed ancestor ($\omega_0:=0$) with severity $R_{n,j,0}^*=R_n^*$. A different set of coefficients is calculated separately for each ancestor $n$ and AST $A_j$. The response functions for the same ancestor event as in Figs.~\ref{fig:spaghetti}b are visualized in Fig. \ref{fig:response}, using (a) the two-dimensional response surfaces, (b) the true vs. fitted response values, (c) the overall slope, measured by the linear coefficient magnitudes, (d) the overall curvature, measured by the eigenvalues of the Hessian of the quadratic fit, and (e) the overall linear and quadratic skills, measured by the coefficient of determination. The response surface gradually transforms from a linear plane, to a curved hilltop, to a saddle, to a jagged landscape, as AST increases. Accordingly, the linear and then the quadratic model lose their skill. The quadratic model is slightly better than the linear model for this particular event, but substantially better when averaged across all events (see the forthcoming Fig. \ref{fig:conditional_severity_distributions}c.i), and so we will use quadratic models only as $\widehat{R}^*$ in the tail estimators. 

\subsection{Conditional severity PDFs: case studies}

Equipped with response functions approximated by quadratic models, we can now construct conditional severity PDFs using Eq.~(\ref{eq:conditional_severity_estimate}), which are displayed in Fig. \ref{fig:conditional_severity_distributions}. For the same ancestor as in Fig. \ref{fig:response} and the same six ASTs, we can see the relationship between actually sampled perturbed severities (red circles and lines), fitted severity PDFs (colored curves, one color for each input scale $s$) evaluated at the bins with lower boundaries $\{\mu[(\frac12)^k]:k=5,\hdots,14\}$, and the climatological PDF (black curves). As AST increases from right to left, the severity PDFs morph from narrow spikes centered at the ancestor severity to long, extended lumps reaching far beyond the ancestor severity, and then recede below the threshold $\mu[(\frac12)^5]$. The PDF's motion resembles a wave crashing onto a shallow beach, blanketing the sand, and then retreating, hitting the true COAST somewhere in the middle stages. But this general behavior is strongly modulated by the choice of scale $s$: red PDFs, representing the smallest scale $s=0.06$, are narrower and located closer to the ancestral severity (horizontal black line) for all ASTs, whereas blue PDFs, representing the largest scale $s=0.9$, spread out further as a result of giving more weight to bigger impulses. This underscores our claim that the input distribution, an arbitrary choice, merits sensitivity analysis, and so we carry it through the remaining steps. 

\subsection{AST selection criteria: case studies}
\begin{figure}
    \includegraphics[width=0.8\linewidth,trim={0cm 0cm 4cm 0cm},clip]{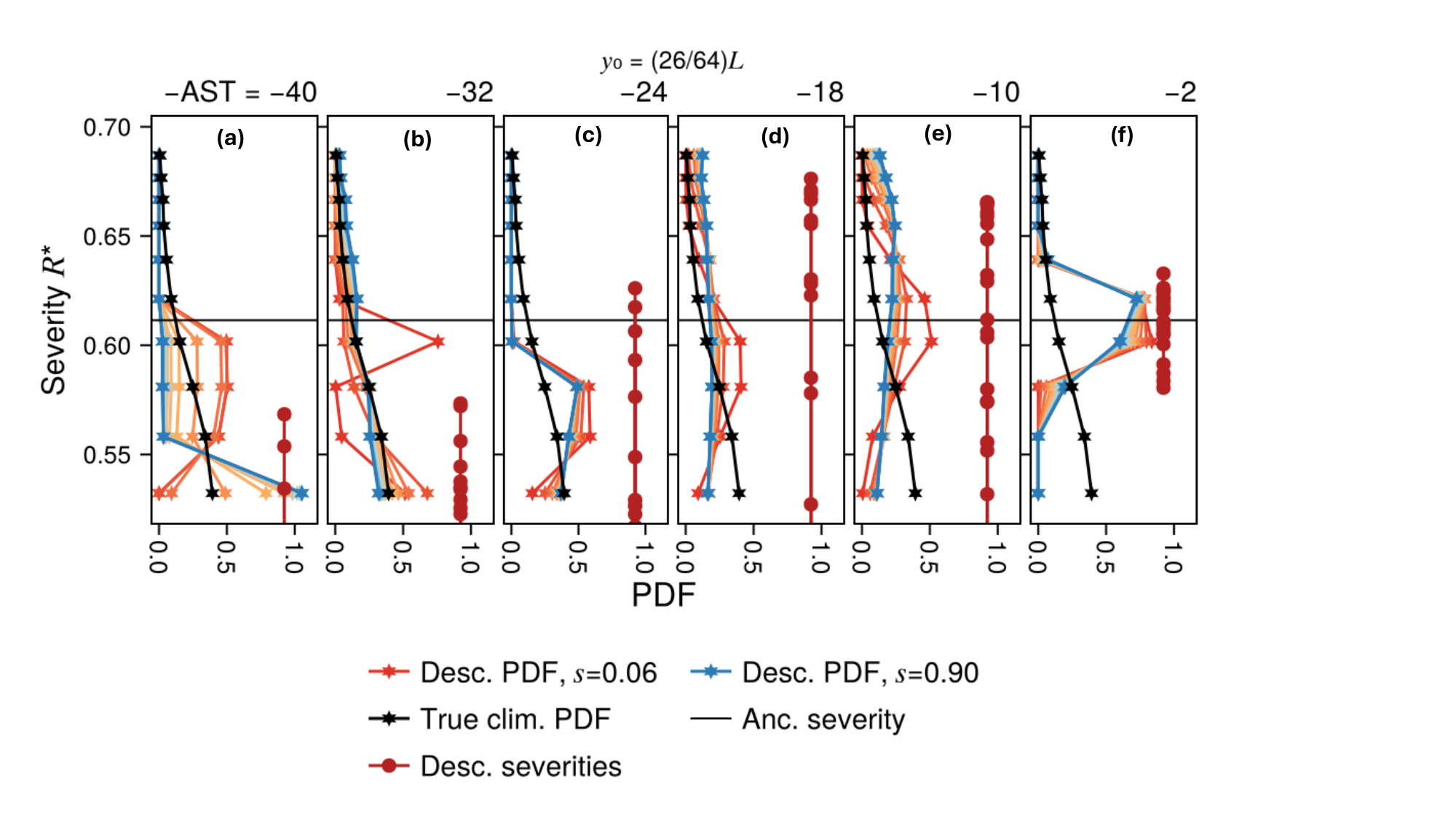}
    \caption{Severities and their conditional distributions for the same case study as Fig. \ref{fig:spaghetti}b. For six ASTs (same as Fig. \ref{fig:response}, decreasing from left to right), perturbed severities are displayed as dark red circles along a vertical line, and the unperturbed (ancestral) severity is marked with a horizontal black line. Colored curves and stars show the severity PDFs above $\mu=0.52$  as inferred from the quadratic regression, for a range of scales $s$ from 0.06 (red) to 0.9 (blue). Note that the longest ASTs (40 and 32 days) show a substantial probability mass beyond the most-extreme sample. This is a sign of poor quadratic fit, which is consistent with Fig. \ref{fig:response}e, and fortunately does not affect the later analysis since optimal ASTs are well short of 32 days. Black curves with stars represent the climatological tail PDF, as inferred from the long DNS, which we will seek to estimate by combining conditional distributions over many ancestors (not just the single ancestor considered here).
    }
    \label{fig:conditional_severity_distributions}
\end{figure}

\begin{figure}
    \includegraphics[width=0.98\linewidth]{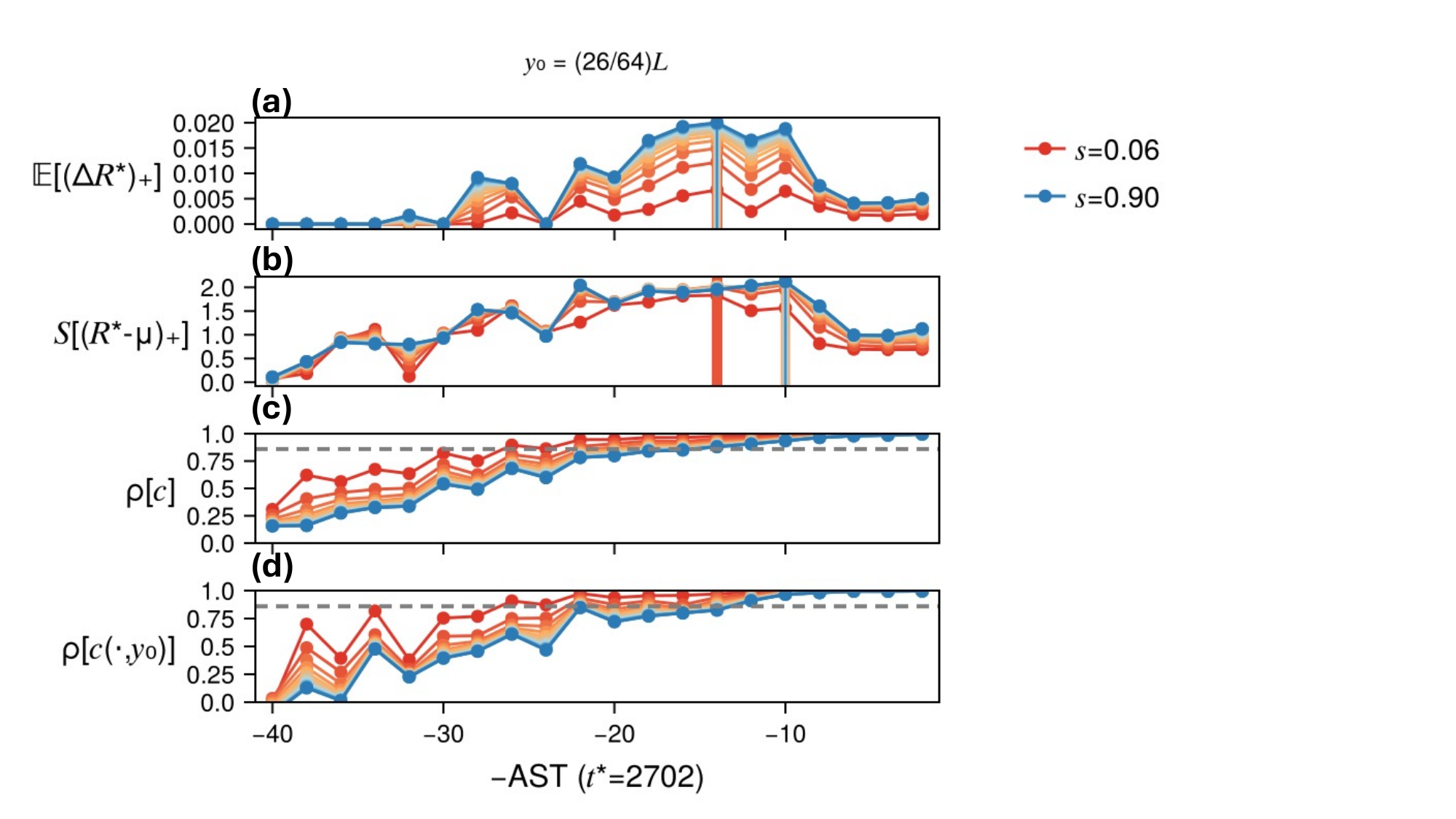}
    \caption{Ensemble dispersion indicators as a function of AST, again for the same case study as Fig. \ref{fig:spaghetti}b: (a) expected improvement EI, (b) thresholded entropy TE, (c) local and (d) global correlations. Colors indicate input scales $s$, from small (red: $s=0.06$) to large (blue: $s=0.9$). In (a,b), vertical bars mark the respective optimal ASTs, which may depend on the scale. In (c,d), horizontal dashed lines are positioned at $1-(\frac38)^2$, corresponding to the rule of thumb from \citet{Finkel2024bringing}.}
    \label{fig:conditional_dispersion_indicators}
\end{figure}

Figure \ref{fig:conditional_dispersion_indicators} display the criteria proposed in Sect. \ref{sec:ast_selection_criteria} that might help determine in which stage of ``wave breaking'' the severity PDF finds the COAST. The EI and TE criteria shown in panels Fig. \ref{fig:conditional_dispersion_indicators}(a,b) both exhibit non-monotonic behavior by design, maximizing at COASTs denoted $A^{\text{EI}}$ and $A^{\text{TE}}$ (see Sect. \ref{sec:ast_selection_criteria}). The AST dependence can be heuristically understood in light of the PDFs in Fig. \ref{fig:conditional_severity_distributions}:
\begin{itemize}
    \item At small AST, the narrow PDFs have a relatively high \emph{probability} of improvement over the ancestor ($\sim\frac12$), but only by small amounts, hence a small EI. By a similar token, the TE terms in Eq.~(\ref{eq:thresholded_entropy}) are almost all positive because the PDF is situated well above $\mu$, but being concentrated in a small number of bins makes its information content low.
    \item At intermediate ASTs of 10-20 days, the PDFs remain roughly centered at the ancestor's severity, meaning that improvements remain highly probable, but are larger when they happen thanks to the long upper tails, contributing to a large EI. Meanwhile, both upper and lower tails contribute to a large TE, which does not directly favor exceptionally high severities but rather \emph{diverse}  severities that are \emph{high enough} to exceed $\mu$.
    \item At large AST past $\sim25$ days, the PDFs have diminishing mass above $\mu$, let alone above the ancestor severity $R_n^*$, which zeros out most of the contributions to both EI and TE.
\end{itemize}
The COAST can change with the scale $s$: even though the overall shapes of TE and EI don't change very much, the location of their maxima might. 
Fortunately, we will find changes in scale for $s\gtrsim0.24$ to have negligible impact.

Fig. \ref{fig:conditional_dispersion_indicators}(c,d) display two versions of pattern correlation $\rho$, defined in Sect. \ref{sec:ast_selection_criteria} for an arbitrary field $F$: the ``global correlation'' $\rho[c]$ uses the whole two-dimensional upper-layer concentration field $F(x,y)=c_1(x,y)$, and the ``local correlation'' $\rho[c(\cdot,y_0)]$ uses only the single-latitude transect $F(x)=c_1(x,y_0)$ at the target latitude $y_0$. Both drop off steadily with AST, although local correlation fluctuates more due to averaging a smaller spatial region. The influence of perturbation scale $s$ enters at the ensemble-averaging step, where the $m$th member's pattern correlation $\rho[F_0,F_m]$ is weighted by $p(\omega_m,s,W)$. Since smaller perturbations take longer to grow, smaller input scales lead to slower dropoff of $\rho$ with $A$---but only at short lead times, where errors are still tiny. Beyond $A\approx6$ and 10 days for global and local correlations respectively, decorrelation proceeds at a similar rate with respect to increasing AST for all scales. The nominal threshold $\rho=\sqrt{1-(\frac38)^2}$ is marked in both, and gives a similar AST for local and global correlations but generally longer than implied by EI or TE. 

\subsection{AST selection criteria: aggregate results}

\begin{figure}
    \includegraphics[width=0.98\linewidth,trim={0cm 0cm 8cm 0cm},clip]{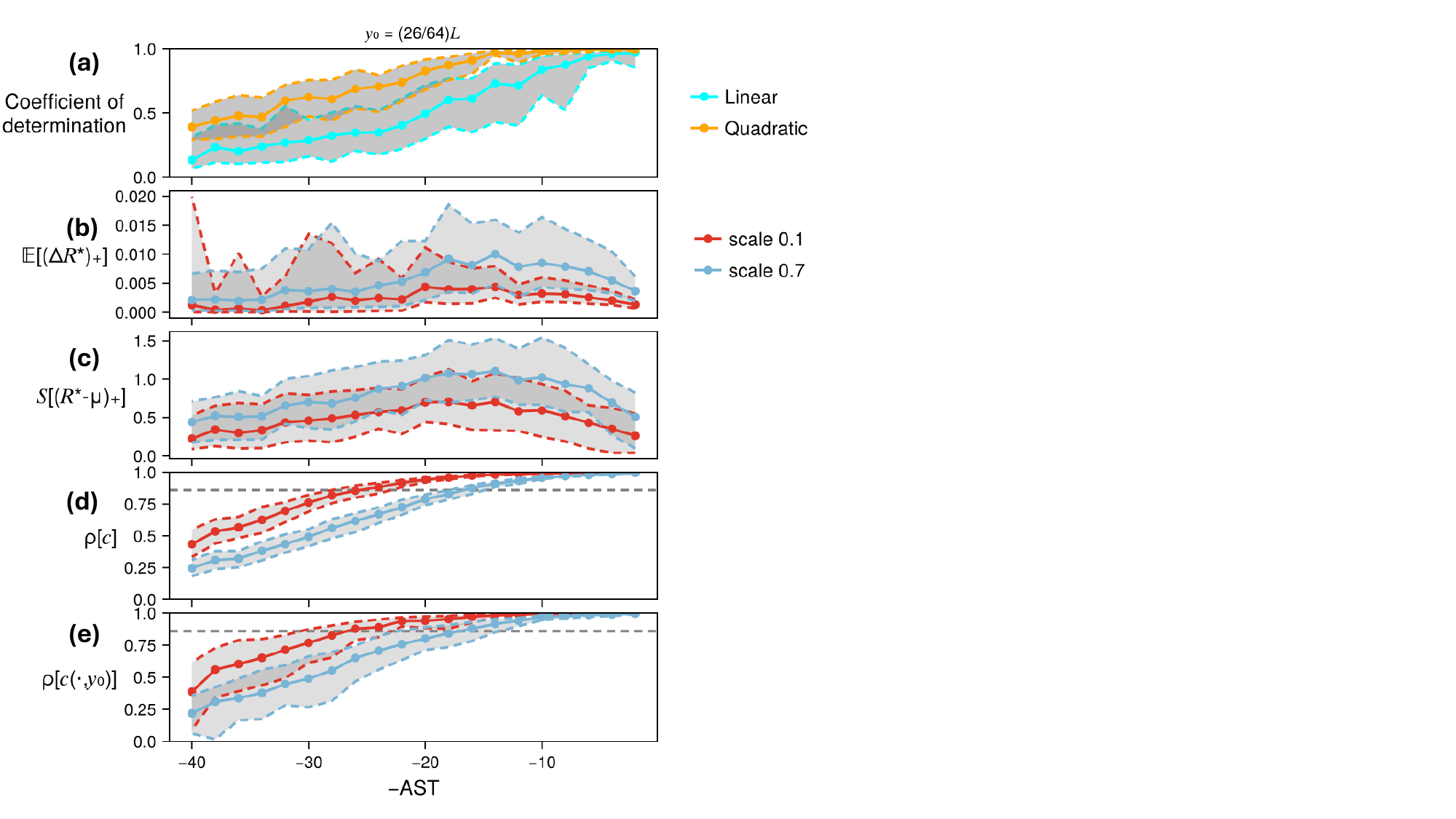}
    \caption{Ensemble dispersion metrics averaged across ancestors at $y_0=26/64 L$. (a) Coefficients of determination for linear (cyan) and quadratic (orange) regressions, averaged across ancestors. (b-e) same quantities as in in Fig. \ref{fig:conditional_dispersion_indicators}(a-d) but averaged across ancestors, with only the largest and smallest scales shown (red: $s=0.06$, blue: $s=0.9$). Shaded regions indicate variation across ancestors, which we quantify using \emph{truncated upper- and lower-means}. For example, the upper truncated mean for correlation $\rho$ is the mean of $\rho$ across ancestors with above-average $\rho$: $\mathbb{E}[\rho|\rho>\mathbb{E}[\rho]]$, separately at each AST. We choose truncated means to avoid the awkward properties of more standard measures of spread: interquartile ranges would be erratic for the relatively small sample size of ancestors, whereas standard deviation envelopes can misleadingly fall outside the bounds $[0,1]$ to which $\rho$ is constrained.}
    \label{fig:average_dispersion_indicators}
\end{figure}

Fig. \ref{fig:average_dispersion_indicators} goes beyond the case study to show dispersion indicators averaged across all ancestors. The coefficients of determination for linear and quadratic models (Fig. \ref{fig:average_dispersion_indicators}a) are farther apart on average than they are for the case study (see Fig.~\ref{fig:response}e), the quadratic model enjoying much higher skill especially during the pivotal 10-20 day range when EI and TE tend to maximize (Fig. \ref{fig:average_dispersion_indicators}b,c). This validates our choice to use the quadratic model. Overall, the EI, TE, global and local correlations (Fig. \ref{fig:average_dispersion_indicators} b-e) are similar on average to the case study, but smoother. 

Note, however, that these averaged dispersion indicators are never used directly in AST selection: the COASTs are chosen separately for each ancestor as the maximizer of its own EI or TE, or at the longest AST such that global or local correlation is above $\rho^{\text{U}}$. This nuance is further illustrated in Fig. \ref{fig:opt_landscapes_y26}(a,b), where (EI, TE) are plotted as joint functions of AST and input scale. Whereas the heatmaps are averages over ancestors of EI and TE just like Fig. \ref{fig:conditional_severity_distributions}c.(ii,iii), the red circles indicate the fraction of ancestors whose EI or TE is maximized at a particular AST for each particular scale. We call the red circle sizes ``COAST frequencies''. For example, at $s=0.24$, the mean EI maximizes at $A=14$ days, and that same AST is the most frequent COAST. However, the second-largest circle indicates that $A=20$ days is a close second-most frequent COAST according to EI. At the same scale, the most frequent COASTs according to TE are $A=18$ and $20$. In general, we gather two patterns from Fig. \ref{fig:opt_landscapes_y26}(a,b): the average EI and TE values (i) are well-correlated with their corresponding COAST frequencies, and (ii) both change rapidly at small scales but stabilize above $s\approx0.24$, at which point the input distributions are close enough to uniform over the $W$-disc. This relative stability is reassuring, but we generally prefer smaller noise which disturbs the model dynamics less. To balance these considerations, we select $s=0.24$ as the nominal scale to examine more closely going forward.  

\section{Results: Climatological severity distributions}
\label{sec:climatological_severity_distributions}

Having explained the construction of conditional distributions, we now aggregate across ancestors using MoCTail and PoPTail estimators to obtain our estimates of the climatological severity distribution from the boosted ensembles. We evaluate the skill of each AST selection rule by the $\chi^2$ divergence of the resulting climatological distribution from ground truth as obtained from the long DNS. We first restrict attention to extremes at $y_0=\frac{26}{64}L$ and then assess a broader swath of latitudes.

\begin{figure}
    \includegraphics[width=0.98\linewidth,trim={0cm 0cm 6cm 0cm},clip]{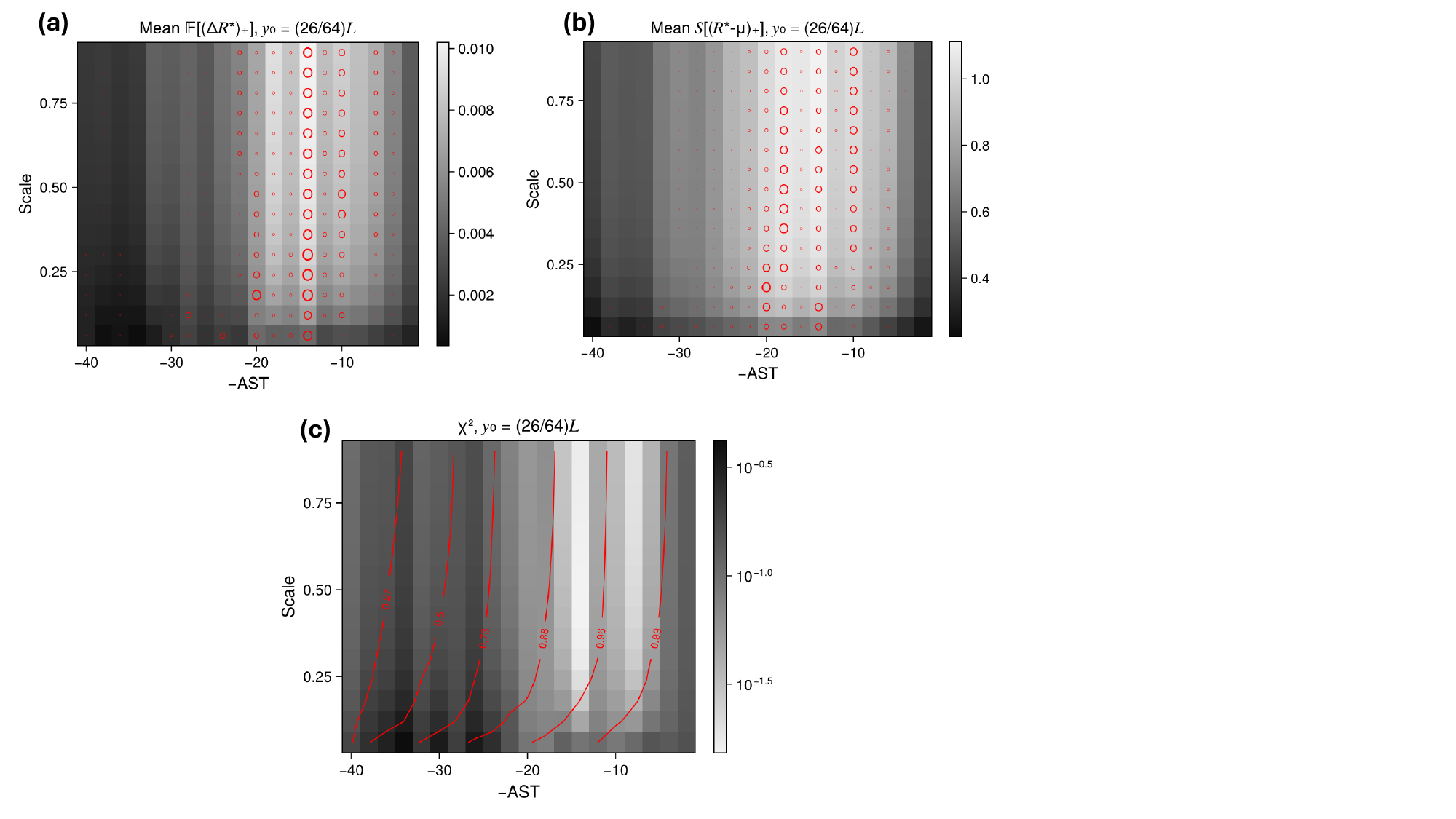}
    \caption{Three optimization landscapes as joint functions of AST and input scale for $y_0=(26/64)L$: (a) expected improvement (EI), (b) thresholded entropy (TE), and (c) $\chi^2$ divergence between the MoCTail and ground truth. Lighter gray indicates better performance---smaller $\chi^2$ divergence or larger EI and TE---and the corresponding ``best'' ASTs consistently fall in the \emph{interior} of the domain, across all scales. Contours of local correlation $\rho[c(y_0,\cdot)]$ are overlaid in (c), giving a rough map of correspondence between correlation levels and AST. The size of red circles in (a,b) indicate the ``COAST frequency'': the fraction of ancestors whose (EI, TE) is maximized at the corresponding AST while holding the scale fixed. Note the multiple local maxima in mean EI and TE (as indicated by the lightness of the gray color in (a,b)), each of which is the global maximum for some significant set of ancestors.  }
    \label{fig:opt_landscapes_y26}
\end{figure}

First, consider the simplest AST selection rule $A=A^{\text{U}}$, a uniform AST over all ancestors. We have no \emph{a priori} principle for $A^{\text{U}}$, so we search through all possible values from 2 to 40 days. Fig. \ref{fig:opt_landscapes_y26}c displays the resulting $\chi^2$ divergence between the MoCTail and ground truth, as a function of $A^{\text{U}}$ and input scale. A clear optimum emerges at $A^{\text{U}}=14$ days and persists for all scales $s\gtrsim0.24$, after rapid changes across smaller scales. Red contours also indicate the local correlation, averaged across ancestors to give a smooth and monotonic function of AST. In terms of correlation, the COAST $A^{\text{U}}=14$ days corresponds to $\rho^{\text{U}}\approx0.92$ depending on the scale, which is slightly above the nominal value $1-(\frac{3}{8})^2=0.86$, meaning one should split a little bit closer to the event than the rule of thumb implies. 

Overall, the $\chi^2$ landscape (inverted) roughly aligns with the EI and TE landscapes, as do their respective optima. This is remarkable and encouraging:  allowing each ancestor to determine its own COAST independently, with no knowledge of the ground truth or even other ancestors' COASTs, leads to a similar solution as the policy of synchronizing them all. Boosting based on EI and TE, therefore, is more parallelizable (optimizations are decoupled across ancestors), extensible (new ancestors can be added without changing the optimal split times for pre-existing ancestors), and interpretable (one can see the optimum clearly based on a case study, without complicated averaging procedures across initial conditions).

\begin{figure}
    \includegraphics[width=0.98\linewidth,trim={0cm 1cm 6cm 0cm},clip]{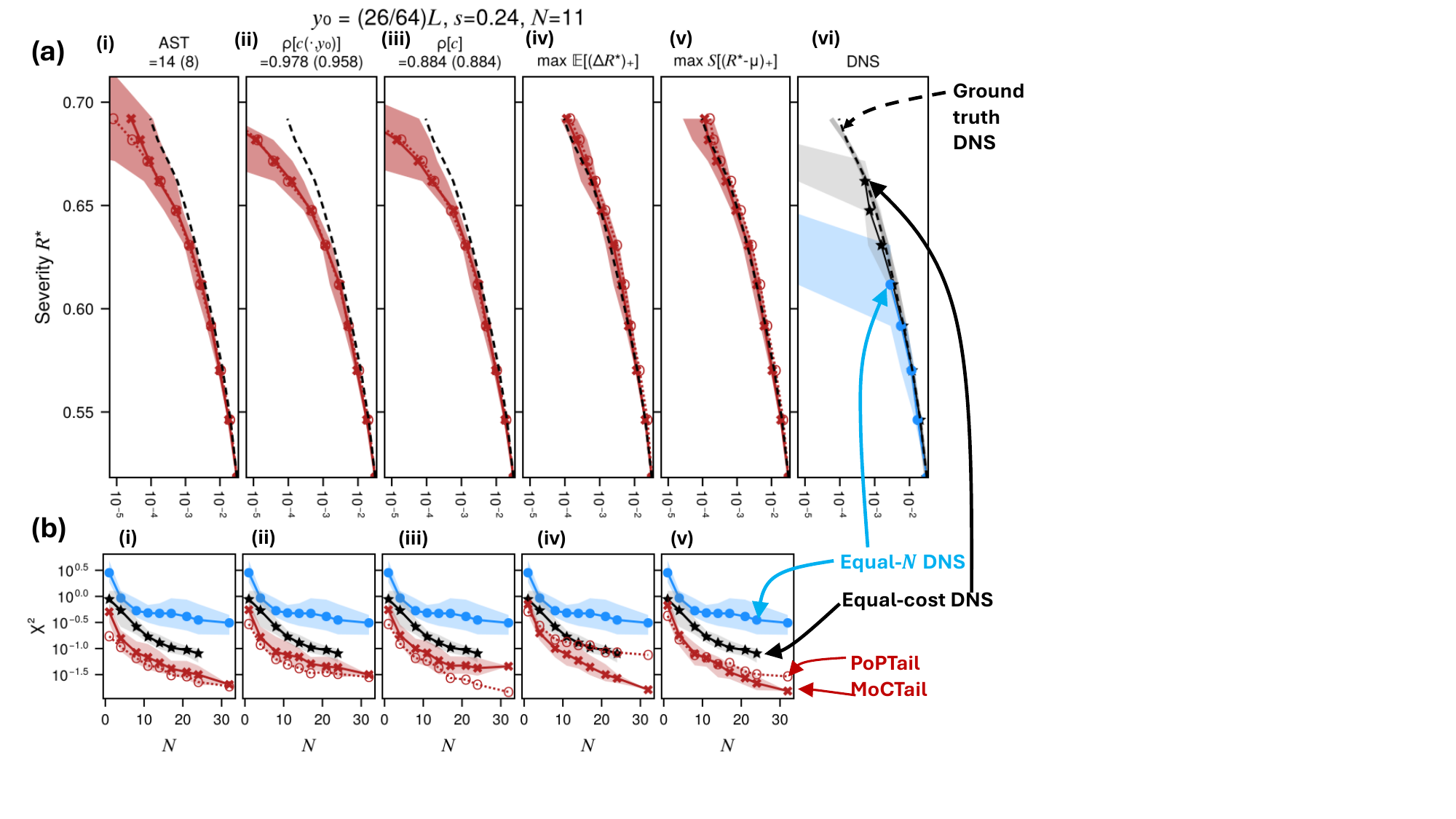}
    \caption{CCDF approximations by various mixing criteria and associated errors, at the latitude $y_0=\frac{26}{64}L$ and input scale choice $s=0.24$. (a.i-v) Tail CCDFs by various estimates using only $N=11$ ancestors, with lines showing medians and bands showing interquartile ranges across many size-11 subsamples of the total set of 32 ancestors. Dotted lines with open circles are PoPTails, while solid lines with crosses are MoCTails. Dashed black lines show the ground truth estimate. Panel a(i) shows the tail approximation using a single uniform AST indicated at the top: 14 days for MoCTail and 8 days (parenthesized) for PoPTail. Panels a.(ii,iii) show the tail approximations using thresholds of (local, global) correlations as AST selection criteria. 
    Panels a.(iv,v) show the tail approximations obtained by maximizing (EI, TE), which unlike the other criteria do not rely on knowing the ground truth to select ancestor-wise ASTs, either directly or through threshold choice.
    (a.vi) also shows estimates from DNS with equal cost to boosting on 11 ancestors (black stars, gray envelope) and DNS from only $N=11$ peaks (brown circles and envelope), in both cases estimating uncertainty by longitudinal rotation. The GPD fit to ground truth is shown as a gray curve.  In a.(i-iii), the thresholds shown at the top (PoPTail thresholds parenthesized) are obtained by using all 32 ancestors, but the CCDFs displayed each choose an AST to minimize $\chi^2$ divergence from ground truth, separately for each subsample. Because this requires ground truth knowledge, the $\chi^2$ divergences must be interpreted as practical lower bounds. The 90\% error bar applies to the MoCTail estimator only, and comes from bootstrapping on entire ``families'' or in other words mixture components (not individual descendants) and choosing the best AST (by $\chi^2$ divergence) for each particular subsample. The error bar widths, too, must then represent lower bounds. 
    (b) $\chi^2$ values for the estimator directly above in each case as a function of $N$, and compared with DNS at equal cost and equal $N$. DNS does not run long enough to equal the total cost accrued by boosting 32 ancestors, so the black curve stops before the others. }
\label{fig:mixture_ccdfs}
\end{figure}

Fig. \ref{fig:mixture_ccdfs} makes a tail-to-tail comparison between all the AST selection rules (a.i-v: $A^{\text{U}},A^{\text{PC}}$ local and global, $A^{\text{EI}},A^{\text{TE}}$), fixing the scale to $s=0.24$ and (in the case of $A^{\text{U}}$ and $A^{\text{PC}}$) selecting \emph{post-hoc} the best-performing threshold to set the COASTs. We used subsets of only 11 of the 32 ancestors, resampling such subsets 64 times to obtain medians (solid) and interquartile ranges (shading) on CCDFs. The numerical values of optimal AST and $\rho$ reported above a.(i-iii), with PoPTail optima parenthesized, are the optima obtained from $N=32$, i.e., the best estimates of the true optima; they don't necessarily correspond to the values used for plotting with $N=11$, which are optimized separately for each resampling.  The brown CCDF in panel (a.vi) is the estimate from the unboosted acestors alone (``equal-$N$''), and the black is the estimate from a larger number of ancestors to equal the cost of boosting. The curves underneath in panel (b) show the rate of improvement of $\chi^2$ with $N$. 

In terms of quantitative improvements in $\chi^2$ for a fixed cost (vertical differences between curves), all the rules considered ($A^{\text{U}},A^{\text{PC}},A^{\text{EI}},A^{\text{TE}}$) improve substantially upon an equal-$N$ DNS and modestly upon an equal-cost DNS. The size of the advantage varies with $N$ in the way that we expect from boosting: substantial improvements in $\chi^2$ with moderate $N$, ($\sim$5-10) when the DNS has sampled the attractor broadly but sparsely and extremes are within reach by perturbation. The advantage might diminish if $N$ increases enough for DNS to see those extremes without perturbation, but we haven't reached that regime yet. MoCTail and PoPTail performances are similar, but not identical: PoPTail seems more suited for threshold-based rules ($A^{\text{U}},A^{\text{PC}}$ local and global in b.(i-iii)), whereas MoCTail seems more suited for optimization-based rules ($A^{\text{EI}},A^{\text{TE}}$ in b.(iv,v)). Another way to measure boosting advantage is by ``speedup'': given a prescribed $\chi^2$ error, how much extra simulation is needed with DNS relative to boosting to achieve it. These are the horizontal distances between curves. Across all AST criteria, speedup varies from $1.5\times$ to $3\times$, and accelerates sharply as the DNS curve flattens around $\chi^2\sim10^{-1}$ while the boosting curves continue to decrease linearly. These are modest speedups compared to other published rare event algorithms, which report between one and four orders of magnitude speedup depending on the event definition and the algorithm \citep[e.g.,][]{Ragone2021rare,Finkel2026rare}, but again, we stress that the computational savings here are only incidental to our main goal of characterizing the COAST. Substantial improvements should be possible by targeted optimization and, potentially, repeated rounds of boosting. 

We selected $N=11$ to display the full CCDFs in Fig. \ref{fig:mixture_ccdfs}(a) as the middle range of values tried, and where enough equal-size ancestor subsets are available for uncertainty quantification by bootstrapping. When comparing with DNS CCDFs, all five rules successfully extend the short, equal-$N$ DNS tail into a longer tail that tracks closer to the ground truth farther into the extreme severity range. They also all find a larger maximum than even the equal-cost DNS found. However, the threshold-based rules exhibit apparent bias, systematically underestimating probabilities for $R^*\gtrsim0.64$, whereas the optimization-based rules are both more accurate and more confident. Our hypothesis for this behavior is that each ancestor has its own predictability timescale, physically linked to the frequency of the wave responsible for that particular event, and that these ancestor-varying timescales cannot all be respected at once by a single, globally imposed time like $A^{\text{U}}$, or even a globally imposed correlation threshold to dictate $A^{\text{PC}}$. The optimization-based criteria $A^{\text{EI}}$ and $A^{\text{TE}}$ are tailored to the ancestor, and might in fact be choosing those predictability timescales implicitly. This is only speculation, however, and must be validated with more detailed analysis than fits in our present scope. 

The COASTs identified by all rules lie strictly between the shortest and longest ASTs considered. For example, $A^{\text{U}}=14$ according to the MoCTail estimator (using all $N=32$ ancestors). By comparing with Fig. \ref{fig:opt_landscapes_y26}c, we recognize 14 as the minimum of the $\chi^2$ landscape for $s=0.24$ (and larger scales), with an approximate local-correlation equivalent of 0.98.

\begin{figure}
    \includegraphics[width=0.98\linewidth,trim={0cm 0cm 0cm 0cm},clip]{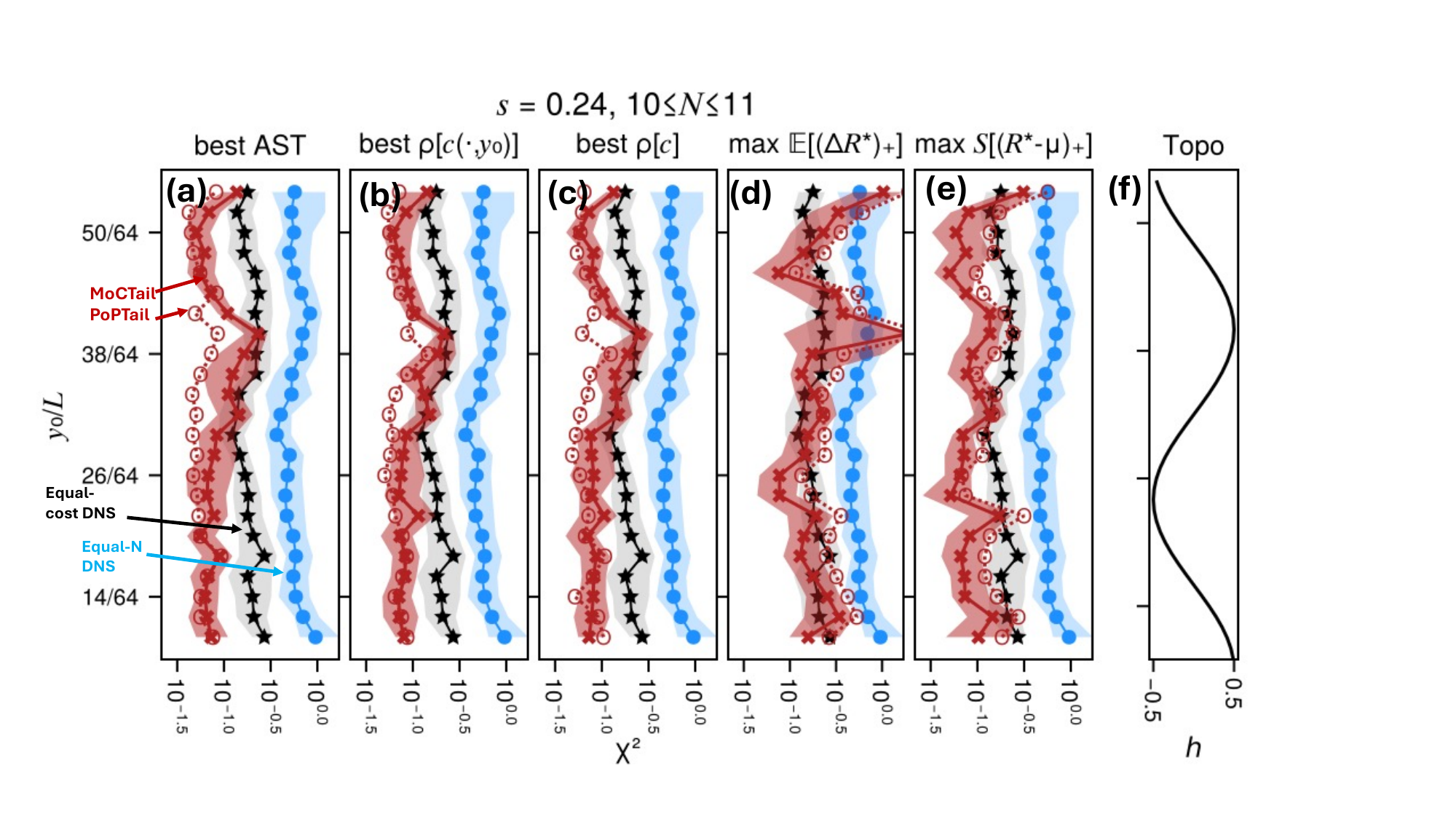}
    \caption{Performance of all AST selection criteria, measured by $\chi^2$ divergence, across all latitudes for $s=0.24$ and $N=10$ or 11, whichever is nearest to 1/3 the number of ancestors found for the latitude in question (sometimes less than 32). Black line and gray envelope represent the error from the short DNS and its 90\% error bar according to quantiles across longitudes. Panels a-e parallel Fig. \ref{fig:mixture_ccdfs}a.(ii-vi). Solid lines and crosses represent the MoCTail estimator, while dotted lines with open circles represent the PoPTail estimator. 
    }
\label{fig:chi2_latdep}
\end{figure}

Similar patterns hold across target latitudes, but with some notable caveats. The $\chi^2$ divergences of each selection rule are plotted in Fig. \ref{fig:chi2_latdep}, of which Fig. \ref{fig:mixture_ccdfs}c is one slice. The most obvious and important point holds: perturbed ensembles improve upon the DNS equal-$N$ estimate, for almost all latitudes and AST selection rules, and they also improve on the equal cost estimate in many cases. But $A^{\text{EI}}$ is less reliable; its favorable performance noted above in Fig. \ref{fig:mixture_ccdfs} is peculiar to the latitude $y_0=\frac{26}{64}L$. At some other latitudes, it is similar or worse in skill than equal-$N$ and even equal-cost DNS. Even so, it tends to fail by \emph{overestimating} severities, which we have confirmed by examining the corresponding CCDFs (not shown), and thus it may serve as a useful upper bound. The MoCTail and PoPTail estimators are similar in quality across latitudes, but as observed in Fig.~\ref{fig:mixture_ccdfs}, PoPTail has an advantage with threshold-based rules ($A^{\text{U}},A^{\text{PC}}$ local and global) whereas MoCTail performs better with optimization-based rules ($A^{\text{EI}},A^{\text{TE}}$). 

\begin{figure}
    \includegraphics[width=0.98\linewidth,trim={1cm 7cm 8cm 0cm},clip]{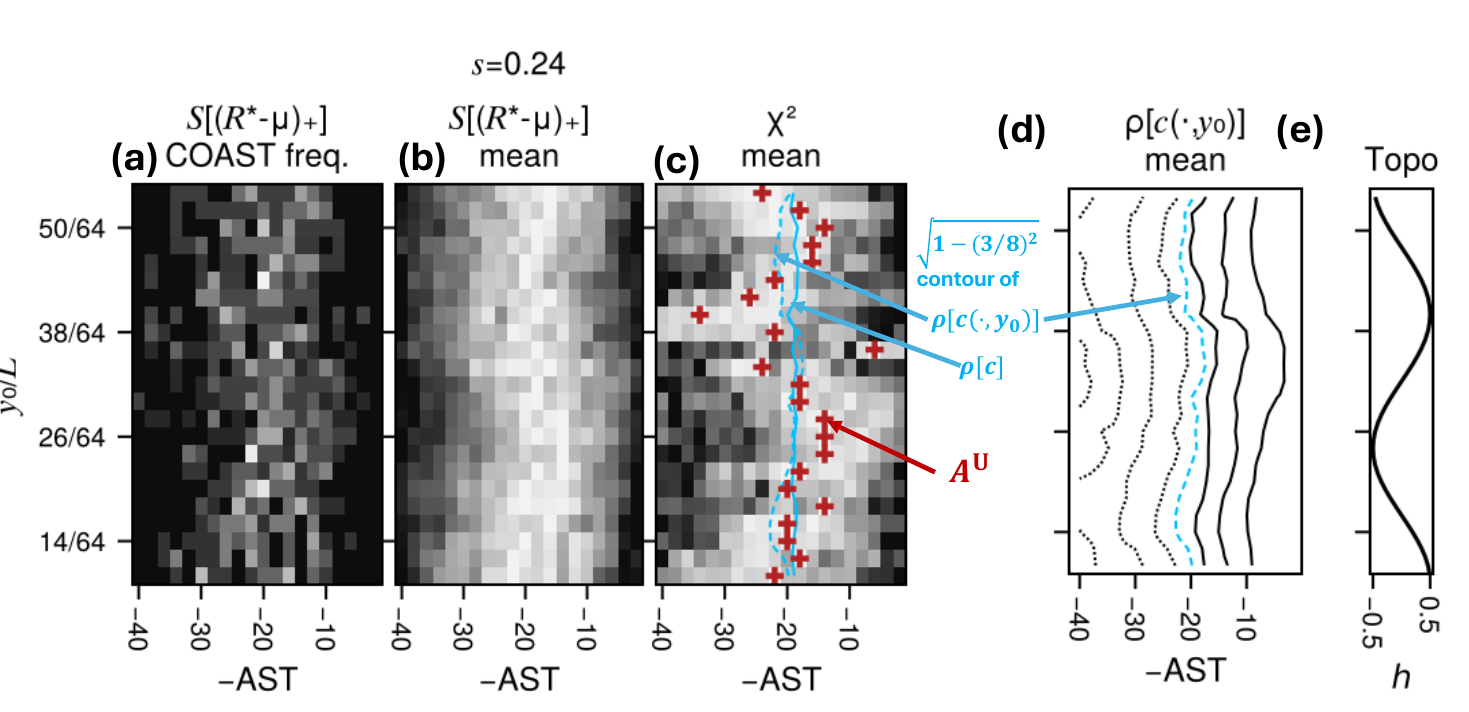}
    \caption{Optimization landscapes and optimal ASTs across latitudes, again fixing the input scale to $s=0.24$. 
    (a) Frequencies of \emph{conditionally} optimal ASTs (COASTs), in the maximum-thresholded entropy sense, at each latitude, with whiter shading indicating higher frequency. E.g., at $y_0/L=26/64$, the two adjacent bright pixels  at AST $=18,20$ indicate that for a large fraction of ancestors, the highest-entropy descendant ensemble is the one launched 18 or 20 days in advance of the peak.
    (b) Thresholded entropy as a function of AST, normalized to the range 0-1 (black-white, so brighter is better) separately at each latitude. This landscape is smoother than $\chi^2$ and varies less dramatically with latitude, but exhibits directionally similar trends.  
    (c) $\chi^2$ divergence as a function of AST and latitude, normalized to the range 0-1 (white-black, so brighter is better) separately at each latitude so that different latitudes are visually comparable. Red crosses mark the optimal AST at each latitude. Cyan (solid, dashed) curves mark the AST at which the (global, local) correlations, averaged across ancestors, reach $1-(\frac38)^2$. This nominal choice is based on \citet{Finkel2024bringing}, and falls squarely in the middle of the latitude-dependent ASTs. 
    (d) Contour map of local correlation, averaged over ancestors, as a function of AST and latitude. The levels range from 0.22 (left-most dotted black curve, fragmented by boundary) to 0.99 (rightmost solid black curve), evenly spaced in a stretched sigmoid scale (levels are shown only for qualitative purposes). The reference level $1-(\frac38)^2$ appears dashed in cyan. (g) Bottom topography for reference.}
    \label{fig:heatmaps_latdep}
\end{figure}

The various estimators and AST selection rules have differences in skill, but a more important commonality: all of them indicate that \emph{an optimal advance split time exists} that is strictly positive, which is not a foregone conclusion in light of standard rare event algorithms like adaptive multilevel splitting \citep[AMS;][]{Lestang2018computing} without ``trying early''. Fig. \ref{fig:opt_landscapes_y26} shows clear intermediate optima when targeting the single latitude $y_0=\frac{26}{64}L$, and Fig. \ref{fig:heatmaps_latdep} extends this result to all latitudes by stacking together cross-sections of the per-latitude counterparts of Fig. \ref{fig:opt_landscapes_y26} at $s=0.24$. The COAST frequency and mean-TE landscapes have broad ridges that meander slowly in AST space with latitude, approximately in phase with topography: smaller ASTs are favored at $y_0\approx\frac{26}{64}L$, where topography is minimized and meridional wind shear is negative, and larger ASTs are favored at $y_0\approx\frac{38}{64}L$, where topography is maximized and meridional wind shear is positive. A similar pattern, but with bigger swings, is seen in the $\chi^2$ landscape. All these patterns are a bit noisy, especially for the COAST frequencies and $\chi^2$-COAST locations, since both come from an inherently unstable ``argmax'' function. Nonetheless, the detailed latitude dependence is only a secondary effect on top of the main point, which is clearly demonstrated: splitting is most effective at intermediate ASTs rather than very short or long ASTs. 

We can also now evaluate the $\frac38$ rule from \citet{Finkel2024bringing} in this broader multi-latitude context, though here we simplify the procedure by first averaging $\rho$ across ancestors and then calculating $A^{\text{U}}$ as a threshold-crossing time of that average, which we call $A_{3/8}^{\text{U}}$, rather than averaging times $A_n^{\text{PC}}[\rho^{\text{U}}=1-(\frac38)^2]$ across ancestors. The same conclusion holds either way. The AST values $A_{3/8}^{\text{U}}$ are overlaid on the $\chi^2$ heatmap (Fig. \ref{fig:heatmaps_latdep}d) as blue curves. The solid curve, representing a level set of ancestor-averaged global correlation, should be constant with latitude and varies only due to sampling errors. Likewise, the dashed curve, representing a level set of ancestor-averaged local correlation, should be approximately symmetric with respect to latitude because of the symmetric tracer boundary conditions and approximate mirror symmetry in velocities, as should all the level sets in panel c. Since the $A^{\text{U}}$ varies differently with latitude, exhibiting roughly odd symmetry about the midline, the $\frac38$ rule cannot possibly be optimal for all latitudes simultaneously. More fundamentally, the COAST depends on more than just a generic metric for ensemble dispersion: it must also depend on the features of the tail being sampled, which in this case is the only possible source of meridional variation (see Fig. \ref{fig:gpd}). 

However, both versions of $A_{3/8}^{\text{U}}$ run right through the mean position of the meandering $\chi^2$ valley and associated COASTs, performing about as well as any such highly-constrained synchronized $A^{\text{U}}$ could do. Thus, the $\frac38$ rule retains its relevance as a starting point for more refined optimization more tailored to the event, at least for this QG system. Whether the $\frac38$ rule generalizes to more heterogeneous systems as the ``optimal synchronized AST'' requires further investigation. We found it provides some guidance for temperature and precipitation extremes in an idealized general circulation model, but overestimated the optimal AST in both cases \citep{Finkel2026rare}.

\section{Conclusion}  
\label{sec:conclusion}

Rare event sampling is a promising strategy to study extreme weather more efficiently with computer models by repeatedly cloning, perturbing, and re-simulating the most extreme events in an ensemble while tracking statistical weights. However, sudden and transient events such as mid-latitude precipitation present a particular challenge for rare event algorithms, leaving ensembles little time to diversify before the event passes by. Ensemble boosting \citep{Gessner2021very,Gessner2022physical,Fischer2023storylines,BloinWibe2025estimating} and ``trying-early adaptive multilevel splitting'' \citep[TEAMS;][]{Finkel2024bringing,Finkel2026rare} get around this problem by perturbing events farther in advance by some \emph{advance split time} (AST) to allow ensembles to spread, but this opens a pivotal question: how should we choose the AST for maximal accuracy and efficiency? If AST is too short, perturbations can't grow enough to give useful samples, and if it is too long, they regress to climatology. To deploy advance-splitting methods at scale, we need more reliable ways to set the AST as well as other hyperparameters. The AST itself may be a property of the physical system, not of algorithmic parameters like ensemble size, which would simplify its optimization while also yielding physical insight into causal mechanisms of the event.

In this paper, we have pursued this hypothesis and 
established the \emph{conditionally optimal advance split time} (COAST) as a quantity more intrinsic to the dynamical system than to the idiosyncrasies of a particular rare event algorithm by removing the confounding effect of randomly selecting ensemble members to split.  
The COAST also depends on the target observable of interest, the imposed distribution over perturbations,
and the initial conditions which may vary in their predictability. We formulate COAST mathematically as the solution to an optimization problem, and through 
a systematic boosting-based sampling and estimation procedure we discern the optimization landscape in the context of an idealized physical model: a baroclinically unstable quasi-geostrophic (QG) flow, with local passive tracer fluctuations as our extreme event of interest. To facilitate more efficient rare event sampling applications, we have further proposed various parsimonious rules for finding the COAST, and evaluated these rules empirically in the QG model. 

We have four conclusions to report:

\begin{enumerate}

    \item A boosting procedure, generated with a suitable AST, can well-approximate a probability distribution's tail using either of two estimators: ``MoCTail'', which we formulate here, and ``PoPTail'', due to \citet{BloinWibe2025estimating}.
    
    \item The optimal AST is strictly greater than zero and varies slowly with latitude, appearing  smaller in regions of negative meridional wind shear (e.g., the northern edges of westerly jets) and larger in regions of positive meridional wind shear (e.g., the southern edges of westerly jets).

    \item Several different rules for selecting the COAST are equally effective. Beyond the simplest option of setting a single fixed AST (called $A^{\text{U}}$), one can set a conditional AST (called $A^{\text{PC}}$) by thresholding on ensemble dispersion. Both $A^{\text{U}}$ and $A^{\text{PC}}$ perform similarly at tail reconstruction, but both unfortunately require a threshold choice, which there is no established method for selecting. Here we selected thresholds \emph{post hoc} with knowledge of the ground truth. The rule proposed in \citet{Finkel2024bringing}---that $A^{\text{U}}\approx$ the time until ensembles disperse to $\frac38$ their saturation value---appears to be a good single choice, but further improvement is possible by tailoring AST to the target location and the initial condition. 

    \item An attractive alternative to thresholding is \emph{optimizing} some functional of the ensemble severity distribution designed to favor both high extremes and wide spread. We have found a suitable functional in \emph{thresholded entropy} (TE), the expected information contained in that part of the ensemble's severity distribution exceeding the pre-selected threshold. Optimization-based AST rules open the door to using Bayesian optimization strategies to home in on the COASTs adaptively during an actual rare event sampling algorithm, avoiding the exhaustive grid searches we have performed here. 
\end{enumerate}

There are many important avenues of research indicated by the present study, both methodology-oriented and science-oriented. On the algorithmic front, it remains to be seen whether thresholded entropy succeeds at matching tail statistics in general systems, but the consistency across different targets within the QG model is encouraging. We suspect that \emph{some} similar objective function over distributions is broadly applicable. Furthermore, the \emph{shape} of perturbations is a possibly very important lever on the potency of perturbations, acting in concert with their timing. While we limited our present study to a two-dimensional perturbation space based on linearized dynamics about a baroclinically unstable background flow, a natural extension would be to use flow-dependent singular vectors as in operational weather forecasting. By design, they effect faster ensemble spread in the small-perturbation regime; however, it must be checked if their advantages carry into the finite-amplitude regime needed for effective rare event sampling. Computational tools such as adjoints, especially in novel machine learning models, invite the use of gradient-based optimization \citep{Wang2020useful,Vonich2024predictability,Whittaker2025constructing}. Since exhaustive grid search over ASTs and perturbation spaces is not an option when deploying rare event algorithms in practice, we are actively pursuing efficient optimization strategies, which are important to make use of this research. 

Intriguing dynamical questions also arise from the latitude dependence of the COAST, which can be seen as a predictability index tailored to extremes: how do the physical parameters such as topography, rotation rate, and the spatial domain affect COAST? Is the effect entirely explainable through the extreme value statistics, as we have speculated, or can two similarly shaped tails belie extremely different COAST behavior? These questions merit further parameter exploration, both within and beyond the quasigeostrophic framework. We expect to draw insight from recent theoretical advances relating extreme value theory to the geometry of chaotic attractors \citep{Lucarini2016extremes}.

In summary, our work makes empirical progress on important theoretical and algorithmic questions regarding the  probabilities of the most extreme weather events. Perturbed ensemble forecasts of individual weather events are distinct from the climatological distribution, but here we have given quantitative evidence for a relationship between the two---so long as the perturbations are well-timed---that can be exploited for efficient risk analysis via judicious perturbed simulations. Our work has elucidated what it means to be ``well-timed'', and furthermore provided quantitative optimization criteria for perturbation timing. Only with this basic pre-requisite information on what to optimize, should we proceed to invest effort into optimizing efficiently.

\codeavailability{
The code to generate all results, including simulation, statistical analysis, and plotting, is available at the Zenodo repository COAST \citep{COAST2025}. J.F. is happy to provide guidance on use and extension of the code upon request.} 





\appendix
\section{Langevin Model}

The schematic in Fig.\ref{fig:schematic} comes from Langevin dynamics, consisting of a single particle moving in one dimension with position $X(t)$ and momentum $Y(t)$ subject to a potential gradient force, friction, and stochastic Gaussian white-noise forcing $W(t)$:

\begin{align}
    dX(t)&=\frac1mY(t)\,dt\\
    dY(t)&=\Big[-V'(X(t))-\gamma Y(t)\Big]\,dt+\sigma\,dW(t) \\ 
    \text{ where the potential function }&V(x)\text{ has a quadratic core and logarithmic wings,}\\
    V(x)&=
    \begin{cases}
       \frac{\alpha+1}{\beta}\Big(\log(\epsilon)+\frac{(x/\epsilon)^2-1}{2}\Big) & |x|\leq\epsilon \\ 
       \frac{\alpha+1}{\beta}\log|x| & |x|>\epsilon,
    \end{cases}
\end{align}
which leads to a heavy-tailed (in $x$) steady-state probability density $p(x,y)\propto\exp\big[-\beta(V(x)+\frac{y^2}{2m})\big]\sim|x|^{-(\alpha+1)}$ for large $|x|$. Constant parameters are $\gamma=0.05$ for friction, $m=1.2$ for mass, $\sigma=0.005$ for stochastic forcing strength, $\epsilon=0.25$ for the extent of the quadratic core of the potential, $\alpha=3.1$ which sets the tail weight, and $\beta=2m\gamma/\sigma^2$ which is the inverse temperature.

\noappendix

\appendixfigures  

\appendixtables   


\authorcontribution{Justin Finkel formulated the initial study, carried out numerical computations, and wrote the initial draft. Paul O'Gorman and Justin Finkel both contributed to refining the methodology and substantially revising the manuscript. } 

\competinginterests{The authors declare no competing interests relevant to this study.} 


\begin{acknowledgements}
We thank Glenn Flierl, Andre Souza, and Talia Tamarin-Brodsky for helpful discussions and advice on theoretical and computational aspects of this work. This research is part of the MIT Climate Grand Challenge on Weather and Climate Extremes. Support was provided by Schmidt Sciences. Computations were performed on the MIT Engaging cluster.

\end{acknowledgements}

\bibliographystyle{copernicus}
\bibliography{references.bib}

@article{Mahesh2024hens1,
      title={Huge Ensembles Part I: Design of Ensemble Weather Forecasts using Spherical Fourier Neural Operators}, 
      author={Ankur Mahesh and William Collins and Boris Bonev and Noah Brenowitz and Yair Cohen and Joshua Elms and Peter Harrington and Karthik Kashinath and Thorsten Kurth and Joshua North and Travis OBrien and Michael Pritchard and David Pruitt and Mark Risser and Shashank Subramanian and Jared Willard},
      year={2024},
      eprint={2408.03100},
      archivePrefix={arXiv},
      primaryClass={physics.ao-ph},
      url={https://arxiv.org/abs/2408.03100}, 
}

@article{Finkel2024bringing,
author = {Finkel, Justin and O’Gorman, Paul A.},
title = {Bringing Statistics to Storylines: Rare Event Sampling for Sudden, Transient Extreme Events},
journal = {Journal of Advances in Modeling Earth Systems},
volume = {16},
number = {6},
pages = {e2024MS004264},
doi = {https://doi.org/10.1029/2024MS004264},
url = {https://agupubs.onlinelibrary.wiley.com/doi/abs/10.1029/2024MS004264},
eprint = {https://agupubs.onlinelibrary.wiley.com/doi/pdf/10.1029/2024MS004264},
note = {e2024MS004264 2024MS004264},
year = {2024}
}

@article{Finkel2023revealing,
author = {Finkel, Justin and Gerber, Edwin P. and Abbot, Dorian S. and Weare, Jonathan},
title = {Revealing the Statistics of Extreme Events Hidden in Short Weather Forecast Data},
journal = {AGU Advances},
volume = {4},
number = {2},
pages = {e2023AV000881},
doi = {https://doi.org/10.1029/2023AV000881},
url = {https://agupubs.onlinelibrary.wiley.com/doi/abs/10.1029/2023AV000881},
eprint = {https://agupubs.onlinelibrary.wiley.com/doi/pdf/10.1029/2023AV000881},
note = {e2023AV000881 2023AV000881},
year = {2023}
}

@article{Lucente2022coupling,
doi = {10.1088/1742-5468/ac7aa7},
url = {https://dx.doi.org/10.1088/1742-5468/ac7aa7},
year = {2022},
month = {aug},
publisher = {IOP Publishing and SISSA},
volume = {2022},
number = {8},
pages = {083201},
author = {Dario Lucente and Joran Rolland and Corentin Herbert and Freddy Bouchet},
title = {Coupling rare event algorithms with data-based learned committor functions using the analogue Markov chain},
journal = {Journal of Statistical Mechanics: Theory and Experiment}
}

@Article{Thompson2017high,
author={Thompson, Vikki
and Dunstone, Nick J.
and Scaife, Adam A.
and Smith, Doug M.
and Slingo, Julia M.
and Brown, Simon
and Belcher, Stephen E.},
title={High risk of unprecedented UK rainfall in the current climate},
journal={Nature Communications},
year={2017},
month={Jul},
day={24},
volume={8},
number={1},
pages={107},
issn={2041-1723},
doi={10.1038/s41467-017-00275-3},
url={https://doi.org/10.1038/s41467-017-00275-3}
}

@article{Mahesh2024hens2,
      title={Huge Ensembles Part II: Properties of a Huge Ensemble of Hindcasts Generated with Spherical Fourier Neural Operators}, 
      author={Ankur Mahesh and William Collins and Boris Bonev and Noah Brenowitz and Yair Cohen and Peter Harrington and Karthik Kashinath and Thorsten Kurth and Joshua North and Travis OBrien and Michael Pritchard and David Pruitt and Mark Risser and Shashank Subramanian and Jared Willard},
      year={2024},
      eprint={2408.01581},
      archivePrefix={arXiv},
      primaryClass={cs.LG},
      url={https://arxiv.org/abs/2408.01581}, 
}

@article{Vonich2024predictability,
author = {Vonich, P. Trent and Hakim, Gregory J.},
title = {Predictability Limit of the 2021 Pacific Northwest Heatwave From Deep-Learning Sensitivity Analysis},
journal = {Geophysical Research Letters},
volume = {51},
number = {19},
pages = {e2024GL110651},
doi = {https://doi.org/10.1029/2024GL110651},
url = {https://agupubs.onlinelibrary.wiley.com/doi/abs/10.1029/2024GL110651},
eprint = {https://agupubs.onlinelibrary.wiley.com/doi/pdf/10.1029/2024GL110651},
note = {e2024GL110651 2024GL110651},
year = {2024}
}

@article{Sapsis2020output,
author = {Sapsis, Themistoklis P. },
title = {Output-weighted optimal sampling for Bayesian regression and rare event statistics using few samples},
journal = {Proceedings of the Royal Society A: Mathematical, Physical and Engineering Sciences},
volume = {476},
number = {2234},
pages = {20190834},
year = {2020},
doi = {10.1098/rspa.2019.0834},

URL = {https://royalsocietypublishing.org/doi/abs/10.1098/rspa.2019.0834},
eprint = {https://royalsocietypublishing.org/doi/pdf/10.1098/rspa.2019.0834}
}

@article{
Mohamad2018sequential,
author = {Mustafa A. Mohamad  and Themistoklis P. Sapsis },
title = {Sequential sampling strategy for extreme event statistics in nonlinear dynamical systems},
journal = {Proceedings of the National Academy of Sciences},
volume = {115},
number = {44},
pages = {11138-11143},
year = {2018},
doi = {10.1073/pnas.1813263115},
URL = {https://www.pnas.org/doi/abs/10.1073/pnas.1813263115},
eprint = {https://www.pnas.org/doi/pdf/10.1073/pnas.1813263115}}

@article{Diaconescu2012singular,
title = {Singular vectors in atmospheric sciences: A review},
journal = {Earth-Science Reviews},
volume = {113},
number = {3},
pages = {161-175},
year = {2012},
issn = {0012-8252},
doi = {https://doi.org/10.1016/j.earscirev.2012.05.005},
url = {https://www.sciencedirect.com/science/article/pii/S0012825212000657},
author = {Emilia Paula Diaconescu and René Laprise}
}

@article{Ragone2021rare,
author = {Ragone, F. and Bouchet, F.},
title = {Rare Event Algorithm Study of Extreme Warm Summers and Heatwaves Over Europe},
journal = {Geophysical Research Letters},
volume = {48},
number = {12},
pages = {e2020GL091197},
doi = {https://doi.org/10.1029/2020GL091197},
url = {https://agupubs.onlinelibrary.wiley.com/doi/abs/10.1029/2020GL091197},
eprint = {https://agupubs.onlinelibrary.wiley.com/doi/pdf/10.1029/2020GL091197},
note = {e2020GL091197 2020GL091197},
year = {2021}
}

@article {Ragone2018computation,
	author = {Ragone, Francesco and Wouters, Jeroen and Bouchet, Freddy},
	title = {Computation of extreme heat waves in climate models using a large deviation algorithm},
	volume = {115},
	number = {1},
	pages = {24--29},
	year = {2018},
	doi = {10.1073/pnas.1712645115},
	publisher = {National Academy of Sciences},
	issn = {0027-8424},
	URL = {https://www.pnas.org/content/115/1/24},
	eprint = {https://www.pnas.org/content/115/1/24.full.pdf},
	journal = {Proceedings of the National Academy of Sciences}
}

@book{Pavliotis2014stochastic,
  title={Stochastic processes and applications: diffusion processes, the Fokker-Planck and Langevin equations},
  author={Pavliotis, Grigorios A},
  volume={60},
  year={2014},
  publisher={Springer}
}

@article{Kahn1951estimation,
  title={Estimation of particle transmission by random sampling},
  author={Kahn, Herman and Harris, Theodore E},
  journal={National Bureau of Standards applied mathematics series},
  volume={12},
  pages={27--30},
  year={1951}
}

@article{Zuckerman2017weighted,
author = {Zuckerman, Daniel M. and Chong, Lillian T.},
title = {Weighted Ensemble Simulation: Review of Methodology, Applications, and Software},
journal = {Annual Review of Biophysics},
volume = {46},
number = {1},
pages = {43-57},
year = {2017},
doi = {10.1146/annurev-biophys-070816-033834},
    note ={PMID: 28301772},

URL = { 
        https://doi.org/10.1146/annurev-biophys-070816-033834
    
},
eprint = { 
        https://doi.org/10.1146/annurev-biophys-070816-033834
    
}
}

@article{Lestang2018computing,
	doi = {10.1088/1742-5468/aab856},
	url = {https://doi.org/10.1088/1742-5468/aab856},
	year = 2018,
	month = {apr},
	publisher = {{IOP} Publishing},
	volume = {2018},
	number = {4},
	pages = {043213},
	author = {Thibault Lestang and Francesco Ragone and Charles-Edouard Br{\'{e}}hier and Corentin Herbert and Freddy Bouchet},
	title = {Computing return times or return periods with rare event algorithms},
	journal = {Journal of Statistical Mechanics: Theory and Experiment}
}

@article{Au2001estimation,
title = {Estimation of small failure probabilities in high dimensions by subset simulation},
journal = {Probabilistic Engineering Mechanics},
volume = {16},
number = {4},
pages = {263-277},
year = {2001},
issn = {0266-8920},
doi = {https://doi.org/10.1016/S0266-8920(01)00019-4},
url = {https://www.sciencedirect.com/science/article/pii/S0266892001000194},
author = {Siu-Kui Au and James L. Beck}
}

@article {Lorenz1998optimal,
      author = "Edward N. Lorenz and Kerry A. Emanuel",
      title = "Optimal Sites for Supplementary Weather Observations: Simulation with a Small Model",
      journal = "Journal of the Atmospheric Sciences",
      year = "1998",
      publisher = "American Meteorological Society",
      address = "Boston MA, USA",
      volume = "55",
      number = "3",
      doi = "10.1175/1520-0469(1998)055<0399:OSFSWO>2.0.CO;2",
      pages=      "399 - 414",
      url = "https://journals.ametsoc.org/view/journals/atsc/55/3/1520-0469_1998_055_0399_osfswo_2.0.co_2.xml"
}

@Article{Wang2020useful,
author={Wang, Qiang
and Mu, Mu
and Sun, Guodong},
title={A useful approach to sensitivity and predictability studies in geophysical fluid dynamics: conditional non-linear optimal perturbation},
journal={National Science Review},
year={2020},
month={Jan},
day={01},
volume={7},
number={1},
pages={214-223},
issn={2095-5138},
doi={10.1093/nsr/nwz039},
url={https://doi.org/10.1093/nsr/nwz039}
}

@Article{Galfi2017convergence,
author={G{\'a}lfi, Vera Melinda
and B{\'o}dai, Tam{\'a}s
and Lucarini, Valerio},
title={Convergence of Extreme Value Statistics in a Two-Layer Quasi-Geostrophic Atmospheric Model},
journal={Complexity},
year={2017},
month={Sep},
day={06},
publisher={Hindawi},
volume={2017},
pages={5340858},
issn={1076-2787},
doi={10.1155/2017/5340858},
url={https://doi.org/10.1155/2017/5340858}
}

@article{Qi2018predicting,
  title={Predicting extreme events for passive scalar turbulence in two-layer baroclinic flows through reduced-order stochastic models},
  author={Qi, Di and Majda, Andrew J},
  journal={Communications in Mathematical Sciences},
  volume={16},
  number={1},
  pages={17--51},
  year={2018},
  publisher={International Press of Boston}
}

@book{Coles2001introduction,
  title={An introduction to statistical modeling of extreme values},
  author={Coles, Stuart},
  series={Springer Series in Statistics},
  year={2001},
  publisher={Springer},
  doi={10.1007/978-1-4471-3675-0},
  isbn={978-1-85233-459-8},
  issn={0172-7397},
  edition={1},
}

@article{Cerou2007adaptive,
author = { Frédéric   Cérou  and  Arnaud   Guyader },
title = {Adaptive Multilevel Splitting for Rare Event Analysis},
journal = {Stochastic Analysis and Applications},
volume = {25},
number = {2},
pages = {417-443},
year  = {2007},
publisher = {Taylor \& Francis},
doi = {10.1080/07362990601139628},

URL = { 
    
        https://doi.org/10.1080/07362990601139628
    
    

},
eprint = { 
    
        https://doi.org/10.1080/07362990601139628
    
    

}
}

@article{Maiocchi2024heterogeneity,
title = {Heterogeneity of the attractor of the Lorenz ’96 model: Lyapunov analysis, unstable periodic orbits, and shadowing properties},
journal = {Physica D: Nonlinear Phenomena},
volume = {457},
pages = {133970},
year = {2024},
issn = {0167-2789},
doi = {https://doi.org/10.1016/j.physd.2023.133970},
url = {https://www.sciencedirect.com/science/article/pii/S016727892300324X},
author = {Chiara Cecilia Maiocchi and Valerio Lucarini and Andrey Gritsun and Yuzuru Sato}
}

@article{Webber2019practical,
author = {Webber,Robert J.  and Plotkin,David A.  and O’Neill,Morgan E  and Abbot,Dorian S.  and Weare,Jonathan },
title = {Practical rare event sampling for extreme mesoscale weather},
journal = {Chaos: An Interdisciplinary Journal of Nonlinear Science},
volume = {29},
number = {5},
pages = {053109},
year = {2019},
doi = {10.1063/1.5081461},

URL = { 
        https://doi.org/10.1063/1.5081461
    
},
eprint = { 
        https://doi.org/10.1063/1.5081461
    
}

}

@article {Haidvogel1980homogeneous,
      author = "Dale B.  Haidvogel and Isaac M.  Held",
      title = "Homogeneous Quasi-Geostrophic Turbulence Driven by a Uniform Temperature Gradient",
      journal = "Journal of Atmospheric Sciences",
      year = "1980",
      publisher = "American Meteorological Society",
      address = "Boston MA, USA",
      volume = "37",
      number = "12",
      doi = "10.1175/1520-0469(1980)037<2644:HQGTDB>2.0.CO;2",
      pages=      "2644 - 2660",
      url = "https://journals.ametsoc.org/view/journals/atsc/37/12/1520-0469_1980_037_2644_hqgtdb_2_0_co_2.xml"
}

@article {Panetta1993zonal,
      author = "R. Lee  Panetta",
      title = "Zonal Jets in Wide Baroclinically Unstable Regions: Persistence and Scale Selection",
      journal = "Journal of Atmospheric Sciences",
      year = "1993",
      publisher = "American Meteorological Society",
      address = "Boston MA, USA",
      volume = "50",
      number = "14",
      doi = "10.1175/1520-0469(1993)050<2073:ZJIWBU>2.0.CO;2",
      pages=      "2073 - 2106",
      url = "https://journals.ametsoc.org/view/journals/atsc/50/14/1520-0469_1993_050_2073_zjiwbu_2_0_co_2.xml"
}

@article {Thompson2010jet,
      author = "Andrew F. Thompson",
      title = "Jet Formation and Evolution in Baroclinic Turbulence with Simple Topography",
      journal = "Journal of Physical Oceanography",
      year = "2010",
      publisher = "American Meteorological Society",
      address = "Boston MA, USA",
      volume = "40",
      number = "2",
      doi = "10.1175/2009JPO4218.1",
      pages=      "257 - 278",
      url = "https://journals.ametsoc.org/view/journals/phoc/40/2/2009jpo4218.1.xml"
}

@Article{Fischer2023storylines,
author={Fischer, E. M.
and Beyerle, U.
and Bloin-Wibe, L.
and Gessner, C.
and Humphrey, V.
and Lehner, F.
and Pendergrass, A. G.
and Sippel, S.
and Zeder, J.
and Knutti, R.},
title={Storylines for unprecedented heatwaves based on ensemble boosting},
journal={Nature Communications},
year={2023},
month={Aug},
day={22},
volume={14},
number={1},
pages={4643},
issn={2041-1723},
doi={10.1038/s41467-023-40112-4},
url={https://doi.org/10.1038/s41467-023-40112-4}
}

@phdthesis{Gessner2022physical,
  title={Physical storylines for very rare climate extremes},
  author={Gessner, Claudia},
  year={2022},
  school={ETH Zurich}
}

@article {Gessner2021very,
      author = "Claudia Gessner and Erich M. Fischer and Urs Beyerle and Reto Knutti",
      title = "Very Rare Heat Extremes: Quantifying and Understanding Using Ensemble Reinitialization",
      journal = "Journal of Climate",
      year = "2021",
      publisher = "American Meteorological Society",
      address = "Boston MA, USA",
      volume = "34",
      number = "16",
      doi = "10.1175/JCLI-D-20-0916.1",
      pages=      "6619 - 6634",
      url = "https://journals.ametsoc.org/view/journals/clim/34/16/JCLI-D-20-0916.1.xml"
}

@article {Berner2015increasing,
      author = "J. Berner and K. R. Fossell and S.-Y. Ha and J. P. Hacker and C. Snyder",
      title = "Increasing the Skill of Probabilistic Forecasts: Understanding Performance Improvements from Model-Error Representations",
      journal = "Monthly Weather Review",
      year = "2015",
      publisher = "American Meteorological Society",
      address = "Boston MA, USA",
      volume = "143",
      number = "4",
      doi = "10.1175/MWR-D-14-00091.1",
      pages=      "1295 - 1320",
      url = "https://journals.ametsoc.org/view/journals/mwre/143/4/mwr-d-14-00091.1.xml"
}

@article{Linz2020framework,
author = {Linz, Marianna and Chen, Gang and Zhang, Boer and Zhang, Pengfei},
title = {A Framework for Understanding How Dynamics Shape Temperature Distributions},
journal = {Geophysical Research Letters},
volume = {47},
number = {4},
pages = {e2019GL085684},
doi = {https://doi.org/10.1029/2019GL085684},
url = {https://agupubs.onlinelibrary.wiley.com/doi/abs/10.1029/2019GL085684},
eprint = {https://agupubs.onlinelibrary.wiley.com/doi/pdf/10.1029/2019GL085684},
note = {e2019GL085684 10.1029/2019GL085684},
year = {2020}
}

@article{Neelin2010long,
author = {Neelin, J. David and Lintner, Benjamin R. and Tian, Baijun and Li, Qinbin and Zhang, Li and Patra, Prabir K. and Chahine, Moustafa T. and Stechmann, Samuel N.},
title = {Long tails in deep columns of natural and anthropogenic tropospheric tracers},
journal = {Geophysical Research Letters},
volume = {37},
number = {5},
pages = {},
doi = {https://doi.org/10.1029/2009GL041726},
url = {https://agupubs.onlinelibrary.wiley.com/doi/abs/10.1029/2009GL041726},
eprint = {https://agupubs.onlinelibrary.wiley.com/doi/pdf/10.1029/2009GL041726},
year = {2010}
}

@article{Jalbert2024extremes,
 title={Extremes.jl: Extreme Value Analysis in Julia},
 volume={109},
 url={https://www.jstatsoft.org/index.php/jss/article/view/v109i06},
 doi={10.18637/jss.v109.i06},
 number={6},
 journal={Journal of Statistical Software},
 author={Jalbert, Jonathan and Farmer, Marilou and Gobeil, Gabriel and Roy, Philippe},
 year={2024},
 pages={1–35}
}

@phdthesis{Noyelle2024statistical,
  TITLE = {{Statistical and dynamical aspects of extreme heatwaves in the mid-latitudes}},
  AUTHOR = {Noyelle, Robin},
  URL = {https://hal.science/tel-04632646},
  NUMBER = {2024UPASJ013},
  SCHOOL = {{Universit{\'e} Paris-Saclay}},
  YEAR = {2024},
  MONTH = Jun,
  TYPE = {Theses},
  HAL_ID = {tel-04632646},
  HAL_VERSION = {v2},
}

@book{Leobacher2014quasi,
  title={Introduction to quasi-Monte Carlo integration and applications},
  author={Leobacher, Gunther and Pillichshammer, Friedrich},
  year={2014},
  publisher={Springer}
}

@book{Lucarini2016extremes,
  title={Extremes and recurrence in dynamical systems},
  author={Lucarini, Valerio and Faranda, Davide and de Freitas, Jorge Miguel Milhazes and Holland, Mark and Kuna, Tobias and Nicol, Matthew and Todd, Mike and Vaienti, Sandro and others},
  year={2016},
  publisher={John Wiley \& Sons}
}

@Article{BloinWibe2025estimating,
AUTHOR = {Bloin-Wibe, L. and Noyelle, R. and Humphrey, V. and Beyerle, U. and Knutti, R. and Fischer, E.},
TITLE = {Estimating return periods for extreme events in climate models through Ensemble Boosting},
JOURNAL = {EGUsphere},
VOLUME = {2025},
YEAR = {2025},
PAGES = {1--40},
URL = {https://egusphere.copernicus.org/preprints/2025/egusphere-2025-525/},
DOI = {10.5194/egusphere-2025-525}
}

@article{Giorgini2024response,
  title = {Response Theory via Generative Score Modeling},
  author = {Giorgini, Ludovico Theo and Deck, Katherine and Bischoff, Tobias and Souza, Andre},
  journal = {Phys. Rev. Lett.},
  volume = {133},
  issue = {26},
  pages = {267302},
  numpages = {7},
  year = {2024},
  month = {Dec},
  publisher = {American Physical Society},
  doi = {10.1103/PhysRevLett.133.267302},
  url = {https://link.aps.org/doi/10.1103/PhysRevLett.133.267302}
}

@article{Farazmand2017variational,
author = {Mohammad Farazmand  and Themistoklis P. Sapsis },
title = {A variational approach to probing extreme events in turbulent dynamical systems},
journal = {Science Advances},
volume = {3},
number = {9},
pages = {e1701533},
year = {2017},
doi = {10.1126/sciadv.1701533},
URL = {https://www.science.org/doi/abs/10.1126/sciadv.1701533},
eprint = {https://www.science.org/doi/pdf/10.1126/sciadv.1701533}}

@article{Yang2022output,
author = {Yang, Yibo  and Blanchard, Antoine  and Sapsis, Themistoklis  and Perdikaris, Paris },
title = {Output-weighted sampling for multi-armed bandits with extreme payoffs},
journal = {Proceedings of the Royal Society A: Mathematical, Physical and Engineering Sciences},
volume = {478},
number = {2260},
pages = {20210781},
year = {2022},
doi = {10.1098/rspa.2021.0781},

URL = {https://royalsocietypublishing.org/doi/abs/10.1098/rspa.2021.0781},
eprint = {https://royalsocietypublishing.org/doi/pdf/10.1098/rspa.2021.0781}
}

@article{Dematteis2019extreme,
author = {Dematteis, Giovanni and Grafke, Tobias and Vanden-Eijnden, Eric},
title = {Extreme Event Quantification in Dynamical Systems with Random Components},
journal = {SIAM/ASA Journal on Uncertainty Quantification},
volume = {7},
number = {3},
pages = {1029-1059},
year = {2019},
doi = {10.1137/18M1211003},

URL = { 
    
        https://doi.org/10.1137/18M1211003
    
    

},
eprint = { 
    
        https://doi.org/10.1137/18M1211003
    
    

}
}

@Article{Yiou2020simulation,
AUTHOR = {Yiou, P. and J\'ez\'equel, A.},
TITLE = {Simulation of extreme heat waves with empirical importance sampling},
JOURNAL = {Geoscientific Model Development},
VOLUME = {13},
YEAR = {2020},
NUMBER = {2},
PAGES = {763--781},
URL = {https://gmd.copernicus.org/articles/13/763/2020/},
DOI = {10.5194/gmd-13-763-2020}
}

@article{Qi2016predicting,
  title={Predicting fat-tailed intermittent probability distributions in passive scalar turbulence with imperfect models through empirical information theory},
  author={Qi, Di and Majda, Andrew J},
  journal={Communications in Mathematical Sciences},
  volume={14},
  number={6},
  pages={1687--1722},
  year={2016},
  publisher={International Press of Boston}
}

@Article{Bourlioux2002elementary,
author={Bourlioux, A.
and Majda, A. J.},
title={Elementary models with probability distribution function intermittency for passive scalars with a mean gradient},
journal={Physics of Fluids},
year={2002},
month={Feb},
day={01},
volume={14},
number={2},
pages={881-897},
issn={1070-6631},
doi={10.1063/1.1430736},
url={https://doi.org/10.1063/1.1430736}
}

@article {Dool1989new,
      author = "H. M.  van den Dool",
      title = "A New Look at Weather Forecasting through Analogues",
      journal = "Monthly Weather Review",
      year = "1989",
      publisher = "American Meteorological Society",
      address = "Boston MA, USA",
      volume = "117",
      number = "10",
      doi = "10.1175/1520-0493(1989)117<2230:ANLAWF>2.0.CO;2",
      pages=      "2230 - 2247",
      url = "https://journals.ametsoc.org/view/journals/mwre/117/10/1520-0493_1989_117_2230_anlawf_2_0_co_2.xml"
}

@inproceedings{Vandal2017deepsd,
author = {Vandal, Thomas and Kodra, Evan and Ganguly, Sangram and Michaelis, Andrew and Nemani, Ramakrishna and Ganguly, Auroop R.},
title = {DeepSD: Generating High Resolution Climate Change Projections through Single Image Super-Resolution},
year = {2017},
isbn = {9781450348874},
publisher = {Association for Computing Machinery},
address = {New York, NY, USA},
url = {https://doi.org/10.1145/3097983.3098004},
doi = {10.1145/3097983.3098004},
booktitle = {Proceedings of the 23rd ACM SIGKDD International Conference on Knowledge Discovery and Data Mining},
pages = {1663–1672},
numpages = {10},
location = {Halifax, NS, Canada},
series = {KDD '17}
}

@article {Penland1993prediction,
      author = "Cécile  Penland and Theresa  Magorian",
      title = "Prediction of Niño 3 Sea Surface Temperatures Using Linear Inverse Modeling",
      journal = "Journal of Climate",
      year = "1993",
      publisher = "American Meteorological Society",
      address = "Boston MA, USA",
      volume = "6",
      number = "6",
      doi = "10.1175/1520-0442(1993)006<1067:PONSST>2.0.CO;2",
      pages=      "1067 - 1076",
      url = "https://journals.ametsoc.org/view/journals/clim/6/6/1520-0442_1993_006_1067_ponsst_2_0_co_2.xml"
}

@article{Saha2024statistical,
author = {Saha, Anamitra and Ravela, Sai},
title = {Statistical-Physical Adversarial Learning From Data and Models for Downscaling Rainfall Extremes},
journal = {Journal of Advances in Modeling Earth Systems},
volume = {16},
number = {6},
pages = {e2023MS003860},
doi = {https://doi.org/10.1029/2023MS003860},
url = {https://agupubs.onlinelibrary.wiley.com/doi/abs/10.1029/2023MS003860},
eprint = {https://agupubs.onlinelibrary.wiley.com/doi/pdf/10.1029/2023MS003860},
note = {e2023MS003860 2023MS003860},
year = {2024}
}

@article{Tebaldi2020emulating,
doi = {10.1088/1748-9326/ab8332},
url = {https://dx.doi.org/10.1088/1748-9326/ab8332},
year = {2020},
month = {jun},
publisher = {IOP Publishing},
volume = {15},
number = {7},
pages = {074006},
author = {Tebaldi, C and Armbruster, A and Engler, H P and Link, R},
title = {Emulating climate extreme indices},
journal = {Environmental Research Letters}
}

@article{Pons2024simulating,
title = {Simulating the Western North America heatwave of 2021 with analogue importance sampling},
journal = {Weather and Climate Extremes},
volume = {43},
pages = {100651},
year = {2024},
issn = {2212-0947},
doi = {https://doi.org/10.1016/j.wace.2024.100651},
url = {https://www.sciencedirect.com/science/article/pii/S2212094724000124},
author = {Flavio Maria Emanuele Pons and Pascal Yiou and Aglaé Jézéquel and Gabriele Messori}
}

@Article{Ghil2011extreme,
AUTHOR = {Ghil, M. and Yiou, P. and Hallegatte, S. and Malamud, B. D. and Naveau, P. and Soloviev, A. and Friederichs, P. and Keilis-Borok, V. and Kondrashov, D. and Kossobokov, V. and Mestre, O. and Nicolis, C. and Rust, H. W. and Shebalin, P. and Vrac, M. and Witt, A. and Zaliapin, I.},
TITLE = {Extreme events: dynamics, statistics and prediction},
JOURNAL = {Nonlinear Processes in Geophysics},
VOLUME = {18},
YEAR = {2011},
NUMBER = {3},
PAGES = {295--350},
URL = {https://npg.copernicus.org/articles/18/295/2011/},
DOI = {10.5194/npg-18-295-2011}
}

@article{Boulaguiem202modeling,
    title={Modeling and simulating spatial extremes by combining extreme value theory with generative adversarial networks},
    volume={1},
    DOI={10.1017/eds.2022.4},
    journal={Environmental Data Science},
    author={Boulaguiem, Younes and Zscheischler, Jakob and Vignotto, Edoardo and van der Wiel, Karin and Engelke, Sebastian},
    year={2022},
    pages={e5}
}

@article{Huser2025modeling,
    title={Modeling of spatial extremes in environmental data science: time to move away from max-stable processes},
    volume={4},
    DOI={10.1017/eds.2024.54},
    journal={Environmental Data Science},
    author={Huser, Raphaël and Opitz, Thomas and Wadsworth, Jennifer L.},
    year={2025},
    pages={e3}
}

@article{Huser2020advances,
author = {Huser, Raphaël and Wadsworth, Jennifer L.},
title = {Advances in statistical modeling of spatial extremes},
journal = {WIREs Computational Statistics},
volume = {14},
number = {1},
pages = {e1537},
doi = {https://doi.org/10.1002/wics.1537},
url = {https://wires.onlinelibrary.wiley.com/doi/abs/10.1002/wics.1537},
eprint = {https://wires.onlinelibrary.wiley.com/doi/pdf/10.1002/wics.1537},
year = {2022}
}

@article{Rampal2025reliable,
author = {Rampal, Neelesh and Gibson, Peter B. and Sherwood, Steven and Abramowitz, Gab and Hobeichi, Sanaa},
title = {A Reliable Generative Adversarial Network Approach for Climate Downscaling and Weather Generation},
journal = {Journal of Advances in Modeling Earth Systems},
volume = {17},
number = {1},
pages = {e2024MS004668},
doi = {https://doi.org/10.1029/2024MS004668},
url = {https://agupubs.onlinelibrary.wiley.com/doi/abs/10.1029/2024MS004668},
eprint = {https://agupubs.onlinelibrary.wiley.com/doi/pdf/10.1029/2024MS004668},
note = {e2024MS004668 2024MS004668},
year = {2025}
}

@article{John2022quantifying,
title = {Quantifying CMIP6 model uncertainties in extreme precipitation projections},
journal = {Weather and Climate Extremes},
volume = {36},
pages = {100435},
year = {2022},
issn = {2212-0947},
doi = {https://doi.org/10.1016/j.wace.2022.100435},
url = {https://www.sciencedirect.com/science/article/pii/S2212094722000238},
author = {Amal John and Hervé Douville and Aurélien Ribes and Pascal Yiou}
}

@article {OGorman2009scaling,
      author = "Paul A. O’Gorman and Tapio Schneider",
      title = "Scaling of Precipitation Extremes over a Wide Range of Climates Simulated with an Idealized GCM",
      journal = "Journal of Climate",
      year = "2009",
      publisher = "American Meteorological Society",
      address = "Boston MA, USA",
      volume = "22",
      number = "21",
      doi = "10.1175/2009JCLI2701.1",
      pages=      "5676 - 5685",
      url = "https://journals.ametsoc.org/view/journals/clim/22/21/2009jcli2701.1.xml"
}

@article{Watt2024generative,
      title={Generative Diffusion-based Downscaling for Climate}, 
      author={Robbie A. Watt and Laura A. Mansfield},
      year={2024},
      eprint={2404.17752},
      archivePrefix={arXiv},
      primaryClass={physics.ao-ph},
      url={https://arxiv.org/abs/2404.17752}, 
}

@article{Sundar2024taudiff,
      title={TAUDiff: Improving statistical downscaling for extreme weather events using generative diffusion models}, 
      author={Rahul Sundar and Nishant Parashar and Antoine Blanchard and Boyko Dodov},
      year={2024},
      eprint={2412.13627},
      archivePrefix={arXiv},
      primaryClass={cs.LG},
      url={https://arxiv.org/abs/2412.13627}, 
}

@article{Breitung2021sorm,
author = {Karl Breitung },
title = {SORM, Design Points, Subset Simulation, and Markov Chain Monte Carlo},
journal = {ASCE-ASME Journal of Risk and Uncertainty in Engineering Systems, Part A: Civil Engineering},
volume = {7},
number = {4},
pages = {04021052},
year = {2021},
doi = {10.1061/AJRUA6.0001166},

URL = {https://ascelibrary.org/doi/abs/10.1061/AJRUA6.0001166},
eprint = {https://ascelibrary.org/doi/pdf/10.1061/AJRUA6.0001166}
}

@Article{Kekem2018wave,
AUTHOR = {van Kekem, D. L. and Sterk, A. E.},
TITLE = {Wave propagation in the Lorenz-96 model},
JOURNAL = {Nonlinear Processes in Geophysics},
VOLUME = {25},
YEAR = {2018},
NUMBER = {2},
PAGES = {301--314},
URL = {https://npg.copernicus.org/articles/25/301/2018/},
DOI = {10.5194/npg-25-301-2018}
}

@Article{Lucarini2020new,
author={Lucarini, Valerio
and Gritsun, Andrey},
title={A new mathematical framework for atmospheric blocking events},
journal={Climate Dynamics},
year={2020},
month={Jan},
day={01},
volume={54},
number={1},
pages={575-598},
issn={1432-0894},
doi={10.1007/s00382-019-05018-2},
url={https://doi.org/10.1007/s00382-019-05018-2}
}

@article{Farrell1996generalized1, 
      author = "Brian F.  Farrell and Petros J.  Ioannou",
      title = "Generalized Stability Theory. Part I: Autonomous Operators",
      journal = "Journal of Atmospheric Sciences",
      year = "1996",
      publisher = "American Meteorological Society",
      address = "Boston MA, USA",
      volume = "53",
      number = "14",
      doi = "10.1175/1520-0469(1996)053<2025:GSTPIA>2.0.CO;2",
      pages=      "2025 - 2040",
      url = "https://journals.ametsoc.org/view/journals/atsc/53/14/1520-0469_1996_053_2025_gstpia_2_0_co_2.xml"
}

@article {Farrell1996generalized2,
      author = "Brian F.  Farrell and Petros J.  Ioannou",
      title = "Generalized Stability Theory. Part II: Nonautonomous Operators",
      journal = "Journal of Atmospheric Sciences",
      year = "1996",
      publisher = "American Meteorological Society",
      address = "Boston MA, USA",
      volume = "53",
      number = "14",
      doi = "10.1175/1520-0469(1996)053<2041:GSTPIN>2.0.CO;2",
      pages=      "2041 - 2053",
      url = "https://journals.ametsoc.org/view/journals/atsc/53/14/1520-0469_1996_053_2041_gstpin_2_0_co_2.xml"
}

@software{Rackauckas2023qmc,
  author       = {Rackauckas, Chris},
  title        = {QuasiMonteCarlo.jl},
  year         = {2023},
  version      = {v0.3.3j},
  url          = {https://github.com/SciML/QuasiMonteCarlo.jl},
  note         = {Accessed: 2025-05-09}
}

@Article{Huang2016estimating,
AUTHOR = {Huang, W. K. and Stein, M. L. and McInerney, D. J. and Sun, S. and Moyer, E. J.},
TITLE = {Estimating changes in temperature extremes from millennial-scale climate simulations using generalized extreme value (GEV) distributions},
JOURNAL = {Advances in Statistical Climatology, Meteorology and Oceanography},
VOLUME = {2},
YEAR = {2016},
NUMBER = {1},
PAGES = {79--103},
URL = {https://ascmo.copernicus.org/articles/2/79/2016/},
DOI = {10.5194/ascmo-2-79-2016}
}

@misc{Zuev2015subset,
      title={Subset Simulation Method for Rare Event Estimation: An Introduction}, 
      author={Konstantin Zuev},
      year={2015},
      eprint={1505.03506},
      archivePrefix={arXiv},
      primaryClass={stat.CO}
}

@article{Baars2021application,
title = {Application of adaptive multilevel splitting to high-dimensional dynamical systems},
journal = {Journal of Computational Physics},
volume = {424},
pages = {109876},
year = {2021},
issn = {0021-9991},
doi = {https://doi.org/10.1016/j.jcp.2020.109876},
url = {https://www.sciencedirect.com/science/article/pii/S0021999120306501},
author = {S. Baars and D. Castellana and F.W. Wubs and H.A. Dijkstra}
}

@article{Finkel2026rare,
author = {Finkel, Justin and O’Gorman, Paul A.},
title = {Rare Event Sampling for Moving Targets: Extremes of Temperature and Daily Precipitation in a General Circulation Model},
journal = {Journal of Advances in Modeling Earth Systems},
volume = {18},
number = {3},
pages = {e2025MS005456},
doi = {https://doi.org/10.1029/2025MS005456},
url = {https://agupubs.onlinelibrary.wiley.com/doi/abs/10.1029/2025MS005456},
eprint = {https://agupubs.onlinelibrary.wiley.com/doi/pdf/10.1029/2025MS005456},
note = {e2025MS005456 2025MS005456},
year = {2026}
}

@Article{Pickering2022discovering,
author={Pickering, Ethan
and Guth, Stephen
and Karniadakis, George Em
and Sapsis, Themistoklis P.},
title={Discovering and forecasting extreme events via active learning in neural operators},
journal={Nature Computational Science},
year={2022},
month={Dec},
day={01},
volume={2},
number={12},
pages={823-833},
issn={2662-8457},
doi={10.1038/s43588-022-00376-0},
url={https://doi.org/10.1038/s43588-022-00376-0}
}

@article{Rolland2022collapse,
    title={Collapse of transitional wall turbulence captured using a rare events algorithm},
    volume={931},
    DOI={10.1017/jfm.2021.957},
    journal={Journal of Fluid Mechanics},
    author={Rolland, Joran}, 
    year={2022}, 
    pages={A22}
}

@ARTICLE{Kabir2018neural,
  author={Kabir, H. M. Dipu and Khosravi, Abbas and Hosen, Mohammad Anwar and Nahavandi, Saeid},
  journal={IEEE Access}, 
  title={Neural Network-Based Uncertainty Quantification: A Survey of Methodologies and Applications}, 
  year={2018},
  volume={6},
  number={},
  pages={36218-36234},
  doi={10.1109/ACCESS.2018.2836917}}

@software{COAST2025,
  author       = {justinfocus12},
  title        = {justinfocus12/COAST: Initial release for submission of BEST COAST paper to NPG},
  month        = oct,
  year         = 2025,
  publisher    = {Zenodo},
  version      = {v0.1},
  doi          = {10.5281/zenodo.17355215},
  url          = {https://doi.org/10.5281/zenodo.17355215},
  swhid        = {swh:1:dir:83fbe67bfe90bd6842c0d21c921d50257ffd0a5b
                   ;origin=https://doi.org/10.5281/zenodo.17355214;vi
                   sit=swh:1:snp:1f0d152c6ee885ee0b70542f7bdf956848e4
                   279c;anchor=swh:1:rel:41633d33a83c5feb9e3450246448
                   1bfcf7ee5773;path=justinfocus12-COAST-8f452aa
                  },
}

@article{Phillips1956general,
author = {Phillips, Norman A.},
title = {The general circulation of the atmosphere: A numerical experiment},
journal = {Quarterly Journal of the Royal Meteorological Society},
volume = {82},
number = {352},
pages = {123-164},
doi = {https://doi.org/10.1002/qj.49708235202},
url = {https://rmets.onlinelibrary.wiley.com/doi/abs/10.1002/qj.49708235202},
eprint = {https://rmets.onlinelibrary.wiley.com/doi/pdf/10.1002/qj.49708235202},
year = {1956}
}

@article {Lapeyre2004role,
      author = "G. Lapeyre and I. M. Held",
      title = "The Role of Moisture in the Dynamics and Energetics of Turbulent Baroclinic Eddies",
      journal = "Journal of the Atmospheric Sciences",
      year = "2004",
      publisher = "American Meteorological Society",
      address = "Boston MA, USA",
      volume = "61",
      number = "14",
      doi = "10.1175/1520-0469(2004)061<1693:TROMIT>2.0.CO;2",
      pages=      "1693 - 1710",
      url = "https://journals.ametsoc.org/view/journals/atsc/61/14/1520-0469_2004_061_1693_tromit_2.0.co_2.xml"
}

@article{Castaing1989scaling,
    title={Scaling of hard thermal turbulence in Rayleigh-Bénard convection},
    volume={204}, 
    DOI={10.1017/S0022112089001643},
    journal={Journal of Fluid Mechanics},
    author={Castaing, Bernard and Gunaratne, Gemunu and Heslot, François and Kadanoff, Leo and Libchaber, Albert and Thomae, Stefan and Wu, Xiao-Zhong and Zaleski, Stéphane and Zanetti, Gianluigi},
    year={1989},
    pages={1–30}
}

@article{Gollub1991fluctuations,
  title = {Fluctuations and transport in a stirred fluid with a mean gradient},
  author = {Gollub, J. P. and Clarke, J. and Gharib, M. and Lane, B. and Mesquita, O. N.},
  journal = {Phys. Rev. Lett.},
  volume = {67},
  issue = {25},
  pages = {3507--3510},
  numpages = {0},
  year = {1991},
  month = {Dec},
  publisher = {American Physical Society},
  doi = {10.1103/PhysRevLett.67.3507},
  url = {https://link.aps.org/doi/10.1103/PhysRevLett.67.3507}
}

@article{Pumir1991exponential,
  title = {Exponential tails and random advection},
  author = {Pumir, Alain and Shraiman, Boris I. and Siggia, Eric D.},
  journal = {Phys. Rev. Lett.},
  volume = {66},
  issue = {23},
  pages = {2984--2987},
  numpages = {0},
  year = {1991},
  month = {Jun},
  publisher = {American Physical Society},
  doi = {10.1103/PhysRevLett.66.2984},
  url = {https://link.aps.org/doi/10.1103/PhysRevLett.66.2984}
}

@article{Norwood2013lyapunov,
doi = {10.1088/1751-8113/46/25/254021},
url = {https://dx.doi.org/10.1088/1751-8113/46/25/254021},
year = {2013},
month = {jun},
publisher = {IOP Publishing},
volume = {46},
number = {25},
pages = {254021},
author = {Adrienne Norwood and Eugenia Kalnay and Kayo Ide and Shu-Chih Yang and Christopher Wolfe},
title = {Lyapunov, singular and bred vectors in a multi-scale system: an empirical exploration of vectors related to instabilities},
journal = {Journal of Physics A: Mathematical and Theoretical}
}

@misc{Whittaker2025constructing,
      title={Constructing Extreme Heatwave Storylines with Differentiable Climate Models}, 
      author={Tim Whittaker and Alejandro Di Luca},
      year={2025},
      eprint={2506.10660},
      archivePrefix={arXiv},
      primaryClass={physics.ao-ph},
      url={https://arxiv.org/abs/2506.10660}, 
}

@article{Tong2021extreme,
  title={Extreme event probability estimation using PDE-constrained optimization and large deviation theory, with application to tsunamis},
  author={Tong, Shanyin and Vanden-Eijnden, Eric and Stadler, Georg},
  journal={Communications in Applied Mathematics and Computational Science},
  volume={16},
  number={2},
  pages={181--225},
  year={2021},
  publisher={Mathematical Sciences Publishers}
}

@article{Blonigan2019are,
  title = {Are extreme dissipation events predictable in turbulent fluid flows?},
  author = {Blonigan, Patrick J. and Farazmand, Mohammad and Sapsis, Themistoklis P.},
  journal = {Phys. Rev. Fluids},
  volume = {4},
  issue = {4},
  pages = {044606},
  numpages = {21},
  year = {2019},
  month = {Apr},
  publisher = {American Physical Society},
  doi = {10.1103/PhysRevFluids.4.044606},
  url = {https://link.aps.org/doi/10.1103/PhysRevFluids.4.044606}
}

\end{document}